\documentclass[iop, apj]{emulateapj}

 \usepackage{graphicx}
\usepackage{natbib}
\usepackage{footnote}
\usepackage{amsmath}
\usepackage{color,hyperref}

\shorttitle{Tidal Dissipation and Obliquity Evolution in Hot Jupiter Systems}
\shortauthors{Valsecchi \& Rasio.}

\begin{document}
   \title{Tidal Dissipation and Obliquity Evolution in Hot Jupiter Systems}

\author{Francesca Valsecchi\altaffilmark{1}, and Frederic A. Rasio\altaffilmark{1}}

\altaffiltext{1}{Center for Interdisciplinary Exploration and Research in Astrophysics (CIERA), and Northwestern University, Department of Physics and Astronomy, 2145 Sheridan Road, Evanston, IL 60208, USA.}


\begin{abstract}
Two formation scenarios have been proposed to explain the tight orbits of hot Jupiters. They could be formed in orbits with a small inclination (with respect to the stellar spin) via disk migration, or in more highly inclined orbits via high-eccentricity migration, where gravitational interactions with a companion and tidal dissipation are at play.
Here we target hot Jupiter systems where the misalignment $\lambda$ has been inferred observationally and we investigate whether their properties are consistent with high-eccentricity migration. Specifically, we study whether stellar tides can be responsible for the observed distribution of $\lambda$ and orbital separations.
Improving on previous studies, we use detailed models for each star, thus accounting for how convection (and tidal dissipation) depends on stellar properties.
In line with observations suggesting that hotter stars have higher $\lambda$, we find that $\lambda$ increases as the amount of stellar surface convection decreases. This trend supports the hypothesis that tides are the mechanism shaping the observed distribution of $\lambda$.
Furthermore, we study the past orbital evolution of five representative systems, chosen to cover a variety of temperatures and misalignments. We consider various initial orbital configurations and integrate the equations describing the coupled evolution of the orbital separation, stellar spin, and misalignment.  We account for stellar tides and wind mass loss, stellar evolution, and magnetic braking.
We show that the current properties of these five representative systems can be explained naturally, given our current understanding of tidal dissipation and with physically motivated assumptions for the effects driving the orbital evolution.
\end{abstract}

\keywords{Planetary Systems: planet-star interactions--planets and satellites: gaseous planets--stars: evolution--stars: general--(stars:) planetary systems}

%

\section{Introduction} \label{Intro}
The plethora of exoplanets discovered in recent years has revealed that planetary systems exist in a much greater variety than we had ever imagined.
To date, more than 1000 exoplanets have been confirmed using different observational techniques. Almost 200 of these planets are similar in mass to Jupiter, but revolve around their parent star every 10\,days or less (NASA Exoplanet Archive), thus challenging our understanding of planet formation and evolution. Different scenarios have been proposed to explain how these so-called {\it hot Jupiters} formed in such tight orbits.
One way to distinguish between these models is to investigate the current properties of the many discovered systems.
Our focus here lies on systems where the obliquity $\lambda$ (the sky-projected angle between the stellar spin and orbital angular momentum vectors) has been constrained observationally.

Two migration models have been invoked to bring gas giants from their birthplace at several AU into the tight orbits we observe today: {\it disk migration} and {\it high-eccentricity migration} (however, see also \citealt{TutukovFedorova2012, Thies+11}). These models predict different orientations of the planet's orbit at present.
In the disk migration scenario, planets could migrate inwards through their interactions with the protoplanetary gas disk \citep{GoldreichTremaine80,Ward97, MurrayHHT98, Lin+96,Guillochon+11}. As the disks tend to damp the orbital inclination \citep{Cresswell+07,XiangPapaloizou14}, this model would naturally lead to nearly circular orbits and small obliquities (e.g., \citealt{GoldreichTremaine80,PapaloizouLarwood00}). In the high-eccentricity migration scenario, gravitational interactions either between several planets or with companion stars could lead to highly eccentric orbits and high obliquities (\citealt{Kozai62, Lidov62, WuLithwick11,Naoz+11,Nagasawa08,FabryckyTremaine07,WuMurray03,RasioFord96, ChatterjeeFMR08}; see also \citealt{PlavchanBilinski13} for empirical evidence).  As tidal dissipation tends to damp the eccentricity while decreasing the orbital separation, close-in planets could then result from tidal circularization. 

\cite{WinnFAJ10} used a sample of 19 systems in which the projected spin-orbit angle $\lambda$ was measured 
via the Rossiter-McLaughlin (RM) effect. The RM effect occurs when a transiting planet blocks the blue- or red-shifted part (or both, depending on the orbital inclination with respect to the stellar spin) of the spinning star as it passes across the stellar disk, thus distorting the star's spectral line profile.
\cite{WinnFAJ10} investigated the behavior of the sky-projected misalignment as a function of the host star's effective temperature ($T_{\rm eff *}$). Their results suggested that the degree of misalignment increases for hotter stars. In particular, a sharp increase in $\lambda$ seems to occur at $T_{\rm eff *}\,\simeq\,$6250~K. These findings were later confirmed by \citeauthor{Albrecht+12} (\citeyear{Albrecht+12}, hereafter A12) with a sample of RM measurements twice as large as the one available to \cite{WinnFAJ10}; see also e.g., \cite{MortonJohnson11}. 
Since $\simeq\,$6250~K is the temperature at which the outer convective zone in main sequence stars starts becoming negligible, \cite{WinnFAJ10} proposed that the mechanism responsible for the trend observed in the data is convective dissipation of tides in the star.  Hot Jupiters could then be produced via a single formation process yielding a broad distribution of obliquities. Later on, tidal dissipation in cool stars would damp the obliquity within a few Gyr. Correspondingly, the high degree of sky-projected misalignment observed in hot stars would result from the inefficiency of tidal dissipation. While its simplicity is appealing, this scenario presents a major weakness. In fact,  tidal dissipation in the star acts both on the misalignment and on the orbital separation, causing orbital decay whenever the stellar spin frequency is lower than the orbital frequency, and it thus fails in explaining the currently observed aligned hot Jupiters (see, e.g., \citealt{RogersLin13}, hereafter R13). 

Possible solutions to this evolutionary conundrum were presented by \cite{WinnFAJ10} and \cite{Lai12}.  \cite{WinnFAJ10}  suggested that, if the star's radiative interior is weakly coupled to the outer convective region and to the planet, then tides would act on the obliquity faster than on the orbital separation (see also A12). However, a large amount of differential rotation inside the star would potentially lead to fluid instabilities, which would tend to quickly re-couple the star's convective and radiative regions. To overcome this problem, \cite{Lai12} presented a different scenario, following the idea that different physical processes dissipate tides with different efficiencies. His model invokes the excitation and damping of inertial waves in a stellar convective zone (see \S~\ref{Tidal Contribution due to inertial wave dissipation} for a summary). These waves are driven by the Coriolis force and are excited only in misaligned systems. In this configuration, the tidal potential to the leading quadrupole order has several terms. Each of these terms generates tidal disturbances with its own dissipation efficiency.
Among these terms, \cite{Lai12} identified a component of the tidal torque which acts only on the misalignment without affecting the orbital separation, thus providing a more efficient mechanism to modify the misalignment.

The validity of this prescription was recently questioned by R13, who  considered a random distribution of initial obliquities for 50 objects (nearly the number of observed systems considered by A12) and integrated the equations derived by \cite{Lai12} forward in time. The authors computed the evolution of the misalignment {\it alone} while keeping the orbit and stellar spin fixed and found that  tides would lead to a nearly equal amount of prograde ($\lambda \textless 90^{o}$),  retrograde ($\lambda \textgreater 90^{o}$), and 90$^{o}$ orbits. This appears inconsistent with the observations, as the majority of observed obliquities are smaller than $90^{o}$.  However, we note that R13's investigation has two major limitations. First, it neglects the simultaneous evolution of the orbital separation, stellar spin, and misalignment, while previous investigations have shown that it is essential to consider the coupled evolution of the orbital elements and spins (e.g., \citealt{Jackson+08,BarkerOgilvie2009,MatsumuraPR2010,Xue+14}). Furthermore, it does not account for the various physical effects that might compete in the evolution of the system (e.g., magnetic braking and the radial expansion of the star as a result of stellar evolution). This last simplification was recently adopted also by \cite{Xue+14}, who integrated the full set of equations presented by \cite{Lai12} and showed that all intermediate states found by R13 eventually evolve towards alignment. While this appears to be consistent with the majority of observed obliquities ($\textless\,90^{o}$), it cannot explain the currently observed intermediate misalignments.  

In this paper, we reconsider tidal dissipation in the star as a possible mechanism responsible for the observed distribution of misalignments and orbital separations. We carefully examine the observed relation between $\lambda$, $T_{\rm eff *}$, and the amount of convection inside the host star. In contrast to previous studies, we use detailed stellar evolution models, thus accounting for the dependence of convection on stellar properties, such as mass, metallicity, effective temperature, and age. Furthermore, we integrate the full set of equations describing the evolution of the orbital separation, stellar spin, and misalignment. We take into account the effects of tidal dissipation in the star, stellar wind mass loss, changes in the star's internal structure as a result of stellar evolution, and magnetic braking. 
The tidal prescription adopted follows \cite{Lai12}, and includes both tides in the weak friction approximation \citep{Zahn1977, Zahn1989} and convective damping of inertial waves.
In the weak-friction regime, a body's response to tides is generally measured via a tidal quality factor $Q$ \citep{GoldreichSoter66}, which parametrizes the efficiency of tidal dissipation. This term measures how a tidally-deformed body undergoing a forced oscillation dissipates part of the associated energy during each oscillation period. It is formally defined as the ratio of the maximum energy stored in the tidal distortion over the energy lost during each cycle.
The value of $Q$ is the result of complex dissipative processes occurring within a body and it thus varies for bodies of different masses and types. Furthermore, $Q$ depends on the tidal forcing frequency and thus on the spins and orbital configuration. As a result, $Q$ is expected to vary by orders of magnitudes \citep{PenevSasselov11} and it is clear that different $Q$ values are needed to explain different systems (e.g., \citealt{MatsumuraPR2010}, hereafter M10). In this work, we prefer not to introduce additional model parameters and instead use a parametrization for tidal dissipation calibrated from observations of binary stars (e.g.,  \citealt{VerbuntPhinney1995,RasioTLL1996,HurleyTP02,BelczynskiStartrack2008}).

For quick reference, the notations adopted in this work for the components and orbital parameters are summarized in Table~\ref{Tab:ParamsDefinition}.

The paper is organized as follows. In \S~\ref{The Sample} we present the sample of hot Jupiters considered in this work. In \S~\ref{Detailed Stellar modeling with MESA} we present the procedure adopted to model in detail the host stars in our sample. In \S~\ref{Orbital Evolution Model}, we summarize the equations that we integrate to study the orbital evolution of misaligned hot Jupiters (tests on the orbital evolution code developed for this work are presented in Appendix~\ref{Tests on the Orbital Evolution code}). 
In \S~\ref{Detailed orbital evolution for four representative systems} we present possible evolutionary sequences of five representative systems: \mbox{HAT-P-6}, WASP-7, 15, 16, and 71 (the results are summarized in Table~\ref{Tab:OrbitalEvolResultsPandTheta}, where we also include a few additional examples, without describing their evolution in detail). We discuss the assumptions adopted in this work in \S~\ref{Discussion}. We summarize and conclude in \S~\ref{Conclusions}.
As mentioned above, we consider tides in the weak friction approximation. Specifically, we account for both convective damping of the equilibrium tide and radiative damping of the dynamical tide. For the latter, we use results of detailed calculations presented by \cite{Zahn1975}, which are valid in the limit of small tidal forcing frequencies (in the weak-friction regime). In Appendix~\ref{Dynamic Tides in WASP-71} we solve the full set of equations describing non-adiabatic non-radial forced stellar oscillations, and we discuss the significance of dynamic tides in the most massive system among those studied in detail in \S~\ref{Detailed orbital evolution for four representative systems} for a wide spectrum of tidal forcing frequencies.
\begin{table}[!h]
\centering
\caption{Definition of the various parameters used in this work.}
\begin{tabular}{lr}
\hline 
{\bf Parameter} & {\bf Definition} \\
\\[-1.0em]
\hline \\[-1.0em]
\smallskip
$M_{*}$, $M_{\rm pl}$, $\Delta\,M_{\rm CZ}$ & Mass\\
\smallskip
$R_{*}$, $\Delta\,R_{\rm CZ}$ & Radius\\
\smallskip
$L_{*}$ & Bolometric luminosity\\
\smallskip
$T_{\rm eff *}$ & Effective temperature\\
\smallskip
Fe/H (or $Z$) &  Metallicity\\
\smallskip
$I_{*}$ &  Moment of inertia\\
\smallskip
$\Omega_{*}$ ($\Omega_{o}$) & Spin (orbital) frequency\\
\smallskip
$i_{*}$ ($i_{o}$) &   Stellar (orbital) inclination\\
\smallskip
$v_{{\rm rot}}{\rm sin}~i_{*}$ & Rotational velocity\\
\smallskip
$a$ & Semimajor axis\\
\smallskip
$P_{\rm orb}$ & Orbital period\\
\smallskip
$e$ & Eccentricity\\
\smallskip
$S$ & Spin angular momentum\\
\smallskip
$L$ & Orbital angular momentum\\
$\lambda$ ($\Theta_{*}$) & Sky-projected (true) misalignment\\
\\[-1.0em]
\hline \\[-1.0em]
\end{tabular}
\label{Tab:ParamsDefinition}
\tablecomments{The subscripts ``*'' and ``pl'' refer to the star and planet, respectively, while ``CZ'' refers to the stellar surface convection zone (see \S~\ref{Detailed Stellar modeling with MESA} for details) . We denote with $i_{*}$  the angle between the stellar spin axis and the line of sight, while $i_{o}$ denotes the angle between the orbital angular momentum and the line of sight.}
\end{table}
\section{The Sample of Hot Jupiter Host Stars}\label{The Sample}
\begin{figure} [!h]
\epsscale{1.1}
\plotone{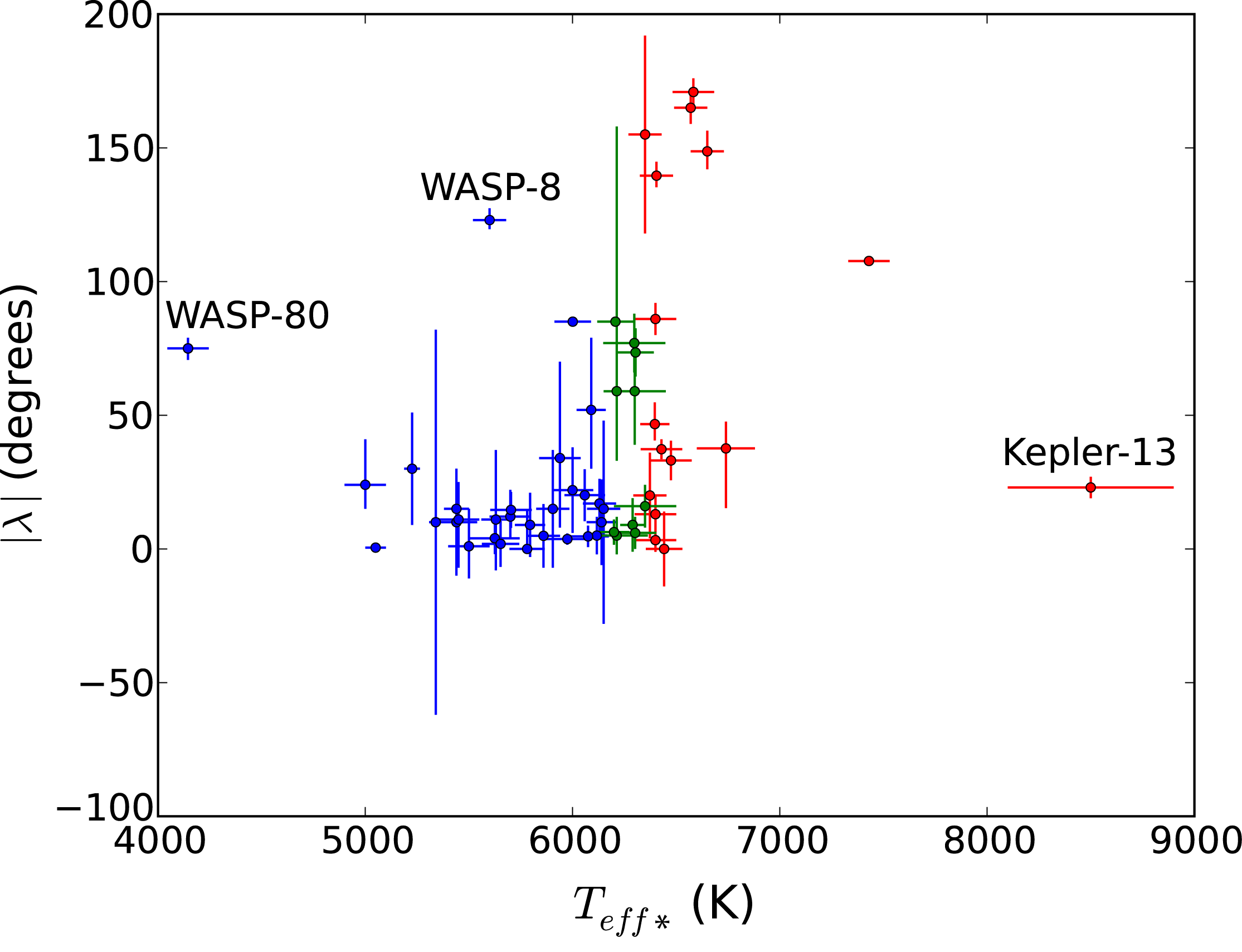}
\caption{Projected obliquities as a function of the effective temperature of the host stars. The color-scheme is as follows. From the mean $T_{\rm eff *}$ and its 1$\sigma$ uncertainties (Table~\ref{Table:hotJupiterSample}), we compute the maximum and minimum values $T_{\rm eff *, max}$ and $T_{\rm eff *, min}$, respectively.  Stars with $T_{\rm eff *, min}~\textgreater~6250~$K ($T_{\rm eff *, max}\textless~6250~$K) are shown with red (blue) symbols, while stars with $T_{\rm eff *, min}\,\leq\,6250\,K\,\leq\,T_{\rm eff *, max}$ are marked with green symbols. Prograde (retrograde) orbits have $|\lambda|\,\textless\,90^\circ$ ($|\lambda|\,\textgreater\,90^\circ$)}.
\label{fig:Obliquity_vs_Teff}
\end{figure}
To test whether tidal dissipation in the star could be responsible for the observed distribution of sky-projected misalignments and effective temperatures, we consider planetary systems hosting giant planets with an observationally inferred best-fit mass and orbital period $M_{\rm pl}~\textgreater~$0.5 $M_{\rm{Jup}}$ and $P_{\rm orb}~\textless~$5~d, respectively. This period range corresponds to the onset of strong tidal dissipation for a giant planet around a solar-like star \citep{RasioTLL1996}. The systems were queried from \href{http://www.openexoplanetcatalogue.com}{\it The Open Exoplanet Catalogue} on 2013 August 31 and we searched the literature to extract the components and orbital properties for each systems (guided by \href{http://exoplanet.eu/catalog/}{http://exoplanet.eu/} for references). 
To the planets thus selected, we followed the catalogue used by A12 and added few more systems that our selection missed because of the cut imposed on the minimum $M_{\rm pl}$ and maximum $P_{\rm orb}$. Specifically, we included HAT-P-2 b,  HAT-P-34 b, WASP-8 b, and WASP-38 b as their orbital period is just above 5$\,$d. We added WASP-17 b and WASP-31 b, as the planet's mass is just below 0.5$\,M_{\rm{Jup}}$.  From the compilation of A12 we excluded HAT-P-11 b and HD 149026 b because of the low planetary mass \citep{LecavelierSGKZ13, Sato+05}, and HD 17156 b and HD 80606 b because of the long orbital period \citep{Nutzman+11, Gillon+09c}. In addition to the systems used by A12, our catalogue includes WASP-52 b, \mbox{WASP-71} b, and WASP-80 b, whose misalignments were measured recently \citep{Hebrard+13,Smith+13, Triaud+13}. Finally, as in A12, we excluded WASP-23 b because the misalignment is not well constrained \citep{Triaud+11}. In Table~\ref{Table:hotJupiterSample} we summarize the systems considered in this study and list some of the orbital and components' properties relevant to our analysis. The systems added from A12's catalogue are listed at the bottom of the Table.

Similarly to Fig.~20 in A12, we show the sky-projected misalignment as a function of the host star's effective temperature in Fig.~\ref{fig:Obliquity_vs_Teff}. Here we show the absolute value for the mean value of the observed misalignment, and note that an orbit is prograde (retrograde) when $|\lambda|\,\textless\,90^\circ$ ($|\lambda|\,\textgreater\,90^\circ$). Our compilation of misaligned hot Jupiters host stars confirms the trend already reported by, e.g., \cite{WinnFAJ10} and A12: higher degrees of misalignment are associated with hotter stars. Three notable exceptions to this trend are WASP-8, WASP-80, and Kepler-13 (see Table~\ref{Table:hotJupiterSample} for references). WASP-8 has been discussed by A12 as one of the systems that least resemble the typical hot Jupiter. It has the longest orbital period among the hot Jupiters considered here ($\simeq\,8\,$d) and the largest ratio between semimajor axis and stellar radius: WASP-8 has $a/R_{*}\simeq\,$18, while the remaining systems all have $a/R_{*}\lesssim\,13$.
These properties result in a longer tidal timescale for alignment (see \S~\ref{Orbital Evolution Model}) and they might reconcile the position of this system in Fig.~\ref{fig:Obliquity_vs_Teff} with the hypothesis of tides being responsible for the observed relation between $\lambda$ and $T_{\rm eff *}$. WASP-80 hosts the least massive star in our sample and it has the second biggest $a/R_{*}$. Whether the system is misaligned is still an open question. In fact, \cite{Triaud+13} report a large discrepancy between the $v_{\rm rot}\,{\rm sin}\,i_{*}$ inferred from the broadening of the star's spectral lines and the observed amplitude of the RM effect. This discrepancy can be explained by either an orbital plane nearly perpendicular to the stellar spin or by an additional source of broadening that was not accounted for. 
Finally, Kepler-13 constitutes one of the most intriguing exoplanet systems given its geometry. Here, different dynamical effects might be at play which affect the observed spin-orbit configuration. The hot Jupiter in this system orbits a rapidly rotating A-star which is the main component of a hierarchical triple system \citep{Santerne+12}. The parent star's rapid rotation leads to oblateness which causes secular variations in the orbital elements \citep{Szabo+12}. 
\section{Detailed Modeling of Host Stars}\label{Detailed Stellar modeling with MESA}
\cite{WinnFAJ10} suggested that hot Jupiters are produced via a single formation mechanism which yields a broad range of obliquities. During their subsequent orbital evolution, tidal dissipation in the star is the main obliquity damping mechanism. If this is the case, we would expect to see a higher degree of sky-projected misalignment in stars where convection is small.  
In this section we describe the procedure we adopted to model the host stars in our sample. Our goal is to verify whether there is a correlation between the observed sky-projected misalignments and the mass fraction and radial extent of the stars' surface convective regions. 

We use MESA (version  4798, \citealt{PBDHLT2011, Paxton+13}) to create a grid of ZAMS stars covering the observed range of $M_{*}$ and Fe/H (see Table~\ref{Table:hotJupiterSample}). Specifically, we create models with metallicity $Z$ between 0.006-0.07 in steps of 0.001 and $M_{*}$ between 0.4$M_{\odot}$-2.2$M_{\odot}$ in steps of 0.005$M_{\odot}$. Next, we evolve each model accounting for stellar wind mass loss. The stellar wind prescriptions adopted follow the test suite example provided with MESA for the evolution of a 1$M_{\odot}$ star (\citealt{Reimers75} and \citealt{Bloecker95} with the $\eta$ parameter entering the two different mass loss prescriptions set to 0.7 in both cases). The results presented in \S~5 show that stellar winds do not play a significant role for the majority of the systems studied in detail here. Indeed, typical hot Jupiter host stars at 0.1$t_{\rm MS}$ (0.9\,$t_{\rm MS}$) have lost $\sim\,$0.01\% ($\sim\,$0.1\,\%) of their mass, where $t_{\rm MS}$ is the star's main sequence lifetime.
The mixing length $\alpha_{\rm MLT}$ parameter was set to 1.918, following the MESA star Standard Solar Model (\citealt{PBDHLT2011}, Table 10). We evolve each model to the end of the Main Sequence, which we take to be the time when the mass fraction of H at the center drops below $10^{-10}$.
Finally, for each host star in our sample, we scan through the grid of evolutionary tracks to find the models that simultaneously match the observed $Z$, $M_{*}$, $R_{*}$, and $T_{\rm eff *}$ within 1$\sigma$\footnote{For the system XO-2 b we can simultaneously match the observed properties of the host star only within 3$\sigma$. This might be attributed to the adopted $\alpha_{\rm MLT}$. This parameter is usually found to vary between 1 and 2 in the literature, depending on the star under investigation (see, e.g., \citealt{PBDHLT2011, Paxton+13} and \S~\ref{Discussion}}). In what follows, we call this a {\it successful model}.
For each successful model, the mass fraction and relative radial extent of the convective envelope or surface convection zone ($\Delta M_{\rm CZ}/M_{*}$ and $\Delta R_{\rm CZ}/R_{*}$, respectively) can be readily extracted from the radial profile of the Brunt-V$\ddot{\rm a}$is$\ddot{\rm a}$l$\ddot{\rm a}$ frequency ($N$). In fact, imaginary values of $N$ ($N^{2}\,\textless\,0$) denote a convective region. As shown below, the amount of surface convection depends, in part (\S~\ref{Discussion}), on the stellar mass, metallicity, temperature, and age.
\begin{figure} [!h]
\epsscale{1.1}
\plotone{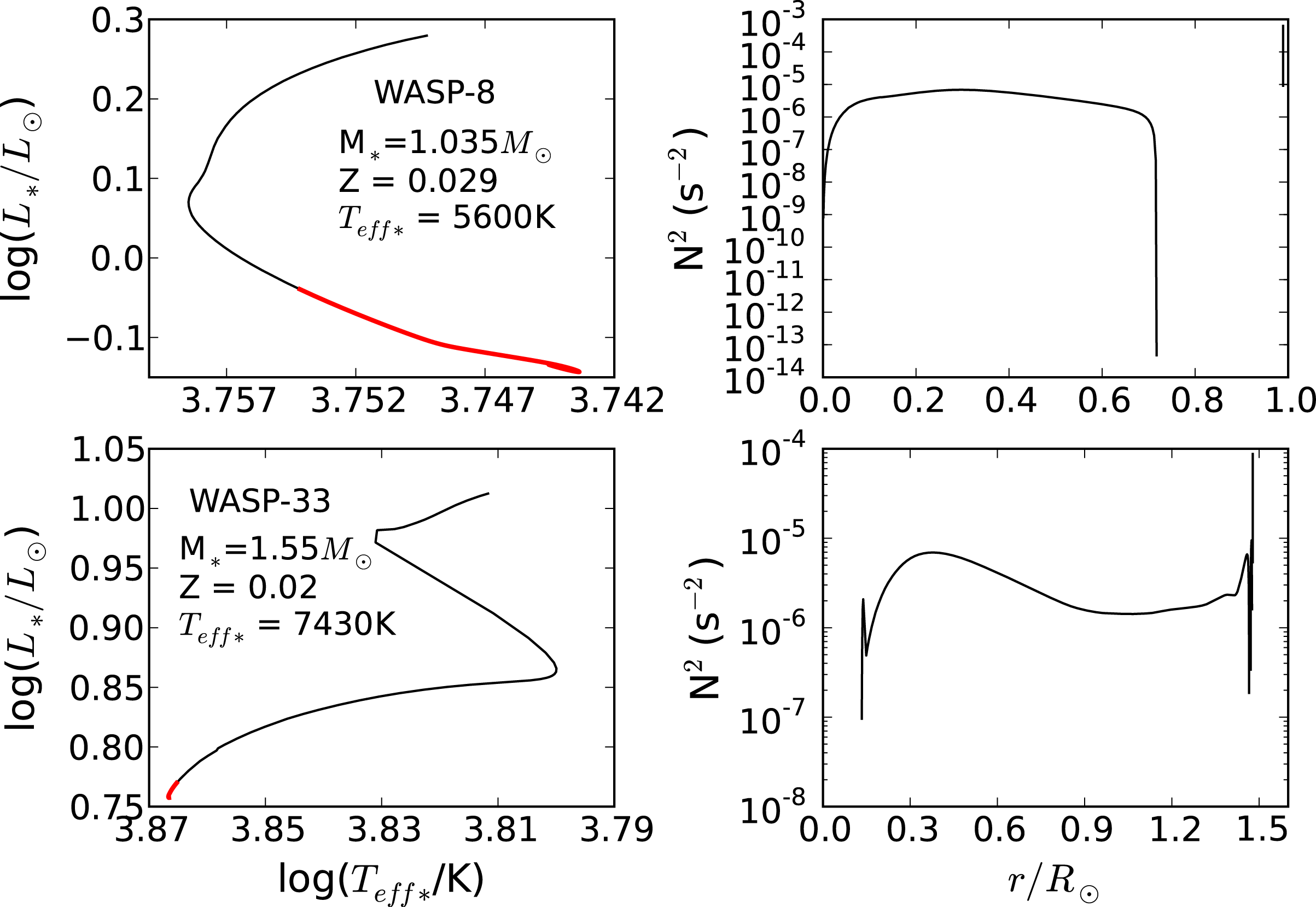}
\caption{HR diagram (left) and radial profile of the Brunt-V$\ddot{\rm a}$is$\ddot{\rm a}$l$\ddot{\rm a}$ frequency squared (right) for a WASP-8- (top) and WASP-33-type (bottom) star evolved on the main sequence. The red solid lines represent the successful models which simultaneously match the observed $Z$, $M_{*}$, $R_{*}$, and $T_{\rm eff *}$ within 1$\sigma$. The Brunt-V$\ddot{\rm a}$is$\ddot{\rm a}$l$\ddot{\rm a}$ frequency squared is shown for two successful models. Imaginary values of $N$ denote a convective region.}
\label{fig:HR_WASP33_WASP8}
\end{figure}

To give a flavor for the different stellar structures for the models considered here, Fig.~\ref{fig:HR_WASP33_WASP8} shows the Hertzsprung-Russell (HR) diagram and the radial profile of $N^{2}$ for a hot and cool star, according to the color-scheme adopted in Fig.~\ref{fig:Obliquity_vs_Teff}. These models represent WASP-8 and WASP-33. The red part of each evolutionary track on the HR diagram represents the models which match the observed $Z$, $M_{*}$, $R_{*}$, and $T_{\rm eff *}$ within 1$\sigma$. We take one of these successful models to plot the radial profile of $N^{2}$. The model representative of WASP-8 has $M_{*}~=~1.035~M_{\odot}$ and $Z~=~0.029$. It is mainly composed of a radiative core and a convective envelope, which extends from $\simeq$0.72$\,R_{\odot}$ to the surface ($R_{*}~\simeq~$0.99$R_{\odot}$). The model representative of WASP-33 has $M_{*}~=~1.55~M_{\odot}$ and $Z~=~0.02$. It is composed of a convective core and a radiative envelope, which extends from $\simeq$0.13$\,R_{\odot}$ to near the surface ($R_{*}~\simeq~$1.48$R_{\odot}$), where thin convective layers are present. When accounting for convection in stars similar to WASP-8, we consider the whole convective envelope. Instead, for \mbox{WASP-33-type} stars, we consider only the surface convective layers, as we expect that most dissipation occurs in these regions \citep{BarkerOgilvie2009}.  We discuss possible uncertainties related to convection within MESA in \S~\ref{Discussion}.

A summary of the stellar properties derived from this modeling is given in Table~\ref{Table:hostStarModeling}.
\subsection{Obliquity vs Convection}\label{Obliquity Vs convection}
\begin{figure} [!h]
\epsscale{1.1}
\plotone{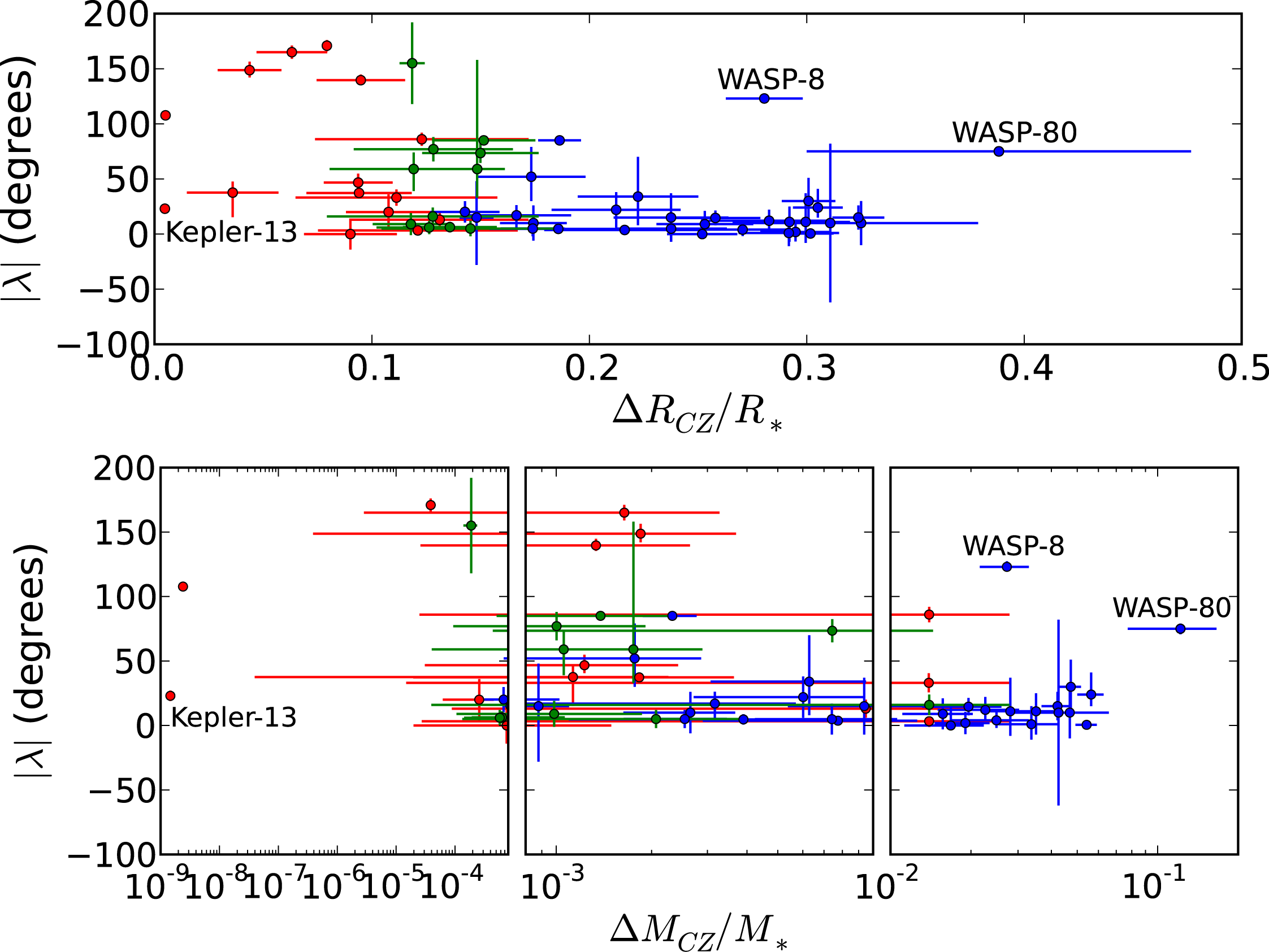}
\caption{Projected obliquities as a function of the fractional radius (top) and mass (bottom) of the star's convection zones. For solar-type stars (e.g. WASP-8 in Fig.~\ref{fig:HR_WASP33_WASP8}), we account for the convective envelope. For stars with predominantly radiative envelopes (e.g. WASP-33 in Fig.~\ref{fig:HR_WASP33_WASP8}) we consider only the surface convective layers, as these are the most dissipative. The color-scheme is as in Fig.~\ref{fig:Obliquity_vs_Teff}. For clarity, we split the x-axis of the bottom plot in three different intervals. For the error bars in $\Delta M_{\rm CZ}/M_{*}$ and $\Delta R_{\rm CZ}/R_{*}$ see Table~\ref{Table:hostStarModeling}.}
\label{fig:obliquity_meanConvection_massAndRadius}
\end{figure}
In Fig.~\ref{fig:obliquity_meanConvection_massAndRadius} we show the behavior of the observationally inferred misalignment as a function of the properties of each star's surface convection zone, as given by our modeling.  Our results point to an increase in $|\lambda|$ as the amount of convection decreases, thus confirming the importance of tidal dissipation in shaping the observed distribution of sky-projected misalignments. The relation between the magnitude of the misalignment and convection is immediately clear if we look at $|\lambda|$ as a function of $\Delta R_{\rm CZ}/R_{*}$ (top plot in Fig.~\ref{fig:obliquity_meanConvection_massAndRadius}). The highest  $|\lambda|$ are associated with $\Delta R_{\rm CZ}/R_{*}\lesssim\,0.1$. As $\Delta R_{\rm CZ}/R_{*}$ increases,  $|\lambda|$ decreases.
This decrease in $|\lambda|$ with decreasing convection is less clear if one considers the mass fraction of the convective regions. In fact, the $\Delta M_{\rm CZ}/M_{*}$ computed for each star varies by several orders of magnitude. Our results suggest that $|\lambda|$ is high in systems where the mass fraction of the star's surface convective region is $\lesssim ~3\times 10^{-3}$ and it decreases as $\Delta M_{\rm CZ}/M_{*}$ increases (the outliers, Kepler-13, WASP-8, and WASP-80, are discussed briefly in \S~\ref{The Sample}). 

This result favors the high-eccentricity migration scenario in which hot Jupiters are formed via tidal circularization of highly eccentric orbits with a broad distribution of $\lambda$. As the subsequent evolution of the orbital separation and misalignment is driven by tidal dissipation in the star (M10; see also \citealt{Jackson+09}), tides in stars with a significant amount of surface convection efficiently decrease the misalignment. We further test this scenario below by performing detailed orbital evolution calculations.

\section{Orbital Evolution Model}\label{Orbital Evolution Model}
\subsection{Assumptions and Basic Equations}\label{Equations describing the orbital evolution of hot Jupiters}
Here we summarize the set of equations that we integrate to study the orbital evolution of misaligned hot Jupiters. We account for tides, stellar wind mass loss, changes in the star's internal structure as a result of stellar evolution, and magnetic braking. In what follows, we denote by {\bf S}\,=\,$S\,\hat{\bf S}$ and {\bf L}\,=\,$L\,\hat{\bf L}$ the spin and orbital angular momentum vectors, respectively. Their magnitudes are given by $S\,=\,I_{*}\Omega_{*}$ and $L\,=\,M_{*}M_{pl}\sqrt{Ga/(M_{*}+M_{\rm pl})}$. The true stellar obliquity $\Theta_{*}$ is related to the projected one $\lambda$ via ${\rm cos}\,\Theta_{*} = {\rm sin\,}i_{*}\,{\rm cos\,}\lambda\,{\rm sin\,}i_{o}+{\rm cos\,}i_{*}\,{\rm cos\,}i_{o}$ \citep{FabryckyWinn09}.

We assume that all effects from distant companions (other planets or stars) can be neglected (the systems studied in detail in \S~\ref{Detailed orbital evolution for four representative systems} do not have observed companions). Furthermore, the evolution of the stellar spin is computed assuming solid-body rotation.
As we apply the tidal prescription proposed by \cite{Lai12} (summarized in \S~\ref{Tidal Contribution due to inertial wave dissipation}), we focus on circular binaries.  

Tides affect $a$, $\Omega_{*}$, and $\Theta_*$.  Here we consider tidal dissipation in the star while neglecting tides in the planet. We discuss this assumption in \S~\ref{Discussion} and note here that it is justified for circular binaries, as stellar tides are expected to largely dominate the evolution of the orbital separation and obliquity (M10).
The tidal evolution is calculated in the standard weak friction approximation (\citealt{Zahn1977, Zahn1989}), following the formalism of \cite{Hut1981}. 
Specifically, we integrate numerically the following differential equations,
\begin{align}
(\dot{a})_{\rm wf} = -\frac{a}{\tau_{\rm wf}}\left(1-\frac{\Omega_{*}}{\Omega_{o}}{\rm cos\,}\Theta_{*}\right), \label{eq:dadtTidesStartrack}
\end{align}
\begin{align}
(\dot{\Omega}_{*})_{\rm wf}=\frac{\Omega_{*}}{\tau_{\rm wf}}\left(\frac{L}{2S}\right)\left[{\rm cos\,}\Theta_{*}-\left(\frac{\Omega_{*}}{2\Omega_{o}}\right)(1+{\rm cos^{2}\,}\Theta_{*})\right], \label{eq:dwdtTidesStartrack}
\end{align} 
\begin{align}
(\dot{\Theta}_{*})_{\rm wf} = -\frac{{\rm sin\,}\Theta_{*}}{\tau_{\rm wf}}\left(\frac{L}{2S}\right)\left[1-\left(\frac{\Omega_{*}}{2\Omega_{o}}\right)\left({\rm cos\,}\Theta_{*}-\frac{S}{L}\right)\right], \label{eq:dThetadtHut}
\end{align}
where the characteristic orbital evolution timescale $\tau_{\rm wf}$ is given by
\begin{align}
\frac{1}{\tau_{\rm wf}} = 6F_{\rm tid}\left(\frac{k}{T}\right)q_{*}(1+q_{*})\left(\frac{R_{*}}{a}\right)^{8} \label{eq:tau_wf}.
\end{align}
The subscript ``wf'' stands for ``weak friction'' and $q_* = M_{\rm pl}/M_{*}$. Eqs.~(\ref{eq:dadtTidesStartrack})-(\ref{eq:dThetadtHut}) are valid for a circular orbit. 
The quantity $k/T$ is the ratio of the apsidal motion constant $k$ over the timescale $T$ of tidal dissipation.
We parametrize the dissipation inefficiency of tides as in \cite{HurleyTP02} and \cite{BelczynskiStartrack2008} and we assume that the only sources of dissipation are eddy viscosity in convective envelopes and radiative damping in radiative envelopes. 
For radiative damping of the dynamical tide we use $F_{\rm tid}\,=\,1$ and, 
\begin{align}
&\left(\frac{k}{T}\right)_{RD} = 1.9782\times 10^4\sqrt{\frac{M_{*}R_{*}^2}{a^5}} (1+q_{*})^{5/6}E_2~{\rm yr^{-1}},\label{eq:kOverTRad}
\end{align}
where $E_2 = 1.592 \times 10^{-9}(M_{*}/M_\odot)^{2.84}$ and the various quantities are expressed in solar units. Instead, for convective damping of the equilibrium tide we use $F_{\rm tid}\,=\,50$ and,
\begin{align}
&\left(\frac{k}{T}\right)_{CD} = \frac{2}{21}\frac{f_{\rm *,conv}}{\tau_{\rm *,conv}}\frac{M_{\rm *, env}}{M_*}~{\rm yr^{-1}}, \label{eq:kOverTConv}
\end{align}
where the subscript ``env'' denotes the convection zone's properties and the various quantities are expressed in solar units. We compute the mass and radius of the convective regions from the radial profile of the Brunt-V$\ddot{\rm a}$is$\ddot{\rm a}$l$\ddot{\rm a}$ frequency as described in \S~\ref{Detailed Stellar modeling with MESA}. The convective turnover timescale $\tau_{\rm *,conv}$ is given by
\begin{align}
&\tau_{\rm *,conv} = 0.431\left[\frac{M_{\rm *, env}R_{\rm *, env}\left(R_* - \frac{R_{\rm *, env}}{2}\right)}{3L_*}\right]^{1/3}~{\rm yr}, 
\end{align}
where $L_{*}$ is the star's bolometric luminosity \citep{RasioTLL1996}. The factor $f_{\rm *,conv}$ represents the reduction in the effectiveness of convective damping when the tidal forcing period is less than the turnover period of the largest eddies \citep{GoldreichNicholson1977}. It is defined as
\begin{align}
&f_{\rm *,conv} = min\left[1, \left(\frac{P_{\rm *,tid}}{2\tau_{\rm *,conv}}\right)^2\right],\label{eq:fConv}
\end{align}
with the tidal pumping timescale $P_{\rm *,tid}$ given by
\begin{align}
P_{\rm *,tid} = \frac{1}{\left|\frac{1}{P_{\rm orb}} - \frac{1}{P_{\rm *, spin}}\right|}.
\end{align}
Eqs.~(\ref{eq:dadtTidesStartrack})-(\ref{eq:dThetadtHut}) reduce to the \cite{Hut1981} equations in the limit of small $\Theta_*$. To Eqs.~(\ref{eq:dwdtTidesStartrack}) and~(\ref{eq:dThetadtHut}), we add the terms derived by \cite{Lai12} and summarized in \S~\ref{Tidal Contribution due to inertial wave dissipation}.
It is important to note that the expression for the term $E_{2}$ entering Eq.~(\ref{eq:kOverTRad}) was fitted by \cite{HurleyTP02}  to values given by \cite{Zahn1975}. The latter investigated the effects of radiative damping of tidally excited gravity modes in massive main sequence binaries in the limit of small tidal forcing frequencies. In this weak-friction regime, the detailed orbital evolution calculations presented here (see \S~\ref{Detailed orbital evolution for four representative systems}) show that radiative damping of the dynamical tide is weaker than convective damping of the equilibrium tide throughout each system's evolution (but see Appendix~\ref{Dynamic Tides in WASP-71}).

Changes in the star's internal structure as a result of stellar evolution affect $\Omega_{*}$. Specifically, we account for changes in the radius and core properties via the evolution of the star's moment of inertia. The corresponding $\dot{\Omega}_{*, \rm evol}$ term is derived considering that changes in $I_{*}$ conserve spin angular momentum. It is given by
\begin{align}
(\dot{\Omega}_{*})_{\rm evol} &= -\Omega_{*}\frac{\dot{I}_*}{I_*}.\label{eq:OmegaDotInertia}
\end{align}
The moment of inertia is computed within MESA from the stellar mass profile at each time step, considering the innermost shell of mass $m_{\rm c}$ and radius $r_{\rm c}$ as a solid sphere with $I_{*}\,=\,\frac{2}{5}\,m_{\rm c}\,r_{\rm c}^{2}$ and adding $\Delta I_{*}\,=\frac{2}{3}\,\Delta m\,r^{2}$ for each spherical shell of mass $\Delta m$ at radius $r$.

Stellar wind mass loss affects the orbital separation and the spin of the star. As in \cite{BelczynskiStartrack2008}, the evolution of the semimajor axis is computed assuming spherically symmetric mass loss, which carries away the specific angular momentum of the mass-losing component (Jeans-mode mass loss). As $a(M_{*}+M_{\rm pl})$ is constant, it is straightforward to derive
\begin{align}
&(\dot{a})_{\rm wind} =- \frac{a}{M_{*}+M_{\rm pl}}\dot{M}_{*}\label{eq:dadtWind}
\end{align}
which is valid for circular orbits.
The evolution of the stellar spin is calculated assuming that mass loss in a wind carries away the angular momentum of the outer shell. Setting $\dot{S} = (2/3)\dot{M}_{*}R_{*}^2\Omega_{*}$, it follows that 
\begin{align}
(\dot{\Omega}_{*})_{\rm wind} &= \frac{2}{3}\frac{\Omega_{*}R_*^2}{I_*}\dot{M}_{*}\label{eq:dOmegadtWind}.
\end{align}

Finally, magnetic braking involves the loss of spin angular momentum through magnetized stellar winds. As in M10, we adopt \citeauthor{Skumanich72}'s (\citeyear{Skumanich72}) law, which is well established for stars with rotational velocities between 1-30\,km\,s$^{-1}$ (like the ones studied in detail here), and use
\begin{align}
(\dot{\Omega}_{*})_{MB} & = -\alpha_{MB}\Omega_{*}^{3}\label{eq:dOmegadtMB}
\end{align} 
where $\alpha_{MB}=1.5\times 10^{-14}~\gamma_{\rm MB}~$yr. Previous studies adopted a value $\gamma_{\rm MB}\,=\,$0.1 for F-dwarfs and $\gamma_{\rm MB}\,=\,$1 for G or K dwarfs (e.g., \citealt{BarkerOgilvie2009,DobbsDixon+04}, M10). In this work we keep $\gamma_{\rm MB}$ as a free parameter and vary it between 0 and 1.
\subsection{Tidal Dissipation of Inertial Waves}\label{Tidal Contribution due to inertial wave dissipation}
To clarify the assumptions adopted in our work, we briefly summarize the tidal prescription proposed by \cite{Lai12}. This recipe is based on tidal dissipation of inertial waves and it is valid for binaries in a circular orbit where the stellar spin and planet's orbital angular momentum are misaligned (see also R13 for a summary). 

In the inertial reference frame centered on $M_{*}$ with the z-axis along the stellar spin angular momentum {\bf S}, the tide-generating potential to the leading quadrupole order can be expanded in terms of spherical harmonics $Y_{2m}(\theta, \phi)$ as
\begin{align}
&U({\bf r}, t) = -\sum_{\substack{mm'}}U_{mm'}(M_{\rm pl}, a, \Theta_{*})r^{2}Y_{2m}(\theta, \phi)e^{-im'\Omega_{o}t}.\label{eq:TidalPotentialAlongS}
\end{align}
Here $\theta$ and $\phi$ are the polar and azimuthal angle, respectively. Moving to a frame co-rotating with the star and introducing the azimuthal angle $\phi_{r}$, it can be shown that each term in the tide-generating potential has the dependence $e^{im\phi_{r}+im\Omega_{*}t-im'\Omega_{o}t}$. Thus, the tidal perturbation from the planet induces in the star a spectrum of forcing angular frequencies $\tilde{\omega}_{mm}'\,=\,m'\Omega_{o}\,-\,m\Omega_{*}$. The corresponding forcing frequency in the inertial frame is $m'\Omega_{o}$. Physically, seven components of the potential contribute to
the transfer of tidal energy between the stellar spin and the orbital angular momentum. Each of these components dissipates this energy with its own quality factor $Q_{mm'}$.

The dispersion relation for an inertial wave is given by \citep{Dejaiffe68}
\begin{align}
\tilde{\omega}^{2}\,=\,(2{\bf \Omega_{*}}\cdot\,{\bf k}/{\bf |k|})^{2},
\end{align}
where $\tilde{\omega}$ is the inertial wave frequency and ${\bf k}$ is its local wavenumber vector. Therefore, these waves exist only when $|\tilde{\omega}|\,\textless\,2\Omega_{*}$. In systems hosting hot Jupiters with $\Omega_{*}\ll\Omega_{o}$ the only  component of the tidal potential whose tidal forcing frequency is small enough to allow the excitation of inertial waves is ($m, m'$)\,=\,(1, 0). While this component acts on $\Omega_{*}$ and $\Theta_{*}$, it does not affect the evolution of the orbital separation $a$.  
In fact,  the (1, 0)-component of the tidal potential is static in the inertial frame. As R13 point out, this might not be the case for stars where $\Omega_{*}\,\textgreater\,\Omega_{o}$, as other components of the tidal response might be relevant which could lead to orbital, spin, and obliquity evolution. Accounting for this additional source of tidal dissipation, the equations describing the evolution of $a$, $\Omega_{*}$ and $\Theta_{*}$ due to tides become
\begin{align}
&(\dot{a})_{\rm tide}  = (\dot{a})_{\rm wf},\label{eq:dadtTideTotal}\\
&(\dot{\Omega}_{*})_{\rm tide}  = (\dot{\Omega}_{*})_{\rm wf}+ (\dot{\Omega}_{*})_{\rm 10}-(\dot{\Omega}_{*})_{\rm 10, wf},\label{eq:58Lai12}\\
&(\dot{\Theta}_{*})_{\rm tide}  = (\dot{\Theta}_{*})_{\rm wf}+ (\dot{\Theta}_{*})_{\rm 10}-(\dot{\Theta}_{*})_{\rm 10, wf}.\label{eq:59Lai12}
\end{align} 
Note that the tidal evolution of the orbital separation only accounts for tides in the weak friction approximation [the terms with subscript ``wf'', which are given in Eqs.~(\ref{eq:dadtTidesStartrack})\,--\,(\ref{eq:dThetadtHut})]. The terms with subscript ``10'' are associated with the ($m, m'$)\,=\,(1, 0) component of the tidal potential in Eq.~(\ref{eq:TidalPotentialAlongS}), written in the frame co-rotating with the star. These are given by
\begin{align}
(\dot{\Omega}_{*})_{10} & = -\frac{\Omega_{*}}{\tau_{\rm 10}}(\rm{sin\,}\Theta_{*}\,\rm{cos\,}\Theta_{*})^{2},\label{eq:51Lai12}\\
(\dot{\Theta}_{*})_{10} & = -\frac{1}{\tau_{\rm 10}}\rm{sin\,}\Theta_{*}\,\rm{cos}^{2}\,\Theta_{*}\,\left(\rm{cos}\,\Theta_{*}+\frac{S}{L}\right),\label{eq:52Lai12}
\end{align} 
where
\begin{align}
\frac{1}{\tau_{\rm 10}} &= \frac{3k_{\rm 10}}{4Q_{\rm 10}}\left(\frac{M_{\rm pl}}{M_{*}}\right)\left(\frac{R_{*}}{a}\right)^{5}\frac{L}{S}\Omega_{o}.\label{eq:53Lai12}
\end{align} 
Here $k_{\rm 10}$ and $Q_{\rm 10}$ are the tidal Love number and quality factor for the ($m, m'$)\,=\,(1,0) component of the tidal potential, respectively. 
The terms with subscript ``10, wf'' are given by
\begin{align}
\frac{\dot{\Theta_{*}}_{\rm 10, wf}}{\dot{\Theta_*}_{\rm 10}} = \frac{\dot{\Omega}_{\rm 10, wf}}{\dot{\Omega}_{\rm 10}} = \frac{\tau_{\rm 10}}{\tau_{\rm wf}}\frac{L}{4S}.\label{eq:CrossLaiTerm}
\end{align} 

As found by \cite{BarkerOgilvie2009}, the efficiency with which inertial waves are dissipated can vary widely between different stars. In fact, it depends on both the stellar spin and the amount of surface convection. The latter, in turn, depends on the stellar mass, metallicity and evolutionary stage. 
\cite{OgilvieLin2007} studied tidal dissipation in rotating sun-like stars. The authors found that the energy dissipation rate in the convection zone is increased by 1-3 orders of magnitude, depending on the stellar spin period, compared to weak-friction tides.
In particular, for the $m\,=\,2$ term in the spherical harmonics expansion of the tidal potential and for a solar-type star at $\tilde{\omega}\,=\,-\Omega_{*}$, the tidal quality factor associated with inertial wave dissipation can increase up to $Q'\simeq\,10^{7}-10^{8}$ for a spin period of $\simeq\,$10~d and $Q'\simeq\,10^{6}-10^{7}$ for a spin period of $\simeq\,$3~d (see Fig.\,3 and Fig.\,6 in their paper). Here $Q'$ is the modified tidal quality factor defined as $Q'\,=\,1.5\,Q/k$. 
Their numerical method was later used by \cite{BarkerOgilvie2009} to investigate the tidal dissipation associated with the $m=$1 components for an F-type star representative of the system XO-3. The authors found that tidal dissipation is significantly enhanced, with $Q'$ increasing up to $\simeq\,10^{6}$ at $\tilde{\omega}\,=\,-\Omega_{*}$ (middle panel Fig.\,7 of their paper). They also investigated how $Q'$ varies for a range of F-type stars with different masses and evolutionary stages and derived values in the range  $\sim\,10^{8}\,-\,10^{13}$ (Fig.\,8 of their paper). The low end of this range was for a 1.2$M_{\odot}$ star at solar metallicity and an age of 1\,Gyr, while the high end was for a 1.5$M_{\odot}$ star at solar metallicity and an age of 0.7\,Gyr. To account for this range of values and given the properties of the stars studied in \S~\ref{Detailed orbital evolution for four representative systems}, we perform detailed orbital evolution calculations considering $Q'_{\rm 10}=10^{6}, 10^{7}, 10^{8}$, and $10^{10}$. It turns out that, for most systems, the results do not change significantly once $Q'_{\rm 10}\textgreater\,10^{6}$. 

\section{Detailed Orbital Evolution of Five Representative Systems}\label{Detailed orbital evolution for four representative systems}
To test whether the tidal prescription proposed by \cite{Lai12} and summarized in \S~\ref{Tidal Contribution due to inertial wave dissipation} can explain the observed distribution of  sky-projected misalignments, we perform detailed orbital evolution calculations for five representative systems. Specifically, we integrate the set of equations presented in \S~\ref{Orbital Evolution Model} and  compute the evolution of $a$, $\Theta_*$, and $\Omega_{*}$ accounting for tides, stellar wind mass loss, changes in the star's moment of inertia, and magnetic braking. 
\subsection{System Selection}\label{System Selection}
\begin{figure} [!h]
\epsscale{1.1}
\plotone{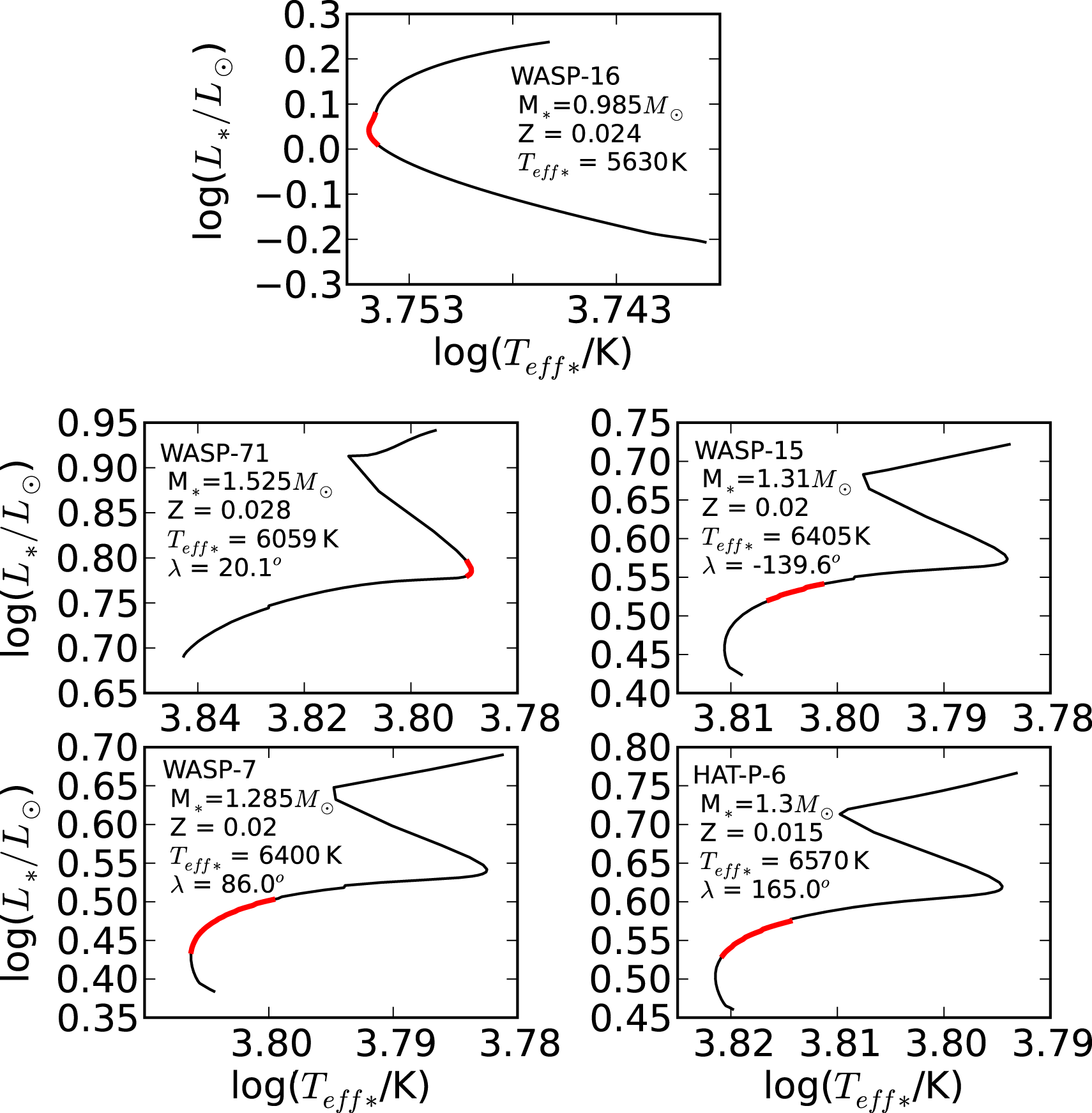}
\caption{HR diagram for the systems selected for our detailed study. The red solid lines represent the successful models which simultaneously match the observed $Z$, $M_{*}$, $R_{*}$, and $T_{\rm eff *}$ within 1$\sigma$. Top: system with $\lambda$ consistent with $0^{o}$ within 1\,$\sigma$ hosting a cool star. In the middle and bottom panels: on the left are systems where $|\lambda|\textless~90^{o}$ with minimum (middle) and maximum (bottom) mean $T_{\rm eff,*}$, while on the right are systems where $|\lambda|\textgreater~90^{o}$ with minimum (middle) and maximum (bottom) mean $T_{\rm eff,*}$. According to our models, the age of WASP-16, 71, 15, 7, and \mbox{HAT-P-6} is in the range $\simeq$\,(0.59\,--\,0.75)\,$t_{\rm MS}$, $\simeq$\,(0.89\,--\,0.92)\,$t_{\rm MS}$, $\simeq$\,(0.44\,--\,0.57)\,$t_{\rm MS}$, $\simeq$\,(0.19\,--\,0.54)\,$t_{\rm MS}$,  and $\simeq$\,(0.30\,--\,0.53)\,$t_{\rm MS}$, respectively, where $t_{\rm MS}$ is the star's main sequence lifetime.}
\label{fig:HR_detailed_orbEv}
\end{figure}
The five representative systems are chosen as follows. As the tidal prescription proposed by \cite{Lai12}  is valid for circular binaries hosting slowly rotating stars, we only consider systems where the measured eccentricity is zero and the star has not been observed to be a fast rotator. For the latter, we compute the stellar spin from the observed $v_{rot}\,{\rm sin\,}i_{*}$ assuming (arbitrarily) $i_{*}=90^{o}$ and eliminate systems where $\Omega_{*}/\Omega_{o}\textgreater 1$. To avoid possible perturbations of the orbital elements due to gravitational interactions with additional companions, we only consider systems where there is no evidence of other bodies. Also, to compute some degree of orbital evolution we require the star to be off its main sequence.
Finally, we consider the following combinations of $\lambda$ and $T_{\rm eff,*}$: $(i)$ to test if the \citeauthor{Lai12}'s (\citeyear{Lai12}) prescription can account for the currently observed {\it aligned} hot Jupiters, we consider the host star with the lowest $T_{\rm eff,*}$ among the systems where $\lambda\,=\,0^{o}$ within 1$\sigma$; $(ii)$ to test if the \citeauthor{Lai12}'s (\citeyear{Lai12}) prescription can account for the currently observed {\it misaligned} hot Jupiters, we pick two systems with $|\lambda|\textgreater~90^{o}$ and two systems with $|\lambda|\textless~90^{o}$ hosting stars with the lowest and highest observed $T_{\rm eff *}$, among the systems where $\lambda\,\neq\,0^{o}$ within 1$\sigma$. 
Selection $(i)$ leaves us with WASP-4, while selection $(ii)$ leaves us with WASP-71, \mbox{WASP-7}, WASP-15, and HAT-P-6. However, we do not consider WASP-4, as it was studied in a companion paper \citep{VR14} using the same numerical approach adopted in this work. Here we only summarize our findings for WASP-4 in Table~\ref{Tab:OrbitalEvolResultsPandTheta} and refer to \cite{VR14} for details. The next system among the aligned ones with the coldest star is WASP-16.
Then, we select one of the successful models for each host star to perform the orbital evolution. For HAT-P-6, WASP-7, 15, and 71, each model is chosen randomly among those whose mass and metallicity ($Fe/H$) differ from the observed mean values by less than 0.05$M_{\odot}$ and 0.01 (to be within the 1$\,\sigma$ errors), respectively, at some point during the evolution of the star. For WASP-16 we chose these same limits on $M_{*}$ and $Fe/H$ to be 0.005$M_{\odot}$ and 0.01. We show the HR diagram for the five stellar models in Fig.~\ref{fig:HR_detailed_orbEv}. According to our detailed modeling, \mbox{WASP-71} is the oldest system among the ones considered here and it has an age $\simeq\,0.9\,t_{\rm MS}$. The age of WASP-16, 15, 7, and \mbox{HAT-P-6} is in the range $\simeq$\,(0.6\,--\,0.7)\,$t_{\rm MS}$, $\simeq$\,(0.4\,--\,0.6)\,$t_{\rm MS}$, $\simeq$\,(0.2\,--\,0.5)\,$t_{\rm MS}$,  and $\simeq$\,(0.3\,--\,0.5)\,$t_{\rm MS}$, respectively. As we show in \S~\ref{WASP-16}, WASP-16's alignment provides a good example of how the inclusion of inertial wave dissipation in the tidal prescription yields a more significant obliquity evolution than orbital separation evolution.
\subsection{Computing the Orbital Evolution}\label{Orbital Evolution Procedure}
We study the past orbital evolution of each system by using the detailed evolution of the host star computed with MESA and by considering a variety of initial configurations. Specifically, we scan the parameter space made of initial orbital periods, degrees of asynchronism between the stellar spin and the planet's orbital frequency ($\Omega_{*}/\Omega_{o}$), misalignments, and $\gamma_{\rm MB}$ values. For each combination of these parameters, we consider 50 values for the stellar inclination $i_{*}$. Below we summarize the initial values considered for $\Theta_{*}$, $\Omega_{*}/\Omega_{o}$, $\gamma_{\rm MB}$, and $i_{*}$.
For the planet's mass and orbital inclination $i_{o}$ we use the observed mean values for each system. 
 
For the misalignment, we consider $\Theta_{*}$ between 0-$180^{o}$ (0$-$-180$^{o}$, depending on the observed misalignment) in steps of 2$^{o}$. 
For the level of asynchronism, as the tidal prescription proposed by \cite{Lai12} is valid for $\Omega_{*}\,\textless\,\Omega_{o}$, we consider initial values of $\Omega_{*}/\Omega_{o}$ between 0\,-\,1 in steps of 0.1 (we discuss cases where $\Omega_{*}/\Omega_{o}\,=\,1$ in the next section). 
For $\gamma_{\rm MB}$, we follow the literature (e.g.,  \citealt{BarkerOgilvie2009,DobbsDixon+04}, M10) and consider $\gamma_{\rm MB}$ values between 0\,-\,1 in steps of 0.1. As far as $i_{*}$ is concerned, the true stellar spin axis orientation is unknown for the systems studied in detail here. This could be estimated from the observationally inferred $v_{\rm rot}{\rm sin\,}i_{*}$ combined with typical rotation rates for a star of the given spectral type and age for each system. However, as pointed out by \cite{FabryckyWinn09}, there are uncertainties related to both spectral type and age and it is possible that the rotation rates of hot Jupiters host stars differ from stars in general (e.g., because of tides). For this reason we use the reasonable assumption that the stellar spin axis angle $i_{*}$ is distributed isotropically and adopt a simple Monte-Carlo approach (similarly to \citealt{Triaud+10}). We draw a random uniform distribution in cos$\,i_{*}$ between 0\,-\,1 and consider 50 values of cos$\,i_{*}$ for each combination of initial $P_{\rm orb}$, $\Theta_{*}$, $\Omega_{*}/\Omega_{o}$, and $\gamma_{\rm MB}$.
As $i_{*}$ is used only to convert the computed $\Omega_{*}$ and $\Theta_{*}$ into a present-day $v_{\rm rot}\,{\rm sin\,}i_{*}$ and $\lambda$, respectively, $i_{*}$ is kept fixed during the integration.

The parameter space mentioned above is scanned for four different values of $Q'_{\rm 10}$. We use the results presented by \cite{OgilvieLin2007} and \cite{BarkerOgilvie2009} (\S~\ref{Tidal Contribution due to inertial wave dissipation}) and consider $Q'_{\rm 10}=10^{6}, 10^{7}, 10^{8}$, and $10^{10}$.

The set of equations presented in  \S~\ref{Orbital Evolution Model} is integrated with a variable-step 4th-order Runge-Kutta integrator with an accuracy requirement of $10^{-12}$.
We stop the integration when the stellar mass, radius, effective temperature, misalignment, and rotational velocity agree with the observationally inferred values within 1$\sigma$, and the orbital period crosses the observed value. 

We first discuss the four misaligned systems. Since \mbox{HAT-P-6}, \mbox{WASP-7}, and \mbox{WASP-15} are similar in stellar mass and evolutionary stage, we summarize their results together, while we reserve \mbox{WASP-71} to a separate section. We discuss WASP-16's alignment at the end of this section.
For each system we discuss whether the initial parameter space can be expanded to $\Omega_{*}/\Omega_{o}\textgreater\,1$.  Here we note that $\Omega_{*}/\Omega_{o}\textless 1$ leads to orbital decay, as tides remove angular momentum from the orbit to spin up the star. Considering rapidly spinning host stars ($\Omega_{*}/\Omega_{o}\,\textgreater\,1$), planets could migrate outward and the overall orbital evolution might change from the one described here (M10; \citealt{DobbsDixon+04}). 
Note that, rather than finding {\it all} possible evolutionary scenarios for these systems, our goal is merely to demonstrate that their current properties can be explained given our current (and limited) understanding of tidal dissipation, and with reasonable assumptions on the physical effects driving the orbital evolution of hot Jupiters in circular and misaligned systems. Even though below we focus on \mbox{HAT-P-6}, \mbox{WASP-7}, 15, 16,  and 71, in Table~\ref{Tab:OrbitalEvolResultsPandTheta} we summarize the results of orbital evolution calculations for a few additional systems, without a detailed discussion in the text. For details about WASP-4 we refer to \cite{VR14}.

In what follows, the subscripts ``in'' and ``pr'' denote initial (at the star's Zero Age Main Sequence) and present values. 
\subsection{HAT-P-6, WASP-7, and WASP-15}\label{HAT-P-6, WASP-7, and WASP-15}
 \begin{deluxetable}{lcccccccc}
\tabletypesize{\footnotesize}
\tablecolumns{5}
 \tablecaption{Orbital Evolution Calculation Results.}
\tablehead{
\\[1.0em]
\colhead{$Q_{\rm 10}'$} & \colhead{$P_{\rm orb, in}$} &\colhead{$\Theta_{\rm *, in}$ } &
\colhead{$\Delta P_{\rm orb}$} & \colhead{$\Delta\Theta_{*}$}
\\
 & \colhead{(d)} &\colhead{(deg) } &
\colhead{(\%)} & \colhead{(\%)}
}
\startdata
\\[0.6em]
\multicolumn{5}{c}{\bf HAT-P-6} \\
\\[0.5em]
\hline
\\[0.1em]
$10^{6}$ & 3.88$\,-\,$3.92 & 112$\,-\,$168 &        0.7$\,-\,$1.7 & -0.2$\,-\,$0.5 \\
\\[0.1em]
$\geq\,10^{7}$ & 3.875$\,-\,$3.92 & 110$\,-\,$168&  0.6$\,-\,$1.7 & $\leq\,$ 0.4 \\
\\[0.1em]
\hline 
\\[0.1em]
\multicolumn{5}{c}{\bf WASP-7} \\
\\[0.1em]
\hline
\\[0.1em]
all & 4.97$\,-\,$5.04 &  80$\,-\,$92 & 0.3$\,-\,$1.7 & 0.2$\,-\,$0.9\\
\\[0.1em]
\hline 
\\[0.1em]
\multicolumn{5}{c}{\bf WASP-15} \\
\\[0.1em]
\hline
\\[0.1em]
all &3.845$\,-\,$3.935 &  -144$\,-\,$-100& 2.4$\,-\,$4.7 &0.4$\,-\,$1.2\\
\\[0.1em]
\hline 
\\[0.1em]
\multicolumn{5}{c}{\bf WASP-71} \\
\\[0.1em]
\hline
\\[0.1em]
$10^{6}$  & 4.04$\,-\,$4.44 &  24$\,-\,$ 70 & 28$\,-\,$35 & 28$\,-\,$61\\
\\[0.1em]
$10^{7}$  & 4.04$\,-\,$4.465 & 14$\,-\,$68 & 28$\,-\,$35 & 20$\,-\,$32\\
\\[0.1em]
$\geq\,10^{8}$  & 4.04$\,-\,$4.465 & 14$\,-\,$68&  28$\,-\,$35 & $\simeq$15$\,-\,$29\\
\\[0.1em]
\hline 
\\[0.1em]
\multicolumn{5}{c}{\bf WASP-4} \\
\\[0.1em]
\hline
\\[0.1em]
$10^{6}$ & 1.5$\,-\,$1.6 & 74$\,-\,$82 &        11$\,-\,$16 & 38$\,-\,$99\\
$10^{7}$ & 1.5 & 2$\,-\,$72 &        11 & 12$\,-\,$85\\
$10^{8}$ & 1.5 & 2$\,-\,$76 &        11 & 7.9$\,-\,$28\\
$10^{10}$ & 1.5 & 2$\,-\,$76 &        11 & 7.7$\,-\,$17\\
\\[0.1em]
\hline 
\\[0.1em]
\multicolumn{5}{c}{\bf WASP-16, $\Theta_{\rm *,in}\,\geq\,0^{o}$} \\
\\[0.1em]
\hline
\\[0.1em]
$10^{6}$ & 3.15$\,-\,$3.4 & 2$\,-\,$74 &        1.0$\,-\,$8.3 & 2.9$\,-\,$36\\
$10^{7}$ & 3.15$\,-\,$3.3 & 2$\,-\,$72 &        1.0$\,-\,$5.5 & 2.1$\,-\,$5.3 \\
$\geq\,10^{8}$ & 3.15$\,-\,$3.3 & 2$\,-\,$72 &        1.0$\,-\,$5.5 & $\simeq$1.0$\,-\,$3.1 \\
\\[0.1em]
\hline 
\\[0.1em]
\multicolumn{5}{c}{\bf WASP-16, $\Theta_{\rm *,in}\,\leq\,0^{o}$} \\
\\[0.1em]
\hline
\\[0.1em]
$10^{6}$ & 3.15$\,-\,$3.2 & -2$\,-\,$-52 & 1.0$\,-\,$2.6 & 8.3$\,-\,$36\\
$10^{7}$ & 3.15$\,-\,$3.2 & -2$\,-\,$-44 & 1.0$\,-\,$2.6 & 3.2$\,-\,$5.3 \\
$10^{8}$ & 3.15$\,-\,$3.2 & -2$\,-\,$-42 & 1.0$\,-\,$2.6 & 1.2$\,-\,$2.6 \\
$10^{10}$ & 3.15                & -2$\,-\,$-30 & 1.0             &    0.8$\,-\,$1.3 \\
\\[0.1em]
\hline 
\\[0.1em]
\multicolumn{5}{c}{\bf HAT-P-8} \\
\\[0.1em]
\hline
\\[0.1em]
$10^{6}$     & 3.4$\,-\,$3.9 & -84$\,-\,$ -20 & 9.5$\,-\,$21 & 14$\,-\,$ 63\\
\\[0.1em]
 $10^{7}$     & 3.4$\,-\,$3.9 & -84$\,-\,$ -10 & 9.5$\,-\,$21 & 14$\,-\,$23\\
\\[0.1em]
 $\geq\,10^{8}$      & 3.4$\,-\,$3.9 & -84$\,-\,$ -10  & $\simeq$9.0$\,-\,$21 & 9.5$\,-\,$20\\
\\[0.1em]
\hline 
\\[0.1em]
\multicolumn{5}{c}{\bf XO-4} \\
\\[0.1em]
\hline
\\[0.1em]
$10^{6}$ & 4.17$\,-\,$4.21 & -68$\,-\,$ -42 & 1.1$\,-\,$2.1 &  2.1$\,-\,$9.3\\
\\[0.1em]
 $\geq\,10^{7}$  & 4.17$\,-\,$4.21 & -68$\,-\,$ -40 & $\simeq$1.1$\,-\,$2.1 & $\simeq$1.5$\,-\,$3.0\\
\\[1.0em]
\enddata
\tablecomments{The percent change in $P_{\rm orb}$ and $\Theta_{*}$ are given by \mbox{$\Delta P_{\rm orb}\,=\,100\,\times\,(P_{\rm orb, in}-P_{\rm orb, pr})/P_{\rm orb, in}$} and \mbox{$\Delta\Theta_{*}\,=\,100\,\times\,(\Theta_{\rm *, in}-\Theta_{\rm *, pr})/\Theta_{\rm *, in}$}, respectively, where the subscripts ``in'' and ``pr'' denote initial and present values. Here we list the full $P_{\rm orb, in}$ interval for which we find solutions. For WASP-16, whose $\lambda$ is consistent with 0$^{\circ}$ within 1\,$\sigma$, we scan the initial parameter space both in $\Theta_{\rm *, in}\geq\,0^{\circ}$ and $\Theta_{\rm *, in}\leq\,0^{\circ}$. \mbox{HAT-P-8} and \mbox{XO-4} are listed as additional examples and are not described in the main text. These represent two of the oldest systems (http://exoplanet.eu/, thus interesting for orbital evolution calculations) among those with $e\,=\,0$, $\Omega_{*}/\Omega_{o}\,\textless\,1$ at present (assuming $i_{*}\,=\,90^{o}$), and $\lambda\,\neq\,0$ within 1$\sigma$ (see \S~\ref{System Selection}). The stellar models representative of \mbox{HAT-P-8} and \mbox{XO-4} are chosen randomly among those whose $M_{*}$ and $Fe/H$ differ from the observed mean values by less than 0.01$M_{\odot}$ and 0.01, respectively, at some point during the evolution of the star. The HAT-P-8- (XO-4-) type star has $M_{*}\,=\,1.2\,M_{\odot}$ (1.33$\,M_{\odot}$) and  $Z\,=\,0.02$ (0.018). For HAT-P-8 (XO-4) we consider $P_{\rm orb, in}$ between 3.3$\,-\,$4.0\,d (4.0$\,-\,$4.5\,d) in steps of 0.05\,d (0.01\,d). For HAT-P-8 (XO-4) we find that $(\Omega_{*}/\Omega_{o})_{\rm in}$ is between 0.2$\,-\,$1 (0.6$\,-\,$1) for any $Q'_{\rm 10}$. For HAT-P-8, we find solutions for $\gamma_{\rm MB}$ between 0$\,-\,$1 for any $Q'_{\rm 10}$, while for XO-4 the allowed $\gamma_{\rm MB}$ is between 0$\,-\,$0.1 for $Q'_{\rm 10}\,\leq\,10^{7}$ and 0$\,-\,$0.2 for $Q'_{\rm 10}\,\geq\,10^{8}$. We derive an age of $\simeq\,3.3\,-\,3.5\,$Gyr and $1.7\,-\,1.9\,$Gyr for HAT-P-8 and XO-4, respectively. These are consistent with the ages quoted in the literature for both systems \citep{Mancini+13, McCullough+08}.}
\label{Tab:OrbitalEvolResultsPandTheta}
\end{deluxetable}

HAT-P-6 harbors a $\simeq\,1.1\,M_{\rm Jup }$ planet orbiting a $\simeq\,1.3\,M_{\odot}$ F star every $\simeq\,3.8\,$d. The stellar metallicity is $Z\,\simeq\,0.015$ and the observed sky-projected misalignment is $\lambda\,=\,165.0^{o}\,\pm\,6^{o}$. The orbital inclination was determined by \cite{Noyes+08} and found to be $i_{o}\simeq\,85.51^{o}$. \mbox{WASP-7} hosts a $\simeq\,1.0\,M_{\rm Jup }$ planet orbiting a $\simeq\,1.3\,M_{\odot}$ F5V \citep{Hellier+09b} star every $\simeq\,4.9\,$d. For this system $Z\,\simeq\,0.02$, $\lambda\,=\,86^{o}\,\pm\,6^{o}$, and $i_{o}\,\simeq\,87.03^{o}$ \citep{Southworth+11b}. \mbox{WASP-15} harbors a $\simeq\,0.6\,M_{\rm Jup }$ planet orbiting a $\simeq\,1.3\,M_{\odot}$ F5 star every $\simeq\,3.8\,$d. The stellar metallicity is $Z\,\simeq\,0.02$ and the observed sky-projected misalignment is $\lambda\,=\,-139.6$$^{o}$$^{+5.2}_{-4.3}$. The orbital inclination was determined by \cite{Southworth+13} and found to be $i_{o}\,\simeq\,85.74^{o}$. There is no evidence of additional companions in these systems \citep{Adams+13, Triaud+10,Bergfors+13}. More references and parameters are in Table~\ref{Table:hotJupiterSample}.

During the scan of the initial parameter space, we consider initial $\Theta_{*}$, $\Omega_{*}/\Omega_{o}$, $\gamma_{\rm MB}$, and $i_{*}$, as described in \S~\ref{Orbital Evolution Procedure}. For \mbox{HAT-P-6}, \mbox{WASP-7}, and \mbox{WASP-15} we consider $P_{\rm orb, in}$ between 3.85\,--\,4.05\,d, 4.9-5.1\,d, and 3.8\,--\,4.1\,d, respectively, in steps of 0.005\,d. During the integration, we compute the term $k/T$ related to convective damping of the equilibrium tide [$(k/T)_{CD}$, Eq.~(\ref{eq:kOverTConv})] and radiative damping of the dynamical tide [$(k/T)_{RD}$, Eq.~(\ref{eq:kOverTRad})] and apply the stronger of the two. At present, $(k/T)_{RD}/(k/T)_{CD}\,\textless\,7\times\,10^{-4}, \textless\,8\times\,10^{-5}$, and $\textless\,2\times\,10^{-4}$ for \mbox{HAT-P-6}, \mbox{WASP-7}, and WASP-15, respectively. The difference between \mbox{HAT-P-6} and \mbox{WASP-7} can be attributed to the different $Z$. Despite the comparable mass, the sub-solar metallicity of \mbox{HAT-P-6} results in less surface convection and thus a higher contribution due to radiative damping. The ratio $(k/T)_{RD}/(k/T)_{CD}$ for \mbox{WASP-15} is in between those for \mbox{WASP-7} and \mbox{HAT-P-6}, as expected given the star's properties. \mbox{WASP-15} has a stronger contribution from radiative damping than \mbox{WASP-7} because its higher mass (and similar $Z$) yields less surface convection. On the other hand, it has a weaker contribution from radiative damping than \mbox{HAT-P-6} because WASP-15's similar mass and higher metallicity yield more surface convection. 

The sequences which at some point during their evolution match each system's observed properties as described in \S~\ref{Orbital Evolution Procedure} are summarized in Table~\ref{Tab:OrbitalEvolResultsPandTheta}. The allowed $P_{\rm orb, in}$ is $\sim$\,4\,d for \mbox{HAT-P-6} and \mbox{WASP-15}, and 5\,d for \mbox{WASP-7}.
The allowed $\Theta_{\rm *, in}$ is between $\sim$110$^{o}$\,--\,170$^{o}$, 80$^{o}$\,--\,90$^{o}$, and -145$^{o}$\,--\,-100$^{o}$ for \mbox{HAT-P-6}, \mbox{WASP-7}, and \mbox{WASP-15}, respectively.

HAT-P-6's inclination is not constrained \citep{Noyes+08} and we find solutions for $i_{*}$ values down to $\sim\,30^{o}$. For \mbox{WASP-7}, \cite{Albrecht+12b} attempted to use the technique of \cite{Schlaufman+10} to estimate $i_{*}$. This technique involves the comparison of the measured $v_{rot}{\rm sin\,}i_{*}$ with the expected value of $v$ for a star of the given mass and age. They found $i_{*}\simeq\,90^{o}$, indicating no evidence of an inclination of the stellar spin axis towards the observed, but they pointed out that \mbox{WASP-7} is at the upper end of the mass range for which \cite{Schlaufman+10} calculated his rotation, mass, age relationship. We find solutions for $i_{*}$ values down to $\simeq\,70^{o}$. 
WASP-15's stellar inclination is not constrained and \cite{Triaud+10} used an approach similar to the one adopted here. They assumed an isotropic distribution for $i_{*}$ to compute the true misalignment from the sky-projected one.
We find solutions for $i_{*}$ values down to $\sim\,20^{o}$.

The age of our systems ranges between $\simeq\,$1\,--\,1.7\,Gyr, $\simeq\,$0.7\,--\,2\, Gyr, and $\simeq\,$1.5\,--\,2\,Gyr for \mbox{HAT-P-6}, \mbox{WASP-7}, and \mbox{WASP-15}, respectively. For \mbox{HAT-P-6}, \cite{Noyes+08} derived an age of 2.3$^{+0.5}_{-0.7}\,$Gyr from evolutionary tracks \citep{Yi+01} and an independent estimate of the age obtained from the Ca$^{+}$\,H and K line emission strength yielded agreement. For \mbox{WASP-7} and \mbox{WASP-15}, predictions from evolutionary models set an age of 2.4$^{+0.8}_{-1.1}\,$Gyr and 2.4$^{+0.6}_{-0.7}\,$Gyr, respectively (\citealt{Southworth+11b, Southworth+13} and references therein). The ages derived with our detailed stellar modeling agree with the ages reported in the literature. Quantitatively, the results presented so far do not depend significantly on the value of $Q'_{\rm 10}$ adopted. Qualitatively, only the evolution of \mbox{HAT-P-6} is affected by $Q'_{\rm 10}$, as discussed below. 
\begin{figure} [!h]
\epsscale{1.1}
\plotone{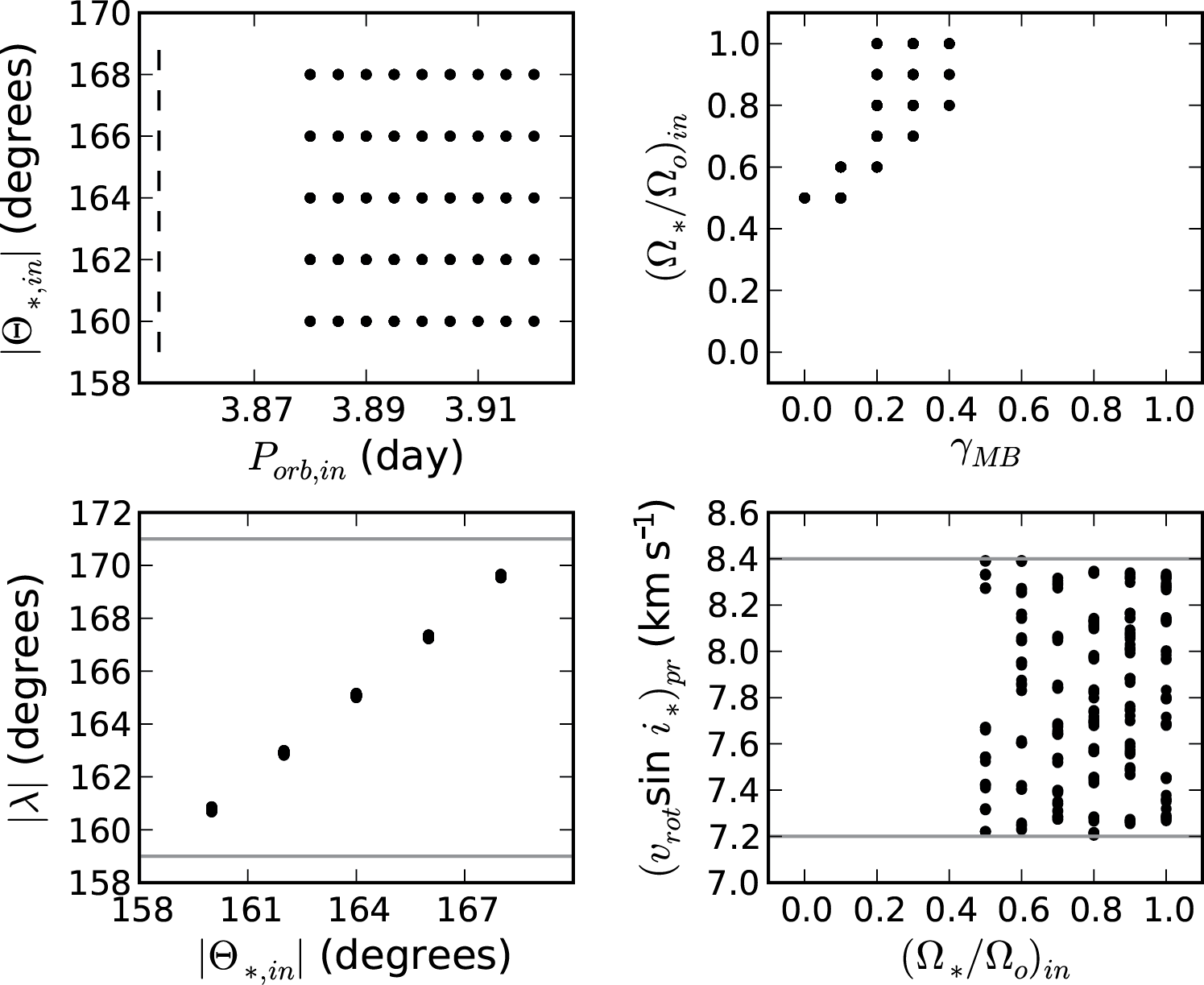}
\caption{Initial parameter space for HAT-P-6-type systems for $i_{*}$ close to 90$^{o}$. Here $Q_{10}'\,=\,10^{7}$, but the results do not change significantly for $Q_{10}'\,=\,10^{6}, 10^{8}$ or $10^{10}$. {Top}: Initial misalignment $\Theta_{*, in}$ as a function of the initial orbital period $P_{\rm orb, in}$ (left) and initial degree of asynchronism between the stellar spin frequency and the planet orbital frequency $(\Omega_{*}/\Omega_{o})_{in}$ as a function of the magnetic braking coefficient $\gamma_{\rm MB}$ (right). {Bottom: } present sky-projected misalignment $\lambda$ as a function of $\Theta_{*, in}$ (left) and present star's rotational velocity $v_{\rm rot}\,{\rm sin\,}i_{*}$ as a function of $(\Omega_{*}/\Omega_{o})_{in}$ (right). The grey solid lines represent the 1$\sigma$ observational constraints at present for a given parameter. To give a sense for the evolution of $P_{\rm orb}$, in the top left panel we mark with a vertical dashed line the mean value of the orbital period {\it at present}. As here $i_{*}$ is close to 90$^{o}$, from $i_{o}$\,=\,85.51$^{o}$ it follows that $\lambda\,\simeq\,\Theta_{*}$.}
\label{fig:orbitalPropertiesEnd_HAT-P-6_Q10_1eM7}
\end{figure}
\begin{figure} [!h]
\epsscale{1.1}
\plotone{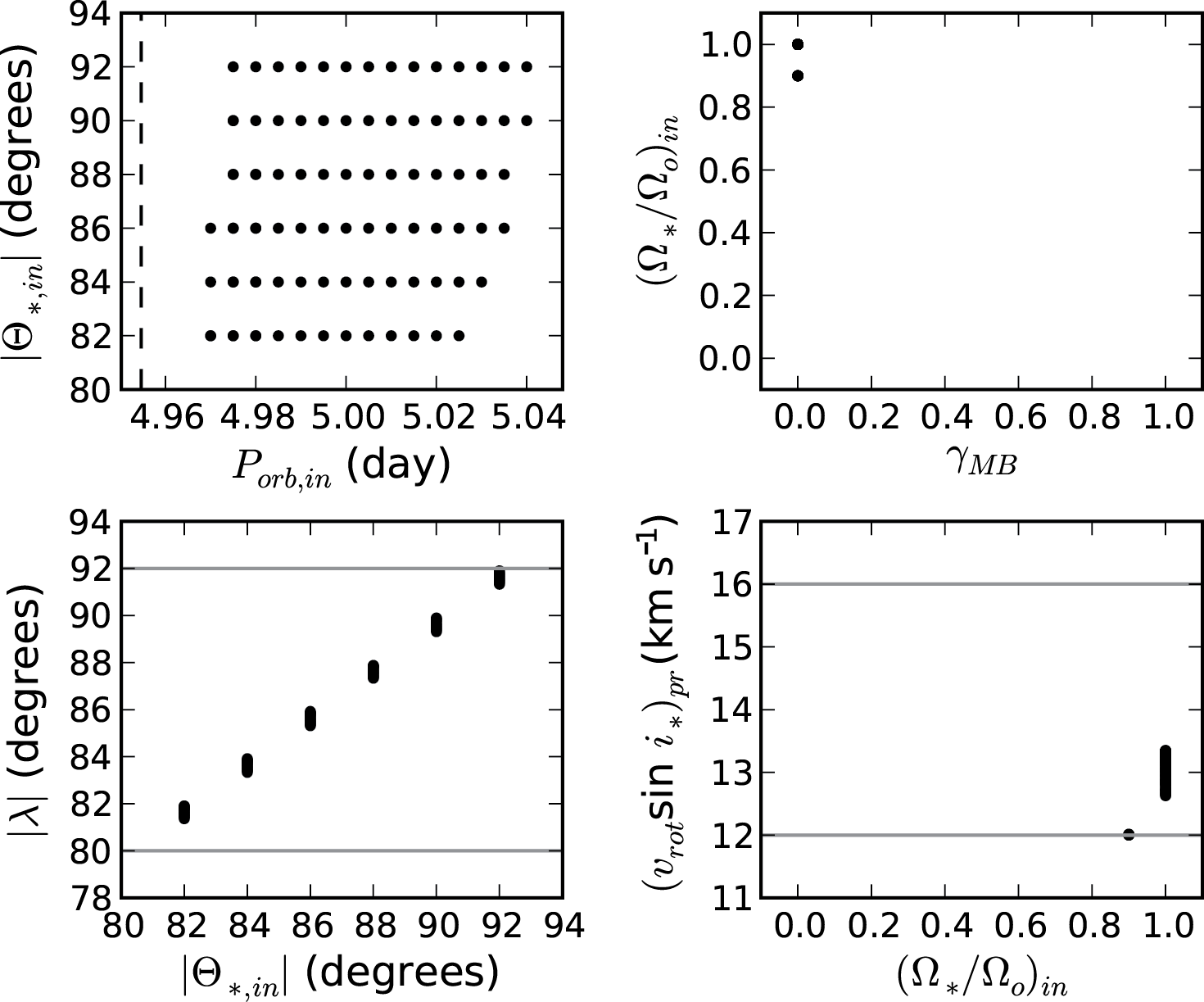}
\caption{Same as Fig.~\ref{fig:orbitalPropertiesEnd_HAT-P-6_Q10_1eM7} but for WASP-7-type systems.}
\label{fig:orbitalPropertiesEnd_WASP-7_Q10_1eM7}
\end{figure}
\begin{figure} [!h]
\epsscale{1.1}
\plotone{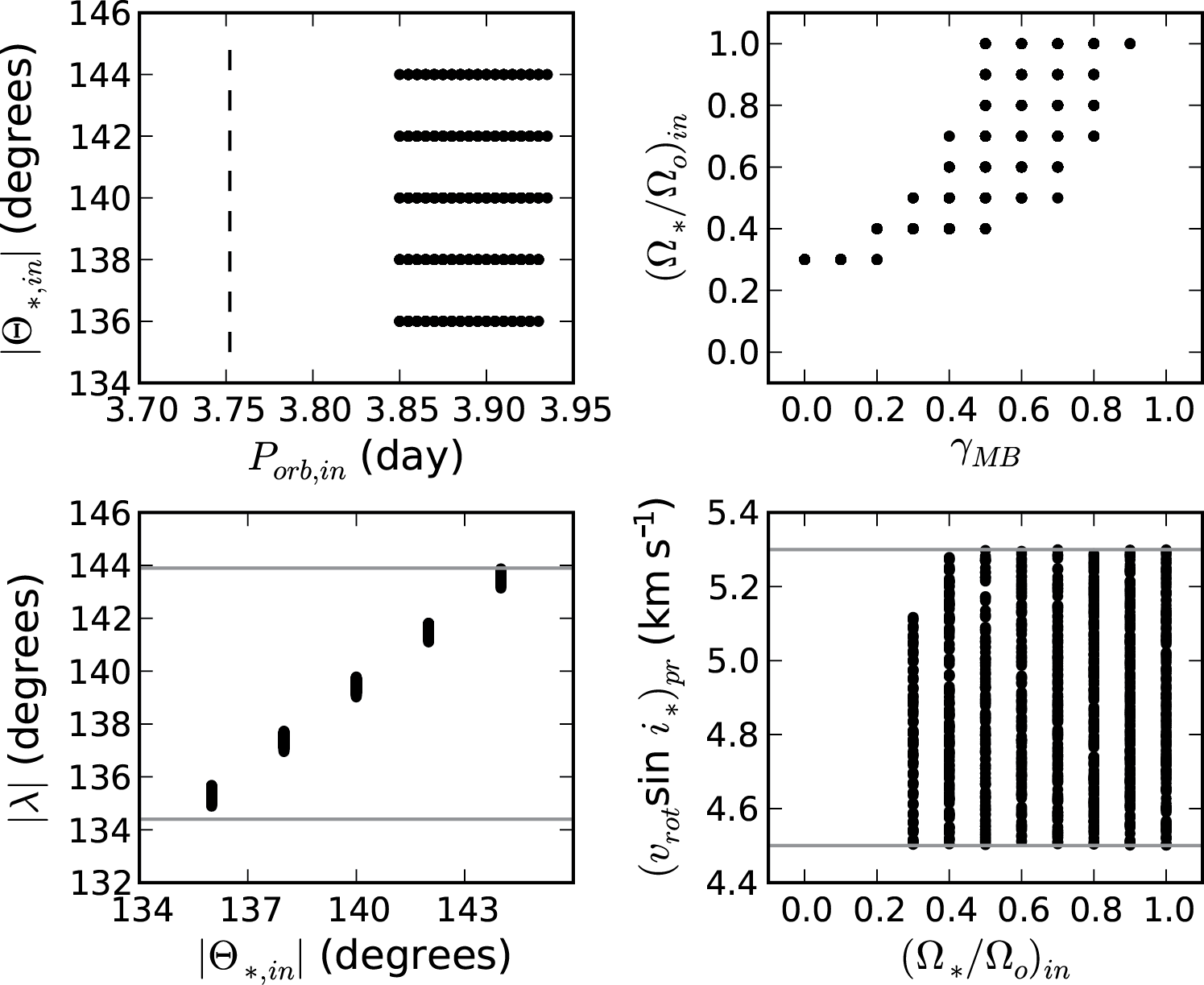}
\caption{Same as Fig.~\ref{fig:orbitalPropertiesEnd_HAT-P-6_Q10_1eM7} but for WASP-15-type systems.}
\label{fig:orbitalPropertiesEnd_WASP-15_Q10_1eM7_inclStar_89to91}
\end{figure}

In Figures.~\ref{fig:orbitalPropertiesEnd_HAT-P-6_Q10_1eM7}, \ref{fig:orbitalPropertiesEnd_WASP-7_Q10_1eM7}, and \ref{fig:orbitalPropertiesEnd_WASP-15_Q10_1eM7_inclStar_89to91} we show the parameter space for \mbox{HAT-P-6}, \mbox{WASP-7}, and WASP-15, respectively. We only display the sequences where the stellar inclination $i_{*}$ is close to 90$^{o}$ and \mbox{$Q'_{\rm 10}\,=\,10^{7}$}, for simplicity. First, we describe the overall parameter space, while we provide detailed examples of orbital evolution sequences at the end of this section.

For each system, the present orbital period (vertical dashed line in the top-left panels) is shorter that any of the allowed $P_{\rm orb, in}$, indicating that tides are the main mechanism driving the evolution of the orbital separation. This effect causes $P_{\rm orb}$ to decrease only slightly (by at most $\sim\,5\%$, see Table~\ref{Tab:OrbitalEvolResultsPandTheta}).
Initial orbital periods shorter (longer) than the values constrained above shrink to the currently observed value before (after) the star reaches (has crossed) the observed $M_{*}$, $R_{*}$, and $T_{\rm eff *}$. WASP-15's bigger change in $P_{\rm orb}$ might be attributed to the star's older age and to its present stellar radius and orbital separation. 
WASP-15 has an age in the range ($0.4\,-\,0.6\,)t_{\rm MS}$, which implies that tides had more time to act. Furthermore, it has the largest $R_{*}/a$ and this results in stronger tides. 


The evolution of $\Theta_{*}$ differs between the three systems. In the case of \mbox{HAT-P-6}, both convective damping of equilibrium tides and dissipation of inertial waves are at play, depending on $Q'_{\rm 10}$. For $Q'_{\rm 10}\,=\,10^{6}$ the misalignment can either increase or decrease slightly (by at most $\simeq\,$0.2\% and 0.5\%, respectively), depending on the system configuration. For higher $Q'_{\rm 10}$ convective damping of the equilibrium tide always causes $\Theta_{*}$ to decrease during the evolution (by at most $\simeq\,0.4$\%). In \mbox{WASP-7} and \mbox{WASP-15} this effect decreases the misalignment very slightly (by at most $\sim\,1\%$) for any $Q'_{\rm 10}$. Comparing \mbox{HAT-P-6} and \mbox{WASP-7} (Table~\ref{Tab:OrbitalEvolResultsPandTheta}), the orbital period change is similar and there is a bigger change in $\Theta_{*}$ for the latter. This is due the allowed $\Theta_{*, \rm in}$ values for \mbox{WASP-7} (close to 90\,$^{o}$) and the dependency of $(\dot{\Theta}_{*})_{\rm wf}$ in Eq.~(\ref{eq:dThetadtHut}) to sin$\,\Theta_{*}$. 

Given the resolution adopted during the scan of the initial parameter space and considering $\Omega_{*}/\Omega_{o}$ between 0\,--\,1, HAT-P-6's (WASP-15's) observed $v_{\rm rot}{\rm sin\,}i_{*}$ can be matched with $\Omega_{*}/\Omega_{o}$ between 0.5\,--\,1 (0.3\,--\,1) and $\gamma_{MB}$ between 0\,--\,0.4 (0\,--\,0.9); see Fig.~\ref{fig:orbitalPropertiesEnd_HAT-P-6_Q10_1eM7} and \ref{fig:orbitalPropertiesEnd_WASP-15_Q10_1eM7_inclStar_89to91}, top-right panel. 
The values of $\gamma_{\rm MB}$ for which we find solutions agree with those adopted in the literature for F-dwarfs (e.g., \citealt{BarkerOgilvie2009, DobbsDixon+04}; M10). 
The lower limit on $\Omega_{*}/\Omega_{o}$ is set by the observed $v_{\rm rot}{\rm sin\,}i_{*}$ and by the stellar spin-down driven by changes in the star's moment of inertia and magnetic braking, if present. This occurs  independently on the $Q'_{\rm, 10}$ adopted. Orbital configurations with $\Omega_{*}/\Omega_{o}$ smaller than the lower limits quoted above, yield a rotation rate for the star smaller than what is observed, even without any magnetic braking. Clearly, as the initial $\Omega_{*}/\Omega_{o}$ increases,  higher values of $\gamma_{\rm MB}$ are allowed to match the present $v_{\rm rot}{\rm sin\,}i_{*}$.
A similar argument holds for \mbox{WASP-7}, but the allowed initial parameter space in $\Omega_{*}/\Omega_{o}$ and $\gamma_{\rm MB}$ is smaller. Specifically, WASP-7's higher $v_{\rm rot}{\rm sin\,}i_{*}$ can be matched with $\Omega_{*}/\Omega_{o}$ between 0.9\,--\,1 and  no magnetic braking (Fig.~\ref{fig:orbitalPropertiesEnd_WASP-7_Q10_1eM7} top-right panel). The stellar spin decreases, driven mainly by changes in $I_{*}$. As a result, initial values of $\Omega_{*}/\Omega_{o}$ below 0.9 yield $v_{\rm rot}{\rm sin\,}i_{*}$ smaller than the observed range. More solutions are found if $(\Omega_{*}/\Omega_{o})_{\rm in}\,=\,$1 and the parameter space would widen considerably allowing $(\Omega_{*}/\Omega_{o})_{\rm in}\,\textgreater\,$1, potentially changing the evolutionary picture presented here (e.g., M10). Even though the prescription proposed by \cite{Lai12} is valid for sub-synchronous stars,  we show below that values of $(\Omega_{*}/\Omega_{o})_{\rm in}\,\geq\,$1 could be justified for the case of \mbox{WASP-7}. In fact, inertial wave dissipation is very inefficient for this system. 
\begin{figure} [!h]
\epsscale{1.1}
\plotone{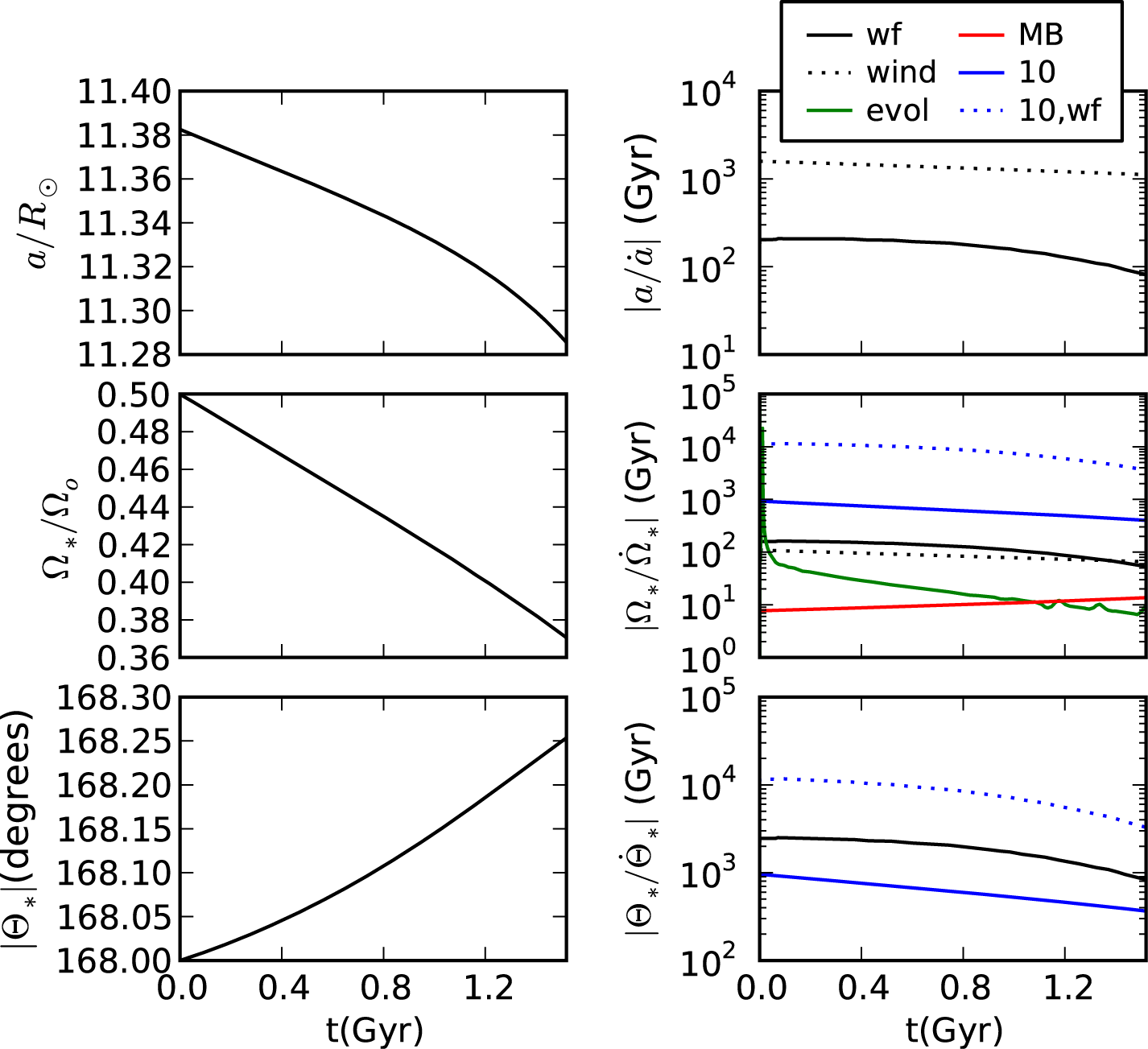}
\caption{Detailed orbital evolution of HAT-P-6-type systems. Left:  evolution of the orbital separation (top), level of asynchronism between the stellar spin and the planet's orbital frequency (middle), and misalignment (bottom). Right:  evolution of the timescales associated with the physical effects considered. Specifically, ``wf'' refers to tides in the weak friction approximation [Eqs.~(\ref{eq:dadtTidesStartrack})-(\ref{eq:dThetadtHut})], ``evol'' refers to changes in the star's moment of inertia due to stellar evolution [Eq.~(\ref{eq:OmegaDotInertia})], ``10'' and ``10,wf'' refer to dissipation of inertial waves [Eqs.~(\ref{eq:51Lai12}), (\ref{eq:52Lai12}), and (\ref{eq:CrossLaiTerm}), respectively], ``wind'' refers to stellar wind mass loss [Eq.~(\ref{eq:dadtWind}) and (\ref{eq:dOmegadtWind})], and ``MB'' refers to magnetic braking [Eq.~(\ref{eq:dOmegadtMB})]. The initial conditions are: $P_{\rm orb} = 3.9\,$d, $\Omega_{*}/\Omega_{o}  = 0.5$, and $\Theta_{*} = 168^{o}$. Furthermore, $\gamma_{MB}\,=\,0.1$, $i_{*}\,=\,88^{o}$, and $Q_{10}'\,=\,10^{6}$.}
\label{fig:DetailedOrbitalEv_HAT-P-6_timescales_ThetaIncrease}
\end{figure}
\begin{figure} [!h]
\epsscale{1.1}
\plotone{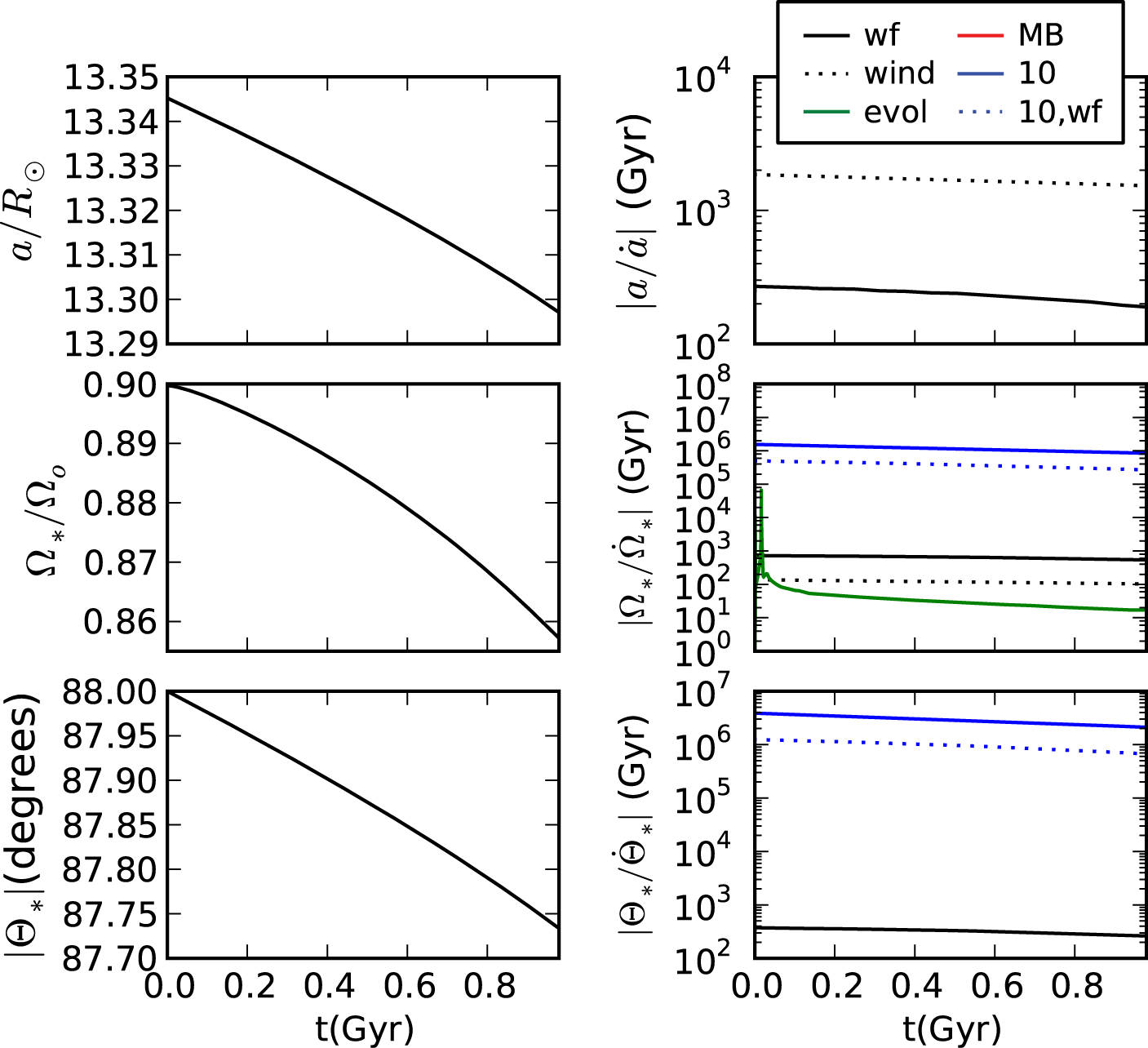}
\caption{Same as Fig.~\ref{fig:DetailedOrbitalEv_HAT-P-6_timescales_ThetaIncrease}, but for WASP-7-type systems. The initial conditions are: $P_{\rm orb} = 4.98\,$d, $\Omega_{*}/\Omega_{o}  = 0.9$, and $\Theta_{*} = 88^{o}$. Furthermore, $\gamma_{MB}\,=\,0$, $i_{*}\,=\,88^{o}$, and $Q_{10}'\,=\,10^{7}$.}
\label{fig:DetailedOrbitalEv_WASP-7_timescales}
\end{figure}
\begin{figure} [!h]
\epsscale{1.1}
\plotone{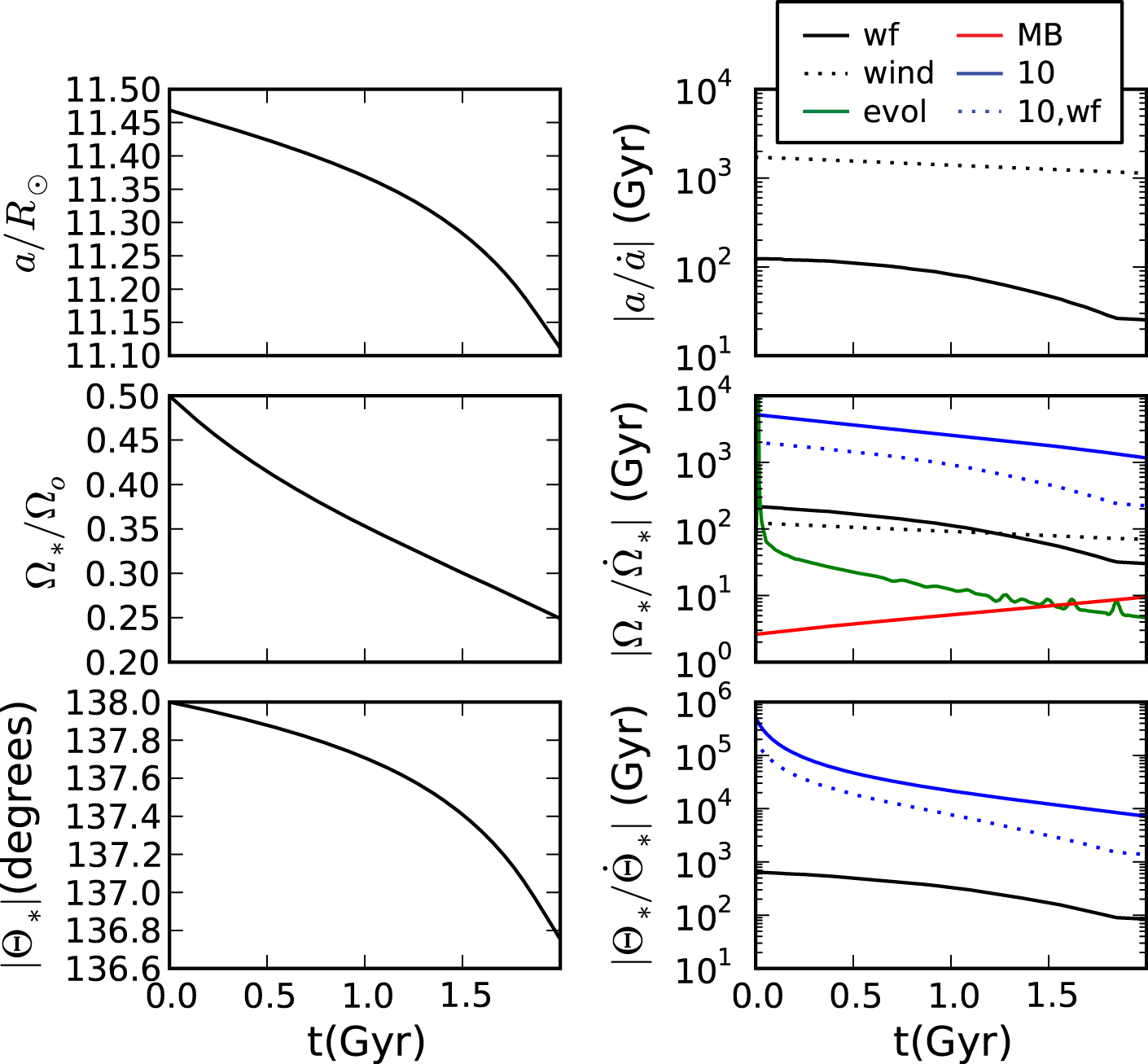}
\caption{Same as Fig.~\ref{fig:DetailedOrbitalEv_HAT-P-6_timescales_ThetaIncrease}, but for WASP-15-type systems. The initial conditions are: $P_{\rm orb} = 3.93\,$d, $\Omega_{*}/\Omega_{o}  = 0.5$, and $\Theta_{*} = -138^{o}$. Furthermore, $\gamma_{MB}\,=\,0.3$, $i_{*}\,=\,88^{o}$ and $Q_{10}'\,=\,10^{7}$. The change in slope in the ``wf'' line at $\simeq\,$1.9\,Gyr is due to the term $f_{\rm *,conv} $ in Eq.~(\ref{eq:fConv}).}
\label{fig:DetailedOrbitalEv_WASP-15_timescales}
\end{figure}

Figures.~\ref{fig:DetailedOrbitalEv_HAT-P-6_timescales_ThetaIncrease}, \ref{fig:DetailedOrbitalEv_WASP-7_timescales} and \ref{fig:DetailedOrbitalEv_WASP-15_timescales} show the detailed orbital evolution of a HAT-P-6-, WASP-7-, and WASP-15-type system, respectively. For \mbox{HAT-P-6} we set $Q'_{\rm 10}\,=\,10^{6}$, while for \mbox{WASP-7} and \mbox{WASP-15} we show $Q'_{\rm 10}\,=\,10^{7}$, as an example. 
In each system, the evolution of $a$ is driven by convective dissipation of equilibrium tides, which tends to decrease the orbital separation (top panels). 
The evolution of WASP-7's spin is driven by changes in the star's moment of inertia (we have no solutions with $\gamma_{\rm MB}\textgreater\,0$, more below). This effect is dominant, together with  magnetic braking in the evolution of HAT-P-6's and \mbox{WASP-15's} spins (middle panels). Finally, the evolution of $\Theta_{*}$ in \mbox{HAT-P-6} is driven by damping of inertial waves which causes the misalignment to increase. This behavior is due to the negative value of (cos$\,\Theta_{*}+S/L)$ in Eq.~\ref{eq:52Lai12} (see also Appendix~\ref{Tests on the Orbital Evolution code}). This effect is inefficient in \mbox{WASP-7} and WASP-15, where convective damping of the equilibrium tide damps the misalignment (bottom panels). For \mbox{WASP-7} and \mbox{WASP-15} this evolutionary picture is not significantly affected by the value of $Q'_{\rm 10}$ adopted, while it varies for HAT-P-6. 
For this system, increasing $Q'_{\rm 10}$ (decreasing the strength of inertial wave dissipation),  the evolution of $\Theta_{*}$ becomes driven by convective damping of equilibrium tides and $\Theta_{*}$ decreases.

For each system, we compute the inefficiency of inertial wave dissipation from the sequences with the maximum and minimum change in ${\Theta}_{*}$.
Specifically, we compare the timescales associated with this dissipation mechanism and those associated with the main driver of spin and misalignment evolution. For $\dot{\Omega}_{*}$ we compare with the timescales associated with changes in $I_{*}$, as this term results from natural stellar evolution and it mostly dominates the evolution of $\Omega_{*}$. For $\dot{\Theta}_{*}$, we compare with the timescales associated with convective damping of equilibrium tides.
For most of HAT-P-6's evolutionary lifetime at $Q'_{\rm 10}\,=\,10^{7}$ (the smallest $Q'_{\rm 10}$ at which inertial waves start becoming non significant), we find that  the timescales associated with this effect for $\dot{\Omega}_{*}$ are $\sim\,20\,-\,500$ times longer than those associated with changes in $I_{*}$. 
For the evolution of $\Theta_{*}$, inertial wave dissipation timescales are $\sim\,1\,-\,10^{2}$ times longer than those associated with convective damping of the equilibrium tide. In \mbox{WASP-7} (WASP-15) and for $Q'_{\rm 10}\,=\,10^{6}$ we find that  the timescales associated with inertial wave dissipation for $\dot{\Omega}_{*}$ are $\sim\,10^{2}\,-\,10^{3}$ ($\sim\,10$) longer than those associated with changes in $I_{*}$ for most of the system's lifetime. For $\dot{\Theta}_{*}$, 
inertial wave dissipation timescales are $\sim\,10^{2}$ ($\sim\,10$) times longer than those associated with convective damping of the equilibrium tide. Clearly, these estimates depend on the orbital configuration.
The difference in the efficiency of inertial wave dissipation between \mbox{WASP-7} and \mbox{HAT-P-6} is due to the allowed $\Theta_{\rm *, in}$ (close to 90$^{o}$ for WASP-7) and the dependency of $(\dot{\Omega})_{\rm 10}$ [Eq.~\ref{eq:51Lai12}] and $(\dot{\Theta}_{*})_{10}$ [Eq.~\ref{eq:52Lai12}] on cos\,$\Theta_{*}$.  

The estimates provided above for the inefficiency of inertial wave dissipation are important in terms of the initial parameter space considered. Our calculations account for initial values of $\Omega_{*}/\Omega_{o}$ up to 1, thus stretching the validity of the tidal prescription proposed by \cite{Lai12}. This recipe might break-down for systems where $\Omega_{*}/\Omega_{o}\geq\,$1, as other components of the tidal response might be relevant which could affect the orbital separation. We showed above that the associated timescales can be orders of magnitude longer than those related to the main drivers of the evolution (depending on $Q'_{\rm 10}$ for \mbox{HAT-P-6}). Thus, it is reasonable to think that this would be the case also for the evolution of $a$. Considering ($\Omega_{*}/\Omega_{o})_{\rm in}\,\textgreater 1$ would considerably expand the allowed initial parameter space for all systems. In particular, it would yield more solutions matching the currently observed $v_{\rm rot}{\rm sin\,}i_{*}$ and higher values of $\gamma_{\rm MB}$ for a WASP-7-type system. 
Configurations with $(\Omega_{*}/\Omega_{o})_{\rm in}\,\textgreater\,1$ would also yield different evolutionary pictures, as in super-synchronous systems tides transfer angular momentum from the spin to the orbit, thus causing orbital expansion. 

\subsection{WASP-71}\label{WASP-71}
WASP-71 harbors a $\simeq\,$2.2$\,M_{\rm Jup }$ planet orbiting an evolved $\simeq\,$1.6$\,M_{\odot}$ F8 star every $\simeq\,$2.9\,d. This system hosts the most massive star and planet among those considered here. The stellar metallicity is $Z\,\simeq\,0.027$ and the observed sky-projected misalignment is $\lambda\,=\,$ 20.1$^{o}\,\pm\,$9.7$^{o}$. The orbital inclination was determined by \cite{Smith+13} and found to be $i_{o}\,\simeq\,$84.9$^{o}$. These authors also found no evidence of other companions in the systems. Additional references and parameters for \mbox{WASP-71} are given in Table~\ref{Table:hotJupiterSample}.

We scan the initial parameter space considering $\Theta_{*}$, $\Omega_{*}/\Omega_{o}$, $\gamma_{\rm MB}$, and $i_{*}$ as outlined in \S~\ref{Orbital Evolution Procedure}. We consider $P_{\rm orb, in}$ between 4\,--\,4.845\,d in steps of 0.01\,d. As for the systems described in the previous section, during the integration we compute $(k/T)_{CD}$ and $(k/T)_{RD}$ and apply the stronger of the two. At present, $(k/T)_{RD}/(k/T)_{CD}\,\textless\,1.5\times\,10^{-3}$. This value is higher than the one computed for \mbox{HAT-P-6}, \mbox{WASP-7}, and \mbox{WASP-15} because of the small (yet significant) amount of convection associated with the higher stellar mass. We show below that this ratio could have been about one order of magnitude higher when the star was on its Zero Age Main Sequence. Furthermore, in Appendix~\ref{Dynamic Tides in WASP-71} we solve the full set of equations describing the non-adiabatic tidal response of the the stellar model considered in detail here, and discuss the effects of dynamic tides for a wide spectrum of tidal forcing frequencies.
\begin{figure} [!h]
\epsscale{1.1}
\plotone{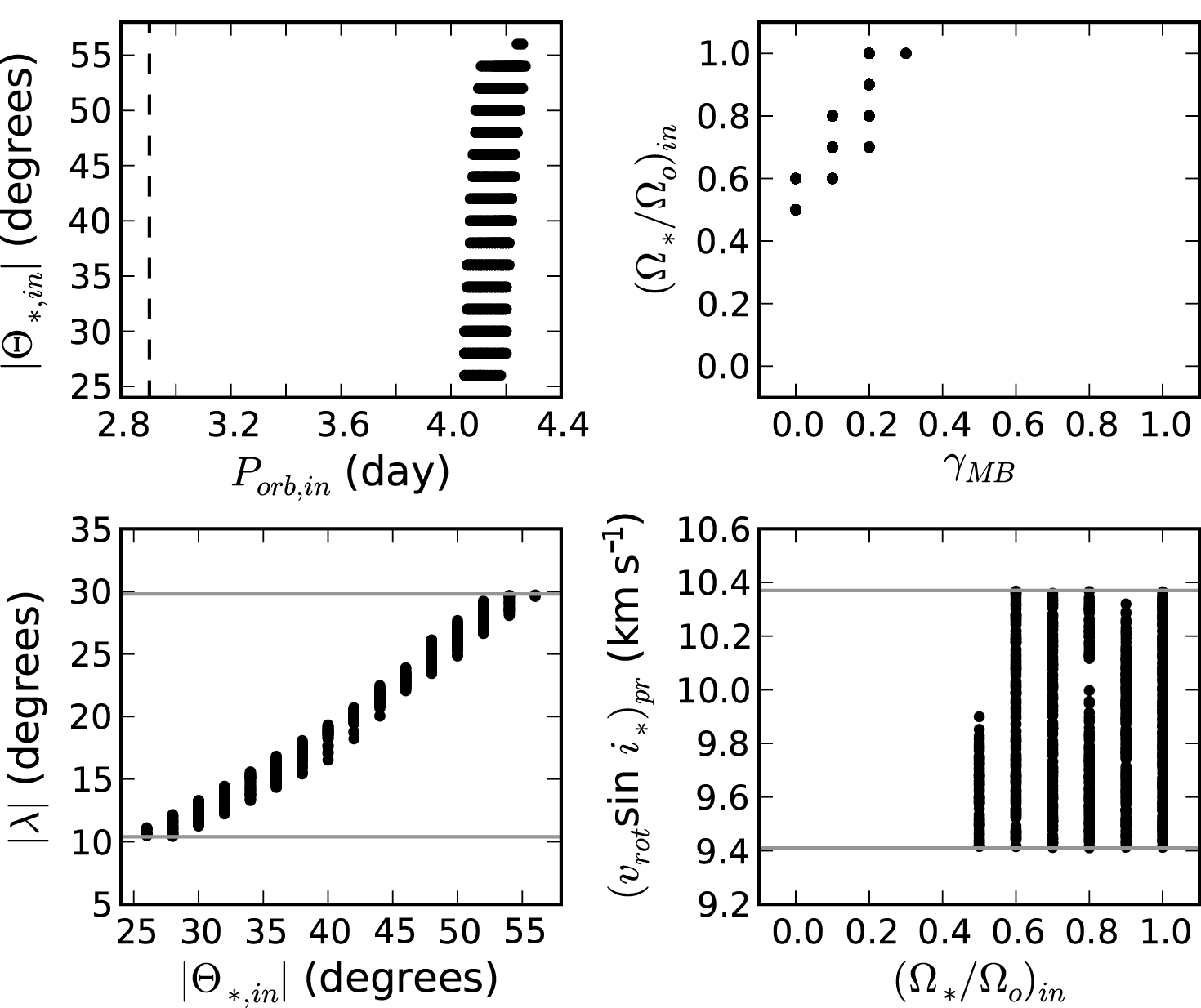}
\caption{Same as Fig.~\ref{fig:orbitalPropertiesEnd_HAT-P-6_Q10_1eM7} but for WASP-71-type systems. Here $Q_{\rm 10}'\,=\,10^{6}$.}
\label{fig:orbitalPropertiesEnd_WASP-71_Q10_1eM6_inclStar_89to91}
\end{figure}

The systems which at some point during their evolution match WASP-71's observational constraints as described in \S~\ref{Orbital Evolution Procedure} are summarized in Table~\ref{Tab:OrbitalEvolResultsPandTheta}. The allowed initial $P_{\rm orb}$ ranges between $\sim$4.0\,--\,4.5\,d for all $Q'_{\rm 10}$ values considered. Instead, the allowed initial misalignments are sensitive to the adopted inertial wave dissipation efficiency. We find that $\Theta_{*, \rm in}$ is between $\sim\,$25$^{o}$\,--\,70$^{o}$ for $Q'_{\rm 10}\,=\,10^{6}$, and $\sim\,$15$^{o}$\,--\,70$^{o}$ for higher $Q'_{\rm 10}$. 
WASP-71's inclination is not constrained \citep{Smith+13} and we find solutions for $i_{*}$ down to $\simeq\,40^{o}$.
The age of our systems ranges between 1.9\,--\,2\,Gyr for all $Q'_{\rm 10}$ values considered. The star's age is uncertain, but from lithium absorption in the spectrum and stellar models it is estimated to be between 2\,--\,3\,Gyr 
(\citealt{Smith+13} and reference therein).

In Fig.~\ref{fig:orbitalPropertiesEnd_WASP-71_Q10_1eM6_inclStar_89to91} we show the parameter space for a WASP-71-type system. We only display the sequences with $i_{*}$ close to 90$^{o}$ and $Q'_{\rm 10}\,=\,10^{6}$, for simplicity. We provide a detailed example of orbital evolution at the end of this section.

The present orbital period (vertical dashed line in the top-left panel of Fig.~\ref{fig:orbitalPropertiesEnd_WASP-71_Q10_1eM6_inclStar_89to91}) is shorter that any of the allowed $P_{\rm orb, in}$, indicating that convective damping of the equilibrium tide is the main mechanism driving the evolution of the orbital separation. This effect causes $P_{\rm orb}$ to decrease during the orbital evolution by about 30\%$\,-\,$35\%. This significant change in orbital period compared to the systems discussed in \S~\ref{HAT-P-6, WASP-7, and WASP-15} can be attributed to the star's evolutionary stage. \mbox{WASP-71} has an age of $\simeq\,0.9\,t_{\rm MS}$ (\S~\ref{Orbital Evolution Procedure} and Fig.~\ref{fig:HR_detailed_orbEv}), which left tides a longer time to significantly affect the orbital evolution. Note also that \mbox{WASP-71} has the largest ratio $R_{*}/a$ at present, which yields the strongest tides.
We show below that stellar wind mass loss might give a small contribution at the beginning of the evolution, but it does not change the orbital configuration significantly. Therefore, the subsequent tidally driven evolution determines the lower and upper limits on the allowed $P_{\rm orb, in}$. 

The evolution of $\Theta_{*}$ is driven by weak-friction tides, with a contribution coming from damping of inertial waves. In fact, the misalignment decreases by about 30\%$\,-\,$60\% and 20\%$\,-\,$30\% for $Q'_{\rm 10}\,=\,10^{6}$ and $10^{7}$, respectively,  and by about 15\%$\,-\,$30\%,  for higher $Q'_{\rm 10}$. We explain why this range decreases as $Q'_{\rm 10}$ increases below. 

Given the resolution adopted during the scan of the initial parameter space and considering $\Omega_{*}/\Omega_{o}$ between 0\,--\,1, WASP-71's observed $v_{\rm rot}{\rm sin\,}i_{*}$ can be matched with $\Omega_{*}/\Omega_{o}$ between 0.5\,--\,1 and $\gamma_{MB}$ between 0\,--\,0.3 (Fig.~\ref{fig:orbitalPropertiesEnd_WASP-71_Q10_1eM6_inclStar_89to91} top-right panel) for all $Q_{\rm 10}'$ values considered. The latter interval contains the value $\gamma_{\rm MB}\,=\,$0.1 that has been previously adopted in the literature for F-dwarfs (e.g.,  \citealt{BarkerOgilvie2009, DobbsDixon+04}; M10). As for \mbox{HAT-P-6}, \mbox{WASP-7}, and WASP-15, the lower limit on $\Omega_{*}/\Omega_{o}$ is set mainly by the observed $v_{\rm rot}{\rm sin\,}i_{*}$ and by the stellar spin-down driven by changes in the star's moment of inertia and magnetic braking, if present. We show below that convective damping of equilibrium tides becomes important towards the very end of the evolution of the stellar spin. 

\begin{figure} [!h]
\epsscale{1.1}
\plotone{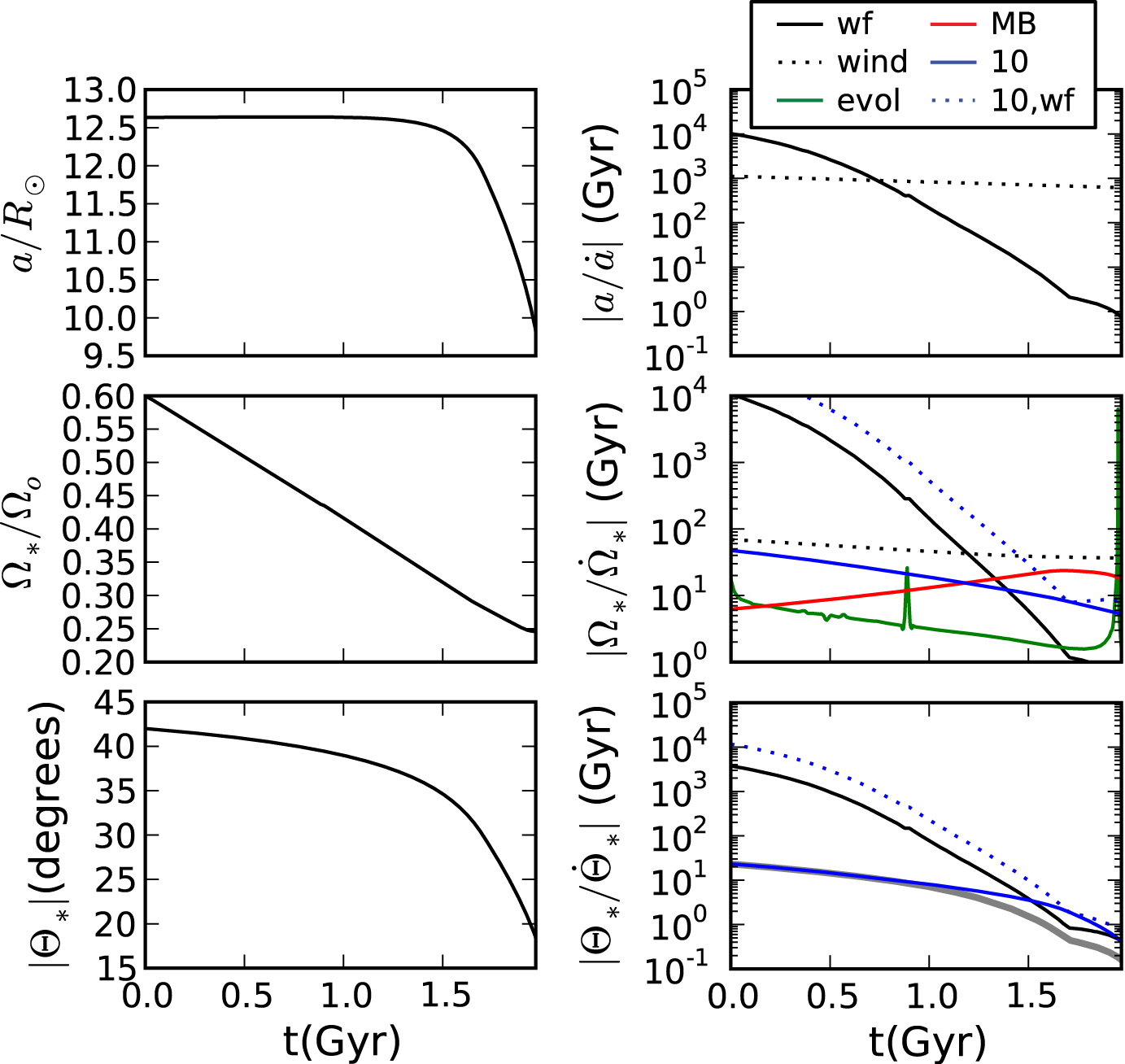}
\caption{Same as Fig.~\ref{fig:DetailedOrbitalEv_HAT-P-6_timescales_ThetaIncrease}, but for WASP-71-type systems. The initial conditions are: $P_{\rm orb} = 4.21\,$d, $\Omega_{*}/\Omega_{o}  = 0.6$, and $\Theta_{*} = 42^{o}$. Furthermore, $\gamma_{MB}\,=\,0.1$, $i_{*}\,=\,88^{o}$, and $Q_{10}'\,=\,10^{6}$. The change in slope in the ``wf'' line at $\simeq\,$1.8\,Gyr is due to the term $f_{\rm *,conv} $ in Eq.~(\ref{eq:fConv}). The thick grey solid line in the bottom-right panel represents the total $|\dot{\Theta}_{*}/\Theta_{*}|$ due to the sum of inertial wave dissipation and weak-friction tides.}
\label{fig:DetailedOrbitalEv_WASP-71_timescales}
\end{figure}
\begin{figure} [!h]
\epsscale{1.1}
\plotone{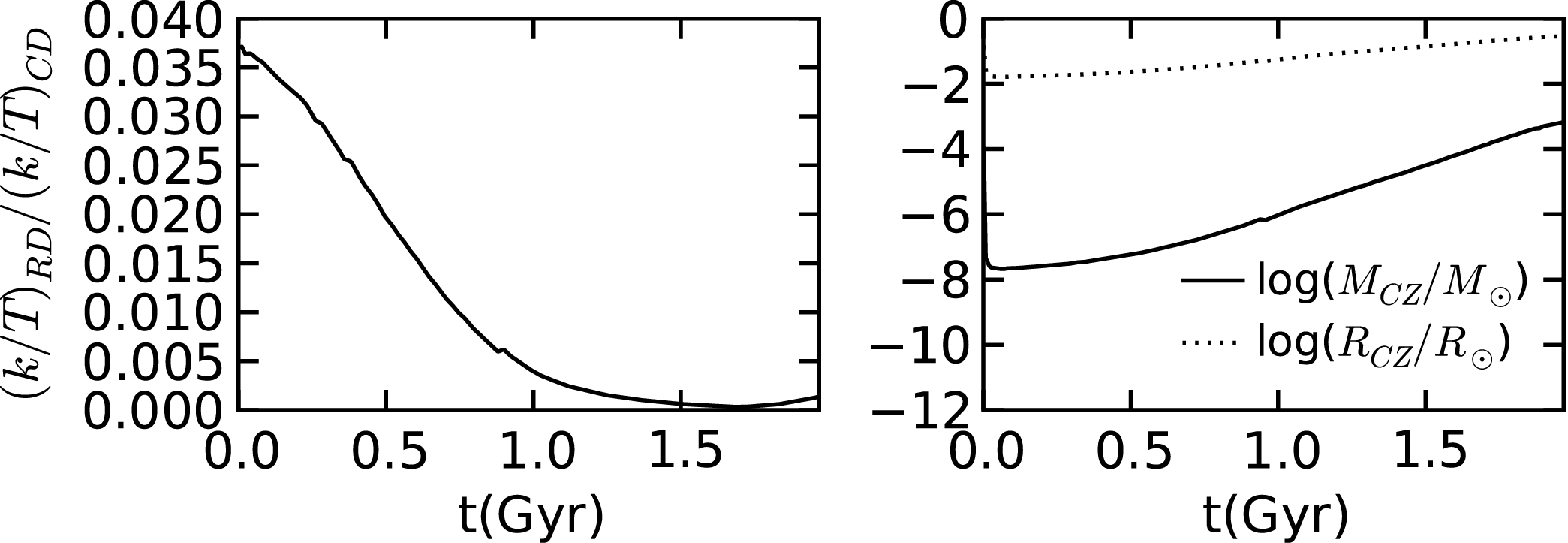}
\caption{Effect of convection for the system presented in Fig.~\ref{fig:DetailedOrbitalEv_WASP-71_timescales}. Left: evolution of the ratio of the apsidal motion constant $k$ over the timescale $T$ of tidal dissipation related to radiative damping of the dynamical tide [$(k/T)_{RD}$] and convective damping of the equilibrium tide [$(k/T)_{CD}$]. Right: mass (solid line) and radial (dotted line) extent of the surface convection zone. The ratio $(k/T)_{RD}/(k/T)_{CD}$ decreases with time as the amount of surface convection increases. }
\label{fig:DetailedOrbitalEv_WASP-71_timescales_convection}
\end{figure}

Fig.~\ref{fig:DetailedOrbitalEv_WASP-71_timescales} shows the detailed orbital evolution of a WASP-71-type system. Here we set $Q'_{\rm 10}\,=\,10^{6}$, as an example. For this system, we show the evolution of $(k/T)_{RD}/(k/T)_{CD}$ and the properties of the surface convection zone in Fig.~\ref{fig:DetailedOrbitalEv_WASP-71_timescales_convection} .
For the first $\simeq\,$0.7\,Gyr, the evolution of $a$ is driven by stellar wind mass loss. This would cause orbital expansion, but the associated timescale is too long to affect the orbit significantly. Eventually, convective dissipation of equilibrium tides becomes dominant. This effect decreases the orbital separation (top panels in Fig.~\ref{fig:DetailedOrbitalEv_WASP-71_timescales}). 
The evolution of the stellar spin is driven mainly by magnetic braking and by changes in the star's moment of inertia, which cause the star to spin down. After $\simeq\,$1.7\,Gyr convective damping of equilibrium tides becomes relevant, preventing $\Omega_{*}$ to decrease further  (middle panels in Fig.~\ref{fig:DetailedOrbitalEv_WASP-71_timescales}). This picture doesn't change significantly with $Q'_{\rm 10}$. For $Q'_{\rm 10}\,=\,10^{6}$, the timescales associated with inertial wave dissipation are at least 1 order of magnitude longer than the timescales associated with the dominant physical effects driving $\dot{\Omega}_{*}$ for most of the evolutionary lifetime. Finally, the evolution of $\Theta_{*}$ is driven by inertial wave dissipation, with a  contribution from convective damping of equilibrium tides towards the end of the evolution. Both effects damp the misalignment on a timescale which, at present, is a factor of $\sim$\,10 shorter than $|\dot{a}/a|$ (bottom panels in Fig.~\ref{fig:DetailedOrbitalEv_WASP-71_timescales}, see also Table~\ref{Tab:OrbitalEvolResultsPandTheta}, where $\Delta\,\Theta_{*}$ can be up to a factor of two bigger than $\Delta\,P_{\rm orb}$ for $Q'_{\rm 10}\,=\,10^{6}$). Increasing $Q'_{\rm 10}$ yields longer timescales for inertial wave dissipation, but it doesn't affect the timescales associated with convective damping of equilibrium tides. This explains why the fractional change in $\Theta_{*}$ decreases as $Q'_{\rm 10}$ increases. Finally, we note that equilibrium tides become more important as the mass and radial extent of the stellar surface convective layers increase during the evolution of the star (Fig.~\ref{fig:DetailedOrbitalEv_WASP-71_timescales_convection})

As for the systems described in the previous section, more solutions could be found considering $(\Omega_{*}/\Omega_{o})_{\rm in}\,\textgreater\,$1. However, as the allowed initial parameter space varies with $Q'_{\rm 10}$, inertial wave dissipation would probably affect the orbital separation for such configurations. 
\subsection{WASP-16}\label{WASP-16}
WASP-16 harbors a $\simeq\,0.8\,M_{\rm Jup }$ planet  in a $\simeq\,3.1\,$d orbit around a $\simeq\,1.0\,M_{\odot}$ G-dwarf . The star has a solar metallicity and its misalignment is $\lambda\,=\,11.0^{o}$$^{+26}_{-19}$, thus consistent with 0$^{\circ}$. The orbital inclination is $i_{o}\simeq\,83.99^{o}$ \citep{Southworth+13}. Additional parameters and references are in Table~\ref{Table:hotJupiterSample}.

We proceed as for the systems described in the previous sections, but considering $P_{\rm orb, in}$ between 3.1\,--\,5\,d in steps of 0.05\,d, and considering both $\Theta_{\rm *, in}\geq\,0^{\circ}$ and $\leq\,0^{\circ}$. The systems which eventually resemble WASP-16 are summarized in Table~\ref{Tab:OrbitalEvolResultsPandTheta}. 
The allowed initial $P_{\rm orb}$ and $\Theta_{*}$ range between $\sim$3.15\,--\,3.4\,d and $\sim\,-52^{\circ}\,-\,74^{\circ}$, respectively, depending on the sign of $\Theta_{\rm *, in}$ and on the $Q'_{\rm 10}$ values considered. 
As for the systems described in the previous sections, $(k/T)_{CD}\,\textgreater\,(k/T)_{RD}$ throughout WASP-16's evolution (as expected, since the stellar $M_{*}$ and $Z$ make of WASP-16 a sun-like star).

\begin{figure} [!h]
\epsscale{1.1}
\plotone{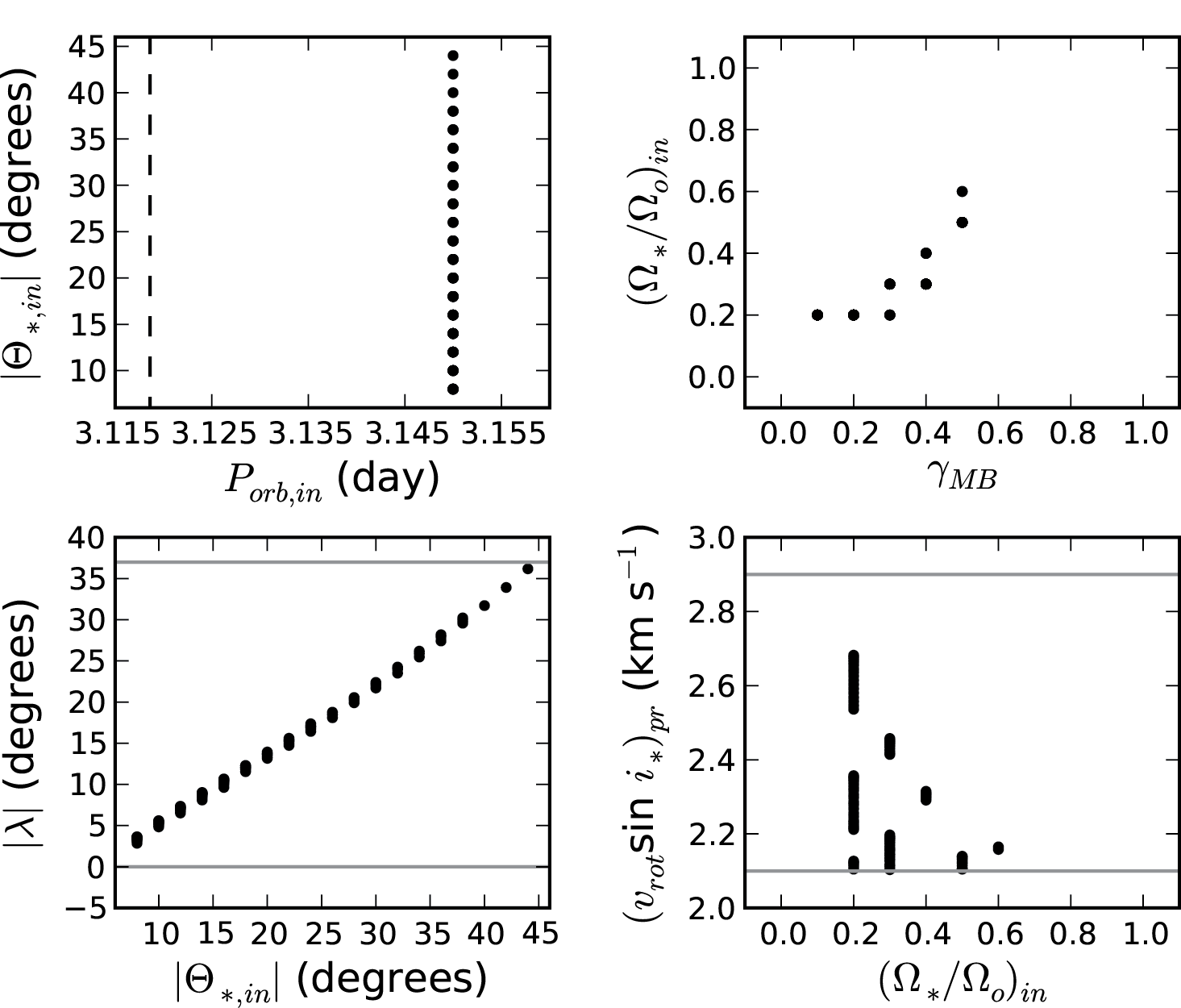}
\caption{Same as Fig.~\ref{fig:orbitalPropertiesEnd_HAT-P-6_Q10_1eM7} but for WASP-16-type systems. Here $Q_{\rm 10}'\,=\,10^{6}$.}
\label{fig:orbitalPropertiesEnd_WASP-16}
\end{figure}
For WASP-16's inclination, we find solutions for $i_{*}$ down to $\simeq\,17^{\circ}$ ($36^{\circ}$) for $\Theta_{\rm *, in}\geq\,0^{\circ} (\leq\,0^{\circ})$.
The age computed with our modeling ranges between 6.6\,--\,8.5\,Gyr (6.6\,--\,8.3\,Gyr) for all $Q'_{\rm 10}$ values considered and $\Theta_{\rm *, in}\geq\,0^{\circ} (\leq\,0^{\circ})$. The stellar age quoted in the literature is uncertain and it comprises values ranging from 2.3$^{+5.8}_{-2.2}$ Gyr (\citealt{Lister+09} and references therein) to 8.6$^{+3.4}_{-2.9}$ Gyr (\citealt{Southworth+13} and reference therein).  

The initial parameter space is displayed in Fig.~\ref{fig:orbitalPropertiesEnd_WASP-16} for $i_{*}$ close to 90$^{o}$ and $Q'_{\rm 10}\,=\,10^{6}$, for simplicity. 

Tides drive the evolution of the orbital separation, causing $P_{\rm orb}$ to decrease during the orbital evolution by $\sim\,$1\%$\,-\,$10\%, depending on $Q'_{\rm 10}$ and the sign of $\Theta_{\rm *, in}$. 
The evolution of $\Theta_{*}$ is driven by inertial wave dissipation alone when $Q'_{\rm 10}\,=\,10^{6}$. This effect decreases the misalignment, potentially by up to $\sim\,$35\%. The decrease in the misalignment becomes less significant as $Q'_{\rm 10}$ is increased.  If $Q'_{\rm 10}\,=\,10^{7}$, the inertial wave dissipation timescales become comparable to the timescales associated to weak-friction tides. Increasing further $Q'_{\rm 10}$, the evolution of $\Theta_{*}$ is driven by weak-friction tides alone. 

As far as $\Omega_{*}$ is concerned, WASP-16's observed $v_{\rm rot}{\rm sin\,}i_{*}$ can be matched with $\Omega_{*}/\Omega_{o}$ between 0.2\,--\,1 for any $Q_{\rm 10}'$, considering $\Theta_{\rm *, in}\,\geq\,0^{\circ}$. Instead,  if $\Theta_{\rm *, in}\,\leq\,0^{\circ}$, the allowed $(\Omega_{*}/\Omega_{o})_{\rm in}$ values range between 0.2\,--\,1 (0.2$\,-\,$0.6) for  $Q'_{\rm 10}\,\leq\,10^{8}$ ($Q'_{\rm 10}\,=\,10^{10}$). The allowed $\gamma_{MB}$ values are between 0\,--\,0.5 (0.1$\,-\,0.5$) for  $\Theta_{\rm *, in}\,\geq\,0^{\circ}$ ($\Theta_{\rm *, in}\,\leq\,0^{\circ}$), while this parameter is normally set to 1 for G-dwarfs (e.g.,  \citealt{BarkerOgilvie2009, DobbsDixon+04}; M10). We attribute this discrepancy to the upper limit imposed on the initial $\Omega_{*}/\Omega_{o}$ values considered. The lower limit on $\Omega_{*}/\Omega_{o}$ is set by the observed $v_{\rm rot}{\rm sin\,}i_{*}$ and by the stellar spin-down driven, as we show below, by magnetic braking. 
\begin{figure} [!h]
\epsscale{1.1}
\plotone{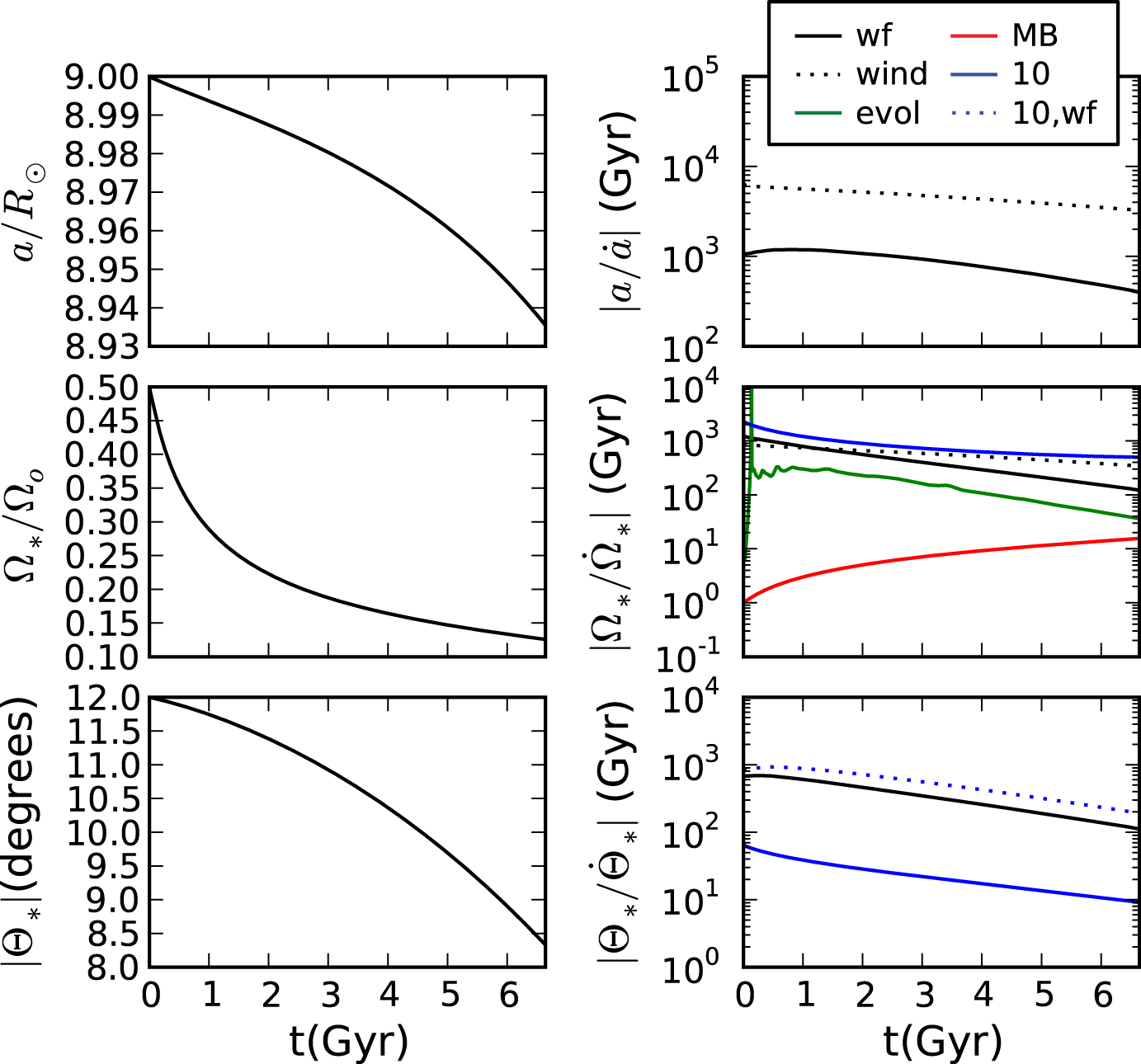}
\caption{Same as Fig.~\ref{fig:DetailedOrbitalEv_HAT-P-6_timescales_ThetaIncrease}, but for WASP-16-type systems. The initial conditions are: $P_{\rm orb} = 3.15\,$d, $\Omega_{*}/\Omega_{o}  = 0.5$, and $\Theta_{*} = 12^{o}$. Furthermore, $\gamma_{MB}\,=\,0.5$, $i_{*}\,=\,88^{o}$, and $Q_{10}'\,=\,10^{6}$. }
\label{fig:DetailedOrbitalEv_WASP-16_timescales}
\end{figure}

Fig.~\ref{fig:DetailedOrbitalEv_WASP-16_timescales} shows the detailed orbital evolution of a WASP-16-type system.
The evolution of $a$ is driven by tides (top panels), while the stellar spin decreases as a result of magnetic braking  (middle panels). This picture doesn't change significantly with $Q'_{\rm 10}$. For $Q'_{\rm 10}\,=\,10^{6}$, the timescales associated with inertial wave dissipation are more than 2 orders of magnitude longer than the timescales associated with the dominant driver of $\Omega_{*}$, for most of the evolutionary lifetime. Finally, the evolution of $\Theta_{*}$ is driven by inertial wave dissipation, which damps the misalignment on a timescale that is 1-2 orders of magnitude shorter than $a/\dot{a}$ (bottom panels). As for WASP-71, increasing $Q'_{\rm 10}$ yields longer timescales for inertial wave dissipation, but it doesn't affect the timescales associated with weak-friction tides. This results in a decreasing $\Delta\,\Theta_{*}$ range as $Q'_{\rm 10}$ increases. 

As for WASP-71, more solutions could be found considering $(\Omega_{*}/\Omega_{o})_{\rm in}\,\textgreater\,$1. However, as inertial wave dissipation can be very significant in the evolution of $\Theta_{*}$, it would probably affect the orbital separation for such configurations.
\section{Discussion}\label{Discussion}
\begin{figure} [!h]
\epsscale{1.15}
\plotone{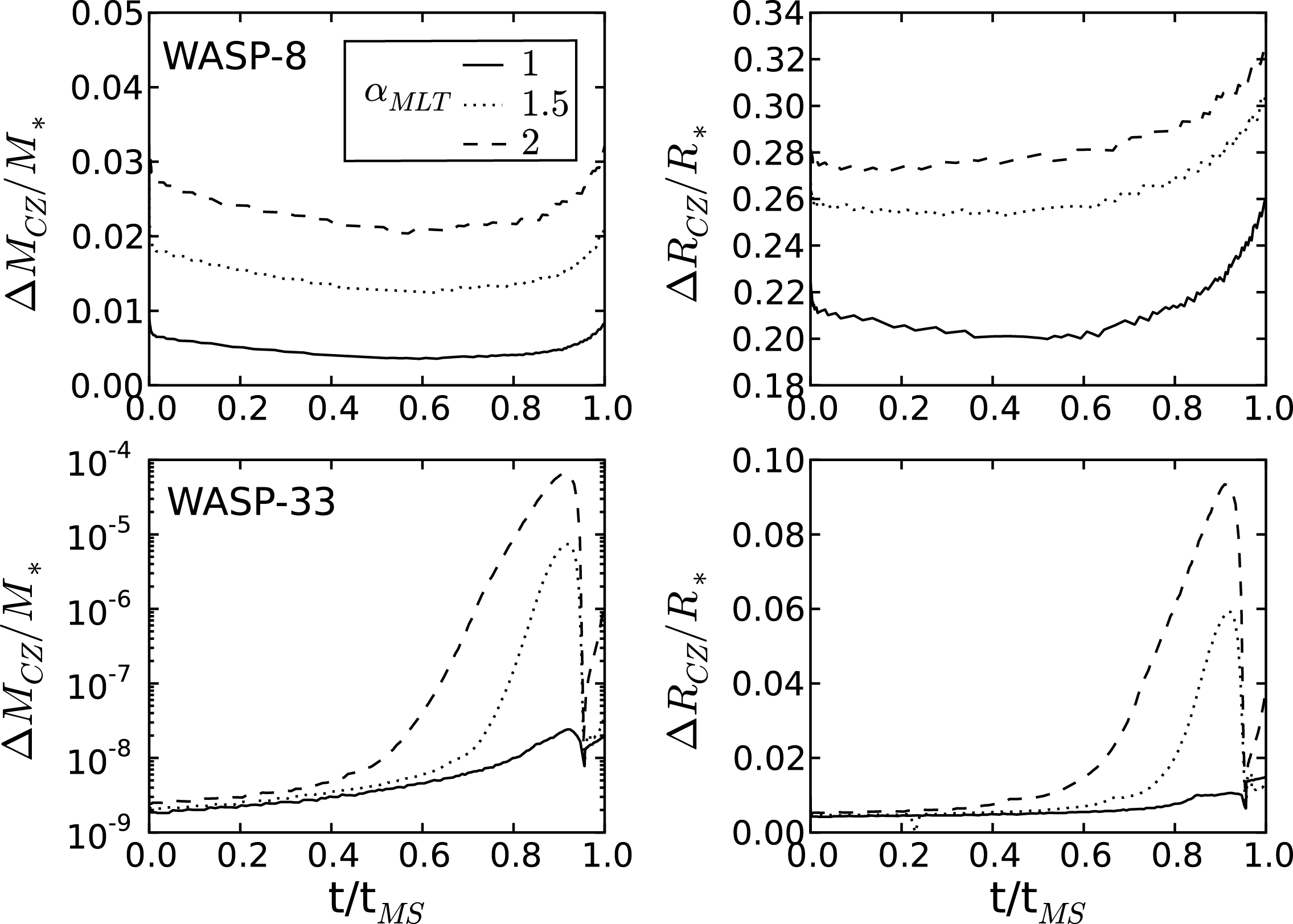}
\caption{Evolution of the fractional mass (left) and radius (right) of the surface convection zone for a WASP-8- (top) and WASP-33-type (bottom) star, and for $\alpha_{\rm MLT}\,=\,1$ (solid), 1.5 (dotted), 2 (dashed). The time is in units of the Main Sequence lifetime.}
\label{fig:ConvectionAndAlphaMLT}
\end{figure}

In \S~\ref{Detailed Stellar modeling with MESA} we showed that our detailed stellar modeling with MESA leads to a trend between the observed sky-projected misalignment and the amount of surface convection inside the host star, especially if one looks at $\lambda$ as a function of $\Delta\,R_{\rm CZ}/R_{*}$. MESA has been calibrated to reproduce the extent of surface convection in the sun only \citep{PBDHLT2011, Paxton+13} and such a calibration for stars with different $M_{*}$ and $Z$ has yet to be performed (Matteo Cantiello, private communication). Apart for $M_{*}$ and $Z$, another parameter which affects the amount of surface convention is the mixing length $\alpha_{\rm MLT}$ parameter. Here we discuss how $\Delta\,M_{\rm CZ}/M_{*}$ and $\Delta\,R_{\rm CZ}/R_{*}$ vary, if $\alpha_{\rm MLT}$ is changed. As test cases, we use WASP-8 and WASP-33, discussed in Fig~\ref{fig:HR_WASP33_WASP8}. For these two stars, the evolution of $\Delta\,M_{\rm CZ}/M_{*}$ and $\Delta\,R_{\rm CZ}/R_{*}$ is displayed in Fig.~\ref{fig:ConvectionAndAlphaMLT} for $\alpha_{\rm MLT}\,=\,1, 1.5,$ and 2 (thus bracketing the values normally adopted in the literature, e.g., \citealt{PBDHLT2011, Paxton+13}). For a solar-type star like WASP-8, $\Delta\,M_{\rm CZ}/M_{*}$ and $\Delta\,R_{\rm CZ}/R_{*}$ increase by a factor of at most $\simeq\,$6 and $\simeq\,$1.5, respectively, if $\alpha_{\rm MLT}$ is varied from 1 to 2. Instead, for a more massive star like WASP-33 younger than 0.5$\,t_{\rm MS}$, $\Delta\,M_{\rm CZ}/M_{*}$ and $\Delta\,R_{\rm CZ}/R_{*}$ increase by a factor of at most $\simeq\,2.5$ and 2, respectively, if $\alpha_{\rm MLT}$ is varied from 1 to 2. However, $\Delta\,M_{\rm CZ}/M_{*}$ and $\Delta\,R_{\rm CZ}/R_{*}$ can increase by up to $\sim\,1000$ and $\sim\,10$ when the star is at the end of its main sequence (0.9$\,t_{\rm MS}$).
Therefore, even though the results presented here might not change significantly for solar-type stars in general, or for more massive stars younger than $\simeq\,$0.5$t_{\rm MS}$, caution should be used when modeling more massive and evolved stars.

In this work we also adopted several simplifying assumptions, which we summarize and discuss below.

Firstly, we followed the regime of validity of the \cite{Lai12} prescription and considered circular systems. As in other investigations of the observed  $\lambda\,-\,T_{\rm eff *}$ correlation (e.g., \citealt{WinnFAJ10}, A12, R13), we neglected the past evolution of the eccentricity. As summarized by \cite{BarkerOgilvie2009}, \cite{Hut1981} considered tides in the star and showed that the misalignment evolves on a longer timescales than the eccentricity in systems where the ratio between stellar spin and orbital angular momentum is smaller than $\simeq\,6$. As this is the case for typical hot Jupiter systems, if we observe a misaligned close-in planet in a circular orbit we cannot rule out a significant eccentricity in the past. 
M10 carried out detailed orbital evolution calculations of hot Jupiters in eccentric and misaligned orbits, accounting for tides both in the star and the planet. They found that stellar tides are expected to largely dominate the evolution of the orbital separation and obliquity. Instead, the rate of circularization depends on the relative efficiency of tidal dissipation inside the star and the planet and, thus, on the stellar and planetary tidal quality factors ($Q_{*}$ and $Q_{\rm pl}$, respectively). Their detailed calculations showed that different $Q_{*}$ and $Q_{\rm pl}$ are needed to explain different systems. Furthermore, depending on $Q_{\rm pl}$, circularization can be achieved much faster than orbital decay. Here we focused on systems where the current eccentricity has been found to be consistent with zero. Therefore, the inclusion of tides in the planet would have amounted to calibrate $Q_{\rm pl}$ for each system, in order to achieve circularization ``fast enough''. M10 showed that orbits can reach circularization in as little as $\sim\,100\,$Myr, while the youngest system studied in detail here has an age of $\sim\,1\,$Gyr (WASP-7). For this work we decided not to introduce additional model parameters and we neglected tides in the planet entirely. A detailed calculation accounting for this effect on the evolution of the eccentricity will be the subject of future work. Such calculation should also take into account the evolution of the host star (neglected by M10) since it might play a significant role, as demonstrated here. The evolution of the star's moment of inertia can significantly decrease the stellar spin. This, in turn, might affect the other parameters entering the problem, if the equations for $\dot{a}$, $\dot{e}$, $\dot{\Omega}_{*}$, and $\dot{\Theta}_{*}$ are solved simultaneously.

Secondly, for stars with predominantly radiative envelopes (e.g., \mbox{WASP-33-type} stars in Fig.~\ref{fig:HR_WASP33_WASP8}) we parametrized the efficiency of radiative damping of the dynamical tide as in \cite{HurleyTP02}. The authors used results presented by \cite{Zahn1975}, who performed detailed calculations of dynamic tides in massive main sequence binaries, in the limit of small tidal forcing frequencies. In Appendix~\ref{Dynamic Tides in WASP-71} we showed that radiative dissipation of the dynamical tide may be significant, once the fully non-adiabatic tidal response of the star for a broad spectrum of tidal forcing frequencies is taken into account. Therefore, host stars with predominantly radiative envelopes could require a more detailed orbital evolution calculation. This should account for the interaction between tides and the star's free oscillation modes, and its effect on the evolution of the orbital separation and stellar spin. Such calculation is computationally expensive and beyond the scope of this work, but we note that \cite{WitteSavonije02} studied the tidal evolution of eccentric binaries hosting a solar~-~type main~-~sequence star and a Jupiter-like planet. Their calculations accounted for the effects of stellar evolution, tidal dissipation in the star in the framework of dynamical tides, and resonant interaction with the g-modes and quasi-toroidal oscillation eigenmodes. The authors found that the dynamical tide with the inclusion of the effects of resonances with the stellar oscillation modes yields more efficient tidal coupling than convective damping of the equilibrium tide.

Finally, here we have assumed a constant value for $Q'_{\rm 10}$ throughout the evolution. \cite{BarkerOgilvie2009} and \cite{OgilvieLin2007} showed that $Q'_{\rm 10}$ can vary greatly between different stars, as it depends not only on the stellar spin, but also on the amount of surface convection. The latter, in turn, depends on the stellar mass, metallicity and evolutionary stage. This implies that assuming a constant value of $Q'_{\rm 10}$ is probably not accurate. However, we note that we could find evolutionary paths for all systems considered and for the full range of $Q'_{\rm 10}$ adopted ($\sim\,10^{6}-10^{10}$). Furthermore,  our results do not change significantly once $Q'_{\rm 10}\,\geq\,10^{7}$ for most systems. These findings  suggest that the evolutionary sequences presented here would not be affected dramatically with a more detailed treatment of inertial waves dissipation.
\section{Summary and Conclusions}\label{Conclusions}
Two formation models have been proposed to explain the tight orbits of hot Jupiters. These giant planets could migrate inward in a disk (in the so-called {\it disk migration} scenario), or they could be formed via tidal circularization of a highly eccentric orbit following gravitational interactions with a companion (in the so-called {\it high-eccentricity migration} scenario). Disk migration yields orbits where the stellar spin and orbital angular momentum vectors are nearly aligned, while high-eccentricity migration results in a broad range of misalignments. Here we targeted the known hot Jupiters where the obliquity has been inferred observationally (following and updating the sample considered by \citealt{Albrecht+12}) and investigated whether their properties are consistent with high-eccentricity migration. 

In contrast to previous studies, we modeled in detail each host star and showed that the observed increase in misalignment $\lambda$ with the star's effective temperature $T_{\rm eff}$ is shaped by the amount of convection inside the host star, as originally suggested by \cite{WinnFAJ10}.
Specifically, higher degrees of misalignment are found in stars with less surface convection, especially if one considers the radial extent of the surface convective region.
This result supports the hypothesis that giant planets are formed with a broad initial distribution of misalignments. During the subsequent orbital evolution, convective dissipation of tides in the star is the mechanism that shapes the observed distribution of misalignments.
To further test this hypothesis, we computed the coupled evolution of the orbital elements and stellar spin of five representative systems: one aligned, two prograde, and two retrograde systems.
We studied in detail \mbox{HAT-P-6}, WASP-7, 15, 16, and 71, and provided results for few more systems, as additional examples. Within the regime of validity of the tidal prescription adopted (\S~\ref{Tidal Contribution due to inertial wave dissipation}), WASP-16 is an aligned star among the coolest ones. Instead, \mbox{WASP-71} and \mbox{WASP-7} (\mbox{WASP-15} and \mbox{HAT-P-6}) are among the coolest and hottest prograde (retrograde) stellar hosts, respectively. 
In contrast to previous studies on the observed  $\lambda\,-\,T_{\rm eff *}$ correlation (e.g., \citealt{Albrecht+12,RogersLin13}), we took into account the combined effects of tides, stellar wind mass loss, magnetic braking, and stellar evolution. The tidal prescription adopted combines tides in the weak friction approximation and inertial wave dissipation, and it was recently proposed by \cite{Lai12} to explain the currently observed aligned hot Jupiters. For the efficiency of inertial wave dissipation, we followed numerical results by \cite{OgilvieLin2007} and \cite{BarkerOgilvie2009} and considered tidal quality factors $Q'_{\rm 10}$ ranging between $10^{6}\,-\,10^{10}$. 
Furthermore, we scanned a broad parameter space in initial orbital periods, misalignments, and degrees of asynchronism between the stellar spin and the planet's orbital frequency ($\Omega_{*}/\Omega_{o}$).

Our results show that, accounting for all the relevant physical mechanisms and considering the simultaneous evolution of the orbital separation, stellar spin, and misalignment, the current properties of the variety of systems considered here can be naturally explained. 

 For HAT-P-6, WASP-7, and 15, with F-dwarfs, we found that both orbital decay and obliquity damping are small even on Gyr timescales. This supports the notion that many of the F stars are presently not capable of either destroying the planet or damping the obliquity, and that the high observed obliquities are the result of the hot-Jupiters formation process.  For WASP-71, we found that both the orbit and obliquity are actively damping, but, for the smallest $Q'_{\rm 10}$ value considered, the obliquity evolves on a timescale that can be almost 1 order of magnitude shorter at present (and up to 3 orders of magnitude shorter in the past, Fig.~\ref{fig:DetailedOrbitalEv_WASP-71_timescales}).  
The same is true for WASP-16, where the obliquity damping timescale is more than 1 order of magnitude shorter than the orbital decay timescale at present (Fig.~\ref{fig:DetailedOrbitalEv_WASP-16_timescales}). This system in particular, together with WASP-4 described in \cite{VR14}, supports the idea that obliquity damping in cool and less massive G-dwarfs can occur more rapidly than orbital decay, provided that inertial wave dissipation is actively driving the misalignment evolution. This mechanism does indeed provide an explanation for the population of currently known aligned hot Jupiters.

Finally, our results show that the physical effects included in this work can all play a significant role in the orbital evolution of misaligned hot Jupiters systems, depending on the properties of the star and planet, as well as the orbital configuration, as summarized below. 

By using detailed stellar models we took into account how changes in the star's moment of inertia affect its spin. Stellar evolution efficiently decreases $\Omega_{*}$ in systems hosting F-dwarfs, where it becomes increasingly significant as the star expands during its main sequence evolution.
Another efficient driver of stellar spin-down is magnetic braking, included here according to \citeauthor{Skumanich72}'s (\citeyear{Skumanich72}) law. We varied its strength by changing the parameter $\gamma_{\rm MB}$ entering the magnetic braking prescription, but the values of $\gamma_{\rm MB}$ which yield \mbox{HAT-P-6-}, \mbox{WASP-15-}, and \mbox{71-}type systems agree with those adopted in the literature. WASP-7, and 16 are exceptions, as their present properties can be matched either {\it without} magnetic braking ($\gamma_{\rm MB}\,=\,0$, for WASP-7, with an F-dwarf) or with $\gamma_{\rm MB}\,\leq\,0.5$ (for WASP-16, with a G-dwarf). However, we argued that more systems resembling WASP-7 with $\gamma_{\rm MB}\,>\,0$ could be found considering higher initial levels of asynchronism between the stellar spin and the planet's orbital frequency.

For tides, the \citeauthor{Lai12}'s (\citeyear{Lai12}) prescription is valid for sub-synchronous systems ($\Omega_{*}\,\textless\,\Omega_{o}$) and inertial wave dissipation acts only on the stellar spin and the misalignment. Instead, in super-synchronous systems ($\Omega_{*}\,\textgreater\,\Omega_{o}$), this dissipation channel might also affect the orbital separation, but we did not explore this possibility and we targeted only configurations where $\Omega_{*}\,\leq\,\Omega_{o}$ (and discussed cases in which $\Omega_{*}\,=\,\Omega_{o}$).
As mentioned above, we showed that inertial wave dissipation can be significant, mainly for the evolution of the misalignment. This dissipation channel can increase the misalignment in systems like \mbox{HAT-P-6}, while it can actively damp it in WASP-16- and 71-type systems. It is unimportant for WASP-15- and 7-type systems and the associated timescales can be orders of magnitude longer than those related to the main drivers of spin and misalignment evolution. For systems like \mbox{WASP-7}, in particular, this finding suggests that the evolution of the orbital separation would not be significantly affected by inertial wave dissipation, if super-synchronous configurations were considered. Expanding the initial parameter space to include $\Omega_{*}/\Omega_{o}\,\textgreater\,1$ would likely yield \mbox{WASP-7-type} systems where the magnetic braking coefficient $\gamma_{\rm MB}$ is consistent with values adopted in the literature.

Tides in the weak friction approximation are the main driver of orbital decay (when $\Omega_{*}\,\textless\,\Omega_{o}$) and they can also affect the misalignment. As weak-friction tides always tend to decrease the misalignment, they can either counteract the effect of inertial wave dissipation (e.g., in \mbox{HAT-P-6-type} systems) or strengthen it (e.g., in \mbox{WASP-71-type} systems), depending on $Q'_{\rm 10}$.
Tides in the weak friction approximation are less important for the evolution of the stellar spin, except for \mbox{WASP-71}: for this system, this mechanism becomes relevant towards the end of the evolution and it counteracts the spin-down driven by stellar evolution and magnetic braking.
 
Finally, stellar wind mass loss is negligible for all systems considered here. In a \mbox{WASP-71-type} system it could have been the main mechanism driving the evolution of the orbital separation early in its past, but the associated timescales were too long to significantly affect it.

To conclude, our detailed stellar modeling, the orbital evolution calculations, and the few more examples summarized in Table~\ref{Tab:OrbitalEvolResultsPandTheta}, provide support to the high-eccentricity migration scenario for the formation of hot Jupiters. A detailed calculation similar to the one presented here on {\it all} misaligned systems is needed to provide a definite answer (this will be the subject of future work, see also \S~\ref{Discussion}). However, here we demonstrated that tidal dissipation in the host star, together with stellar wind mass loss, magnetic braking, and stellar evolution, can account for the observed distribution of misalignments (and alignments) and orbital separations found around stars of different temperatures. In particular, the orbital configuration of five representative hot Jupiter systems covering a variety of misalignments and stellar temperatures can be explained with reasonable assumptions on the physical effects driving the orbital evolution in circular and misaligned systems, despite the limited parameter space considered.

\begin{acknowledgements}
      This work was supported by NASA Grant NNX12AI86G at Northwestern University. We thank the reviewer Josh Winn for his review and appreciate the comments and suggestions, which significantly contributed to improving the manuscript. We thank Tassos Fragos for providing the routine to compute the star's moment of inertia within MESA. We also thank Adrian Barker, Nick Cowan, Matteo Cantiello, Aaron Geller, and Meagan Morsher for useful discussions during the development of this project. Computational resources supporting this work were provided by the Northwestern University ÒGrailÓ cluster, purchased with a National Science Foundation Major Research Instrumentation award (PHY-1126812), and by the Northwestern University Quest High Performance Computing (HPC) cluster. This research has made use of the NASA Exoplanet Archive, which is operated by the California Institute of Technology, under contract with the National Aeronautics and Space Administration under the Exoplanet Exploration Program.

\end{acknowledgements}
\newpage
\clearpage

\bibliographystyle{apj}

\begin{thebibliography}{}
\expandafter\ifx\csname natexlab\endcsname\relax\def\natexlab#1{#1}\fi

\bibitem[{{Adams} {et~al.}(2013){Adams}, {Dupree}, {Kulesa}, \&
  {McCarthy}}]{Adams+13}
{Adams}, E.~R., {Dupree}, A.~K., {Kulesa}, C., \& {McCarthy}, D. 2013, \aj,
  146, 9

\bibitem[{{Albrecht} {et~al.}(2012{\natexlab{a}}){Albrecht}, {Winn}, {Butler},
  {Crane}, {Shectman}, {Thompson}, {Hirano}, \& {Wittenmyer}}]{Albrecht+12b}
{Albrecht}, S., {Winn}, J.~N., {Butler}, R.~P., {et~al.} 2012{\natexlab{a}},
  \apj, 744, 189

\bibitem[{{Albrecht} {et~al.}(2011){Albrecht}, {Winn}, {Johnson}, {Butler},
  {Crane}, {Shectman}, {Thompson}, {Narita}, {Sato}, {Hirano}, {Enya}, \&
  {Fischer}}]{Albrecht+11}
{Albrecht}, S., {Winn}, J.~N., {Johnson}, J.~A., {et~al.} 2011, \apj, 738, 50

\bibitem[{{Albrecht} {et~al.}(2012{\natexlab{b}}){Albrecht}, {Winn}, {Johnson},
  {Howard}, {Marcy}, {Butler}, {Arriagada}, {Crane}, {Shectman}, {Thompson},
  {Hirano}, {Bakos}, \& {Hartman}}]{Albrecht+12}
---. 2012{\natexlab{b}}, \apj, 757, 18

\bibitem[{{Alonso} {et~al.}(2008){Alonso}, {Auvergne}, {Baglin}, {Ollivier},
  {Moutou}, {Rouan}, {Deeg}, {Aigrain}, {Almenara}, {Barbieri}, {Barge},
  {Benz}, {Bord{\'e}}, {Bouchy}, {de La Reza}, {Deleuil}, {Dvorak}, {Erikson},
  {Fridlund}, {Gillon}, {Gondoin}, {Guillot}, {Hatzes}, {H{\'e}brard},
  {Kabath}, {Jorda}, {Lammer}, {L{\'e}ger}, {Llebaria}, {Loeillet}, {Magain},
  {Mayor}, {Mazeh}, {P{\"a}tzold}, {Pepe}, {Pont}, {Queloz}, {Rauer},
  {Shporer}, {Schneider}, {Stecklum}, {Udry}, \& {Wuchterl}}]{Alonso+08}
{Alonso}, R., {Auvergne}, M., {Baglin}, A., {et~al.} 2008, \aap, 482, L21

\bibitem[{{Ammler-von Eiff} {et~al.}(2009){Ammler-von Eiff}, {Santos}, {Sousa},
  {Fernandes}, {Guillot}, {Israelian}, {Mayor}, \& {Melo}}]{AmmlerVE+09}
{Ammler-von Eiff}, M., {Santos}, N.~C., {Sousa}, S.~G., {et~al.} 2009, \aap,
  507, 523

\bibitem[{{Anderson} {et~al.}(2011{\natexlab{a}}){Anderson}, {Collier Cameron},
  {Gillon}, {Hellier}, {Jehin}, {Lendl}, {Queloz}, {Smalley}, {Triaud}, \&
  {Vanhuysse}}]{Anderson+11}
{Anderson}, D.~R., {Collier Cameron}, A., {Gillon}, M., {et~al.}
  2011{\natexlab{a}}, \aap, 534, A16

\bibitem[{{Anderson} {et~al.}(2011{\natexlab{b}}){Anderson}, {Collier Cameron},
  {Gillon}, {Hellier}, {Jehin}, {Lendl}, {Queloz}, {Smalley}, {Triaud}, \&
  {Vanhuysse}}]{Anders+11}
---. 2011{\natexlab{b}}, \aap, 534, A16

\bibitem[{{Anderson} {et~al.}(2011{\natexlab{c}}){Anderson}, {Smith},
  {Lanotte}, {Barman}, {Collier Cameron}, {Campo}, {Gillon}, {Harrington},
  {Hellier}, {Maxted}, {Queloz}, {Triaud}, \& {Wheatley}}]{Anderson+11b}
{Anderson}, D.~R., {Smith}, A.~M.~S., {Lanotte}, A.~A., {et~al.}
  2011{\natexlab{c}}, \mnras, 416, 2108

\bibitem[{{Anderson} {et~al.}(2011{\natexlab{d}}){Anderson}, {Collier Cameron},
  {Hellier}, {Lendl}, {Lister}, {Maxted}, {Queloz}, {Smalley}, {Smith},
  {Triaud}, {West}, {Brown}, {Gillon}, {Pepe}, {Pollacco}, {S{\'e}gransan},
  {Street}, \& {Udry}}]{Anderson+11d}
{Anderson}, D.~R., {Collier Cameron}, A., {Hellier}, C., {et~al.}
  2011{\natexlab{d}}, \aap, 531, A60

\bibitem[{{Bakos} {et~al.}(2009){Bakos}, {Howard}, {Noyes}, {Hartman},
  {Torres}, {Kov{\'a}cs}, {Fischer}, {Latham}, {Johnson}, {Marcy}, {Sasselov},
  {Stefanik}, {Sip{\H o}cz}, {Kov{\'a}cs}, {Esquerdo}, {P{\'a}l},
  {L{\'a}z{\'a}r}, {Papp}, \& {S{\'a}ri}}]{Bakos+09}
{Bakos}, G.~{\'A}., {Howard}, A.~W., {Noyes}, R.~W., {et~al.} 2009, \apj, 707,
  446

\bibitem[{{Bakos} {et~al.}(2011){Bakos}, {Hartman}, {Torres}, {Latham},
  {Kov{\'a}cs}, {Noyes}, {Fischer}, {Johnson}, {Marcy}, {Howard}, {Kipping},
  {Esquerdo}, {Shporer}, {B{\'e}ky}, {Buchhave}, {Perumpilly}, {Everett},
  {Sasselov}, {Stefanik}, {L{\'a}z{\'a}r}, {Papp}, \& {S{\'a}ri}}]{Bakos+11}
{Bakos}, G.~{\'A}., {Hartman}, J., {Torres}, G., {et~al.} 2011, \apj, 742, 116

\bibitem[{{Bakos} {et~al.}(2012){Bakos}, {Hartman}, {Torres}, {B{\'e}ky},
  {Latham}, {Buchhave}, {Csubry}, {Kov{\'a}cs}, {Bieryla}, {Quinn},
  {Szklen{\'a}r}, {Esquerdo}, {Shporer}, {Noyes}, {Fischer}, {Johnson},
  {Howard}, {Marcy}, {Sato}, {Penev}, {Everett}, {Sasselov}, {F{\H u}r{\'e}sz},
  {Stefanik}, {L{\'a}z{\'a}r}, {Papp}, \& {S{\'a}ri}}]{Bakos+12}
{Bakos}, G.~{\'A}., {Hartman}, J.~D., {Torres}, G., {et~al.} 2012, \aj, 144, 19

\bibitem[{{Barge} {et~al.}(2008){Barge}, {Baglin}, {Auvergne}, {Rauer},
  {L{\'e}ger}, {Schneider}, {Pont}, {Aigrain}, {Almenara}, {Alonso},
  {Barbieri}, {Bord{\'e}}, {Bouchy}, {Deeg}, {La Reza}, {Deleuil}, {Dvorak},
  {Erikson}, {Fridlund}, {Gillon}, {Gondoin}, {Guillot}, {Hatzes}, {Hebrard},
  {Jorda}, {Kabath}, {Lammer}, {Llebaria}, {Loeillet}, {Magain}, {Mazeh},
  {Moutou}, {Ollivier}, {P{\"a}tzold}, {Queloz}, {Rouan}, {Shporer}, \&
  {Wuchterl}}]{Barge2008}
{Barge}, P., {Baglin}, A., {Auvergne}, M., {et~al.} 2008, \aap, 482, L17

\bibitem[{{Barker} \& {Ogilvie}(2009)}]{BarkerOgilvie2009}
{Barker}, A.~J., \& {Ogilvie}, G.~I. 2009, \mnras, 395, 2268

\bibitem[{{Barnes} {et~al.}(2011){Barnes}, {Linscott}, \&
  {Shporer}}]{Barnes+11}
{Barnes}, J.~W., {Linscott}, E., \& {Shporer}, A. 2011, \apjs, 197, 10

\bibitem[{{Barros} {et~al.}(2011){Barros}, {Faedi}, {Collier Cameron},
  {Lister}, {McCormac}, {Pollacco}, {Simpson}, {Smalley}, {Street}, {Todd},
  {Triaud}, {Boisse}, {Bouchy}, {H{\'e}brard}, {Moutou}, {Pepe}, {Queloz},
  {Santerne}, {Segransan}, {Udry}, {Bento}, {Butters}, {Enoch}, {Haswell},
  {Hellier}, {Keenan}, {Miller}, {Moulds}, {Norton}, {Parley}, {Skillen},
  {Watson}, {West}, \& {Wheatley}}]{Barros+11}
{Barros}, S.~C.~C., {Faedi}, F., {Collier Cameron}, A., {et~al.} 2011, \aap,
  525, A54

\bibitem[{{Belczynski} {et~al.}(2008){Belczynski}, {Kalogera}, {Rasio}, {Taam},
  {Zezas}, {Bulik}, {Maccarone}, \& {Ivanova}}]{BelczynskiStartrack2008}
{Belczynski}, K., {Kalogera}, V., {Rasio}, F.~A., {et~al.} 2008, \apjs, 174,
  223

\bibitem[{{Bergfors} {et~al.}(2013){Bergfors}, {Brandner}, {Daemgen}, {Biller},
  {Hippler}, {Janson}, {Kudryavtseva}, {Gei{\ss}ler}, {Henning}, \&
  {K{\"o}hler}}]{Bergfors+13}
{Bergfors}, C., {Brandner}, W., {Daemgen}, S., {et~al.} 2013, \mnras, 428, 182

\bibitem[{{Blecic} {et~al.}(2011){Blecic}, {Harrington}, {Madhusudhan},
  {Stevenson}, {Hardy}, {Campo}, {Bowman}, {Nymeyer}, {Cubillos}, \&
  {Anderson}}]{Blecic+11}
{Blecic}, J., {Harrington}, J., {Madhusudhan}, N., {et~al.} 2011, ArXiv
  e-prints, arXiv:1111.2363

\bibitem[{{Bloecker}(1995)}]{Bloecker95}
{Bloecker}, T. 1995, \aap, 297, 727

\bibitem[{{Bonomo} {et~al.}(2012){Bonomo}, {H{\'e}brard}, {Santerne}, {Santos},
  {Deleuil}, {Almenara}, {Bouchy}, {D{\'{\i}}az}, {Moutou}, \&
  {Vanhuysse}}]{Bonomo+12}
{Bonomo}, A.~S., {H{\'e}brard}, G., {Santerne}, A., {et~al.} 2012, \aap, 538,
  A96

\bibitem[{{Bouchy} {et~al.}(2005){Bouchy}, {Udry}, {Mayor}, {Moutou}, {Pont},
  {Iribarne}, {da Silva}, {Ilovaisky}, {Queloz}, {Santos}, {S{\'e}gransan}, \&
  {Zucker}}]{Bouchy+05}
{Bouchy}, F., {Udry}, S., {Mayor}, M., {et~al.} 2005, \aap, 444, L15

\bibitem[{{Bouchy} {et~al.}(2008){Bouchy}, {Queloz}, {Deleuil}, {Loeillet},
  {Hatzes}, {Aigrain}, {Alonso}, {Auvergne}, {Baglin}, {Barge}, {Benz},
  {Bord{\'e}}, {Deeg}, {de La Reza}, {Dvorak}, {Erikson}, {Fridlund},
  {Gondoin}, {Guillot}, {H{\'e}brard}, {Jorda}, {Lammer}, {L{\'e}ger},
  {Llebaria}, {Magain}, {Mayor}, {Moutou}, {Ollivier}, {P{\"a}tzold}, {Pepe},
  {Pont}, {Rauer}, {Rouan}, {Schneider}, {Triaud}, {Udry}, \&
  {Wuchterl}}]{Bouchy+08}
{Bouchy}, F., {Queloz}, D., {Deleuil}, M., {et~al.} 2008, \aap, 482, L25

\bibitem[{{Brown} {et~al.}(2012){Brown}, {Cameron}, {Anderson}, {Enoch},
  {Hellier}, {Maxted}, {Miller}, {Pollacco}, {Queloz}, {Simpson}, {Smalley},
  {Triaud}, {Boisse}, {Bouchy}, {Gillon}, \& {H{\'e}brard}}]{Brown+12}
{Brown}, D.~J.~A., {Cameron}, A.~C., {Anderson}, D.~R., {et~al.} 2012, \mnras,
  423, 1503

\bibitem[{{Buchhave} {et~al.}(2010){Buchhave}, {Bakos}, {Hartman}, {Torres},
  {Kov{\'a}cs}, {Latham}, {Noyes}, {Esquerdo}, {Everett}, {Howard}, {Marcy},
  {Fischer}, {Johnson}, {Andersen}, {F{\H u}r{\'e}sz}, {Perumpilly},
  {Sasselov}, {Stefanik}, {B{\'e}ky}, {L{\'a}z{\'a}r}, {Papp}, \&
  {S{\'a}ri}}]{Buchhave+10}
{Buchhave}, L.~A., {Bakos}, G.~{\'A}., {Hartman}, J.~D., {et~al.} 2010, \apj,
  720, 1118

\bibitem[{{Burke} {et~al.}(2007){Burke}, {McCullough}, {Valenti},
  {Johns-Krull}, {Janes}, {Heasley}, {Summers}, {Stys}, {Bissinger}, {Fleenor},
  {Foote}, {Garc{\'{\i}}a-Melendo}, {Gary}, {Howell}, {Mallia}, {Masi},
  {Taylor}, \& {Vanmunster}}]{Burke+07}
{Burke}, C.~J., {McCullough}, P.~R., {Valenti}, J.~A., {et~al.} 2007, \apj,
  671, 2115

\bibitem[{{Cantiello} {et~al.}(2009){Cantiello}, {Langer}, {Brott}, {de Koter},
  {Shore}, {Vink}, {Voegler}, {Lennon}, \& {Yoon}}]{CantielloLBdKSVVLY2009}
{Cantiello}, M., {Langer}, N., {Brott}, I., {et~al.} 2009, \aap, 499, 279

\bibitem[{{Chan} {et~al.}(2011){Chan}, {Ingemyr}, {Winn}, {Holman},
  {Sanchis-Ojeda}, {Esquerdo}, \& {Everett}}]{Chan+11}
{Chan}, T., {Ingemyr}, M., {Winn}, J.~N., {et~al.} 2011, \aj, 141, 179

\bibitem[{{Chatterjee} {et~al.}(2008){Chatterjee}, {Ford}, {Matsumura}, \&
  {Rasio}}]{ChatterjeeFMR08}
{Chatterjee}, S., {Ford}, E.~B., {Matsumura}, S., \& {Rasio}, F.~A. 2008, \apj,
  686, 580

\bibitem[{{Ciceri} {et~al.}(2013){Ciceri}, {Mancini}, {Southworth}, {Nikolov},
  {Bozza}, {Bruni}, {Calchi Novati}, {D'Ago}, \& {Henning}}]{Ciceri+13}
{Ciceri}, S., {Mancini}, L., {Southworth}, J., {et~al.} 2013, \aap, 557, A30

\bibitem[{{Collier Cameron} {et~al.}(2010{\natexlab{a}}){Collier Cameron},
  {Bruce}, {Miller}, {Triaud}, \& {Queloz}}]{CollierC+10b}
{Collier Cameron}, A., {Bruce}, V.~A., {Miller}, G.~R.~M., {Triaud},
  A.~H.~M.~J., \& {Queloz}, D. 2010{\natexlab{a}}, \mnras, 403, 151

\bibitem[{{Collier Cameron} {et~al.}(2010{\natexlab{b}}){Collier Cameron},
  {Guenther}, {Smalley}, {McDonald}, {Hebb}, {Andersen}, {Augusteijn},
  {Barros}, {Brown}, {Cochran}, {Endl}, {Fossey}, {Hartmann}, {Maxted},
  {Pollacco}, {Skillen}, {Telting}, {Waldmann}, \& {West}}]{CollierC+10}
{Collier Cameron}, A., {Guenther}, E., {Smalley}, B., {et~al.}
  2010{\natexlab{b}}, \mnras, 407, 507

\bibitem[{{Cresswell} {et~al.}(2007){Cresswell}, {Dirksen}, {Kley}, \&
  {Nelson}}]{Cresswell+07}
{Cresswell}, P., {Dirksen}, G., {Kley}, W., \& {Nelson}, R.~P. 2007, \aap, 473,
  329

\bibitem[{{Deleuil} {et~al.}(2008){Deleuil}, {Deeg}, {Alonso}, {Bouchy},
  {Rouan}, {Auvergne}, {Baglin}, {Aigrain}, {Almenara}, {Barbieri}, {Barge},
  {Bruntt}, {Bord{\'e}}, {Collier Cameron}, {Csizmadia}, {de La Reza},
  {Dvorak}, {Erikson}, {Fridlund}, {Gandolfi}, {Gillon}, {Guenther}, {Guillot},
  {Hatzes}, {H{\'e}brard}, {Jorda}, {Lammer}, {L{\'e}ger}, {Llebaria},
  {Loeillet}, {Mayor}, {Mazeh}, {Moutou}, {Ollivier}, {P{\"a}tzold}, {Pont},
  {Queloz}, {Rauer}, {Schneider}, {Shporer}, {Wuchterl}, \&
  {Zucker}}]{Deleuil+08}
{Deleuil}, M., {Deeg}, H.~J., {Alonso}, R., {et~al.} 2008, \aap, 491, 889

\bibitem[{{D{\'e}sert} {et~al.}(2011){D{\'e}sert}, {Charbonneau}, {Demory},
  {Ballard}, {Carter}, {Fortney}, {Cochran}, {Endl}, {Quinn}, {Isaacson},
  {Fressin}, {Buchhave}, {Latham}, {Knutson}, {Bryson}, {Torres}, {Rowe},
  {Batalha}, {Borucki}, {Brown}, {Caldwell}, {Christiansen}, {Deming},
  {Fabrycky}, {Ford}, {Gilliland}, {Gillon}, {Haas}, {Jenkins}, {Kinemuchi},
  {Koch}, {Lissauer}, {Lucas}, {Mullally}, {MacQueen}, {Marcy}, {Sasselov},
  {Seager}, {Still}, {Tenenbaum}, {Uddin}, \& {Winn}}]{Desert+11}
{D{\'e}sert}, J.-M., {Charbonneau}, D., {Demory}, B.-O., {et~al.} 2011, \apjs,
  197, 14

\bibitem[{{Dobbs-Dixon} {et~al.}(2004){Dobbs-Dixon}, {Lin}, \&
  {Mardling}}]{DobbsDixon+04}
{Dobbs-Dixon}, I., {Lin}, D.~N.~C., \& {Mardling}, R.~A. 2004, \apj, 610, 464

\bibitem[{{Doyle} {et~al.}(2013){Doyle}, {Smalley}, {Maxted}, {Anderson},
  {Cameron}, {Gillon}, {Hellier}, {Pollacco}, {Queloz}, {Triaud}, \&
  {West}}]{Doyle+13}
{Doyle}, A.~P., {Smalley}, B., {Maxted}, P.~F.~L., {et~al.} 2013, \mnras, 428,
  3164

\bibitem[{{Enoch} {et~al.}(2011){Enoch}, {Cameron}, {Anderson}, {Lister},
  {Hellier}, {Maxted}, {Queloz}, {Smalley}, {Triaud}, {West}, {Brown},
  {Gillon}, {Hebb}, {Lendl}, {Parley}, {Pepe}, {Pollacco}, {Segransan},
  {Simpson}, {Street}, \& {Udry}}]{Enoch+11}
{Enoch}, B., {Cameron}, A.~C., {Anderson}, D.~R., {et~al.} 2011, \mnras, 410,
  1631

\bibitem[{{Fabrycky} \& {Tremaine}(2007)}]{FabryckyTremaine07}
{Fabrycky}, D., \& {Tremaine}, S. 2007, \apj, 669, 1298

\bibitem[{{Fabrycky} \& {Winn}(2009)}]{FabryckyWinn09}
{Fabrycky}, D.~C., \& {Winn}, J.~N. 2009, \apj, 696, 1230

\bibitem[{{Gillon}(2009)}]{Gillon+09c}
{Gillon}, M. 2009, ArXiv e-prints, arXiv:0906.4904

\bibitem[{{Gillon} {et~al.}(2009{\natexlab{a}}){Gillon}, {Anderson}, {Triaud},
  {Hellier}, {Maxted}, {Pollaco}, {Queloz}, {Smalley}, {West}, {Wilson},
  {Bentley}, {Collier Cameron}, {Enoch}, {Hebb}, {Horne}, {Irwin}, {Joshi},
  {Lister}, {Mayor}, {Pepe}, {Parley}, {Segransan}, {Udry}, \&
  {Wheatley}}]{Gillon+09b}
{Gillon}, M., {Anderson}, D.~R., {Triaud}, A.~H.~M.~J., {et~al.}
  2009{\natexlab{a}}, \aap, 501, 785

\bibitem[{{Gillon} {et~al.}(2009{\natexlab{b}}){Gillon}, {Smalley}, {Hebb},
  {Anderson}, {Triaud}, {Hellier}, {Maxted}, {Queloz}, \& {Wilson}}]{Gillon+09}
{Gillon}, M., {Smalley}, B., {Hebb}, L., {et~al.} 2009{\natexlab{b}}, \aap,
  496, 259

\bibitem[{{Gillon} {et~al.}(2010){Gillon}, {Lanotte}, {Barman}, {Miller},
  {Demory}, {Deleuil}, {Montalb{\'a}n}, {Bouchy}, {Collier Cameron}, {Deeg},
  {Fortney}, {Fridlund}, {Harrington}, {Magain}, {Moutou}, {Queloz}, {Rauer},
  {Rouan}, \& {Schneider}}]{Gillon+10}
{Gillon}, M., {Lanotte}, A.~A., {Barman}, T., {et~al.} 2010, \aap, 511, A3

\bibitem[{{Goldreich} \& {Nicholson}(1977)}]{GoldreichNicholson1977}
{Goldreich}, P., \& {Nicholson}, P.~D. 1977, Icarus, 30, 301

\bibitem[{{Goldreich} \& {Soter}(1966)}]{GoldreichSoter66}
{Goldreich}, P., \& {Soter}, S. 1966, Icarus, 5, 375

\bibitem[{{Goldreich} \& {Tremaine}(1980)}]{GoldreichTremaine80}
{Goldreich}, P., \& {Tremaine}, S. 1980, \apj, 241, 425

\bibitem[{{Greenspan}(1968)}]{Dejaiffe68}
{Greenspan}, H.~P. 1968, The theory of Rotating Fluids (Cambridge: Cambridge
  Univ. Press)

\bibitem[{{Guenther} {et~al.}(2012){Guenther}, {D{\'{\i}}az}, {Gazzano},
  {Mazeh}, {Rouan}, {Gibson}, {Csizmadia}, {Aigrain}, {Alonso}, {Almenara},
  {Auvergne}, {Baglin}, {Barge}, {Bonomo}, {Bord{\'e}}, {Bouchy}, {Bruntt},
  {Cabrera}, {Carone}, {Carpano}, {Cavarroc}, {Deeg}, {Deleuil}, {Dreizler},
  {Dvorak}, {Erikson}, {Ferraz-Mello}, {Fridlund}, {Gandolfi}, {Gillon},
  {Guillot}, {Hatzes}, {Havel}, {H{\'e}brard}, {Jehin}, {Jorda}, {Lammer},
  {L{\'e}ger}, {Moutou}, {Nortmann}, {Ollivier}, {Ofir}, {Pasternacki},
  {P{\"a}tzold}, {Parviainen}, {Queloz}, {Rauer}, {Samuel}, {Santerne},
  {Schneider}, {Tal-Or}, {Tingley}, {Weingrill}, \& {Wuchterl}}]{Guenther+12}
{Guenther}, E.~W., {D{\'{\i}}az}, R.~F., {Gazzano}, J.-C., {et~al.} 2012, \aap,
  537, A136

\bibitem[{{Guillochon} {et~al.}(2011){Guillochon}, {Ramirez-Ruiz}, \&
  {Lin}}]{Guillochon+11}
{Guillochon}, J., {Ramirez-Ruiz}, E., \& {Lin}, D. 2011, \apj, 732, 74

\bibitem[{{Hartman} {et~al.}(2011){Hartman}, {Bakos}, {Torres}, {Latham},
  {Kov{\'a}cs}, {B{\'e}ky}, {Quinn}, {Mazeh}, {Shporer}, {Marcy}, {Howard},
  {Fischer}, {Johnson}, {Esquerdo}, {Noyes}, {Sasselov}, {Stefanik},
  {Fernandez}, {Szklen{\'a}r}, {L{\'a}z{\'a}r}, {Papp}, \&
  {S{\'a}ri}}]{Hartman+11}
{Hartman}, J.~D., {Bakos}, G.~{\'A}., {Torres}, G., {et~al.} 2011, \apj, 742,
  59

\bibitem[{{Hebb} {et~al.}(2010){Hebb}, {Collier-Cameron}, {Triaud}, {Lister},
  {Smalley}, {Maxted}, {Hellier}, {Anderson}, {Pollacco}, {Gillon}, {Queloz},
  {West}, {Bentley}, {Enoch}, {Haswell}, {Horne}, {Mayor}, {Pepe}, {Segransan},
  {Skillen}, {Udry}, \& {Wheatley}}]{Hebb+10}
{Hebb}, L., {Collier-Cameron}, A., {Triaud}, A.~H.~M.~J., {et~al.} 2010, \apj,
  708, 224

\bibitem[{{H{\'e}brard} {et~al.}(2011){H{\'e}brard}, {Evans}, {Alonso},
  {Fridlund}, {Ofir}, {Aigrain}, {Guillot}, {Almenara}, {Auvergne}, {Baglin},
  {Barge}, {Bonomo}, {Bord{\'e}}, {Bouchy}, {Cabrera}, {Carone}, {Carpano},
  {Cavarroc}, {Csizmadia}, {Deeg}, {Deleuil}, {D{\'{\i}}az}, {Dvorak},
  {Erikson}, {Ferraz-Mello}, {Gandolfi}, {Gibson}, {Gillon}, {Guenther},
  {Hatzes}, {Havel}, {Jorda}, {Lammer}, {L{\'e}ger}, {Llebaria}, {Mazeh},
  {Moutou}, {Ollivier}, {Parviainen}, {P{\"a}tzold}, {Queloz}, {Rauer},
  {Rouan}, {Santerne}, {Schneider}, {Tingley}, \& {Wuchterl}}]{Hebrard+11}
{H{\'e}brard}, G., {Evans}, T.~M., {Alonso}, R., {et~al.} 2011, \aap, 533, A130

\bibitem[{{H{\'e}brard} {et~al.}(2013){H{\'e}brard}, {Collier Cameron},
  {Brown}, {D{\'{\i}}az}, {Faedi}, {Smalley}, {Anderson}, {Armstrong},
  {Barros}, {Bento}, {Bouchy}, {Doyle}, {Enoch}, {G{\'o}mez Maqueo Chew},
  {H{\'e}brard}, {Hellier}, {Lendl}, {Lister}, {Maxted}, {McCormac}, {Moutou},
  {Pollacco}, {Queloz}, {Santerne}, {Skillen}, {Southworth}, {Tregloan-Reed},
  {Triaud}, {Udry}, {Vanhuysse}, {Watson}, {West}, \& {Wheatley}}]{Hebrard+13}
{H{\'e}brard}, G., {Collier Cameron}, A., {Brown}, D.~J.~A., {et~al.} 2013,
  \aap, 549, A134

\bibitem[{{Hellier} {et~al.}(2009{\natexlab{a}}){Hellier}, {Anderson}, {Collier
  Cameron}, {Gillon}, {Hebb}, {Maxted}, {Queloz}, {Smalley}, {Triaud}, {West},
  {Wilson}, {Bentley}, {Enoch}, {Horne}, {Irwin}, {Lister}, {Mayor}, {Parley},
  {Pepe}, {Pollacco}, {Segransan}, {Udry}, \& {Wheatley}}]{Hellier+09}
{Hellier}, C., {Anderson}, D.~R., {Collier Cameron}, A., {et~al.}
  2009{\natexlab{a}}, \nat, 460, 1098

\bibitem[{{Hellier} {et~al.}(2009{\natexlab{b}}){Hellier}, {Anderson},
  {Gillon}, {Lister}, {Maxted}, {Queloz}, {Smalley}, {Triaud}, {West},
  {Wilson}, {Alsubai}, {Bentley}, {Collier Cameron}, {Hebb}, {Horne}, {Irwin},
  {Kane}, {Mayor}, {Pepe}, {Pollacco}, {Skillen}, {Udry}, {Wheatley},
  {Christian}, {Enoch}, {Haswell}, {Joshi}, {Norton}, {Parley}, {Ryans},
  {Street}, \& {Todd}}]{Hellier+09b}
{Hellier}, C., {Anderson}, D.~R., {Gillon}, M., {et~al.} 2009{\natexlab{b}},
  \apjl, 690, L89

\bibitem[{{Hoyer} {et~al.}(2012){Hoyer}, {Rojo}, \&
  {L{\'o}pez-Morales}}]{Hoyer+12}
{Hoyer}, S., {Rojo}, P., \& {L{\'o}pez-Morales}, M. 2012, \apj, 748, 22

\bibitem[{{Hoyer} {et~al.}(2013){Hoyer}, {L{\'o}pez-Morales}, {Rojo},
  {Nascimbeni}, {Hidalgo}, {Astudillo-Defru}, {Concha}, {Contreras},
  {Servajean}, \& {Hinse}}]{Hoyer+13}
{Hoyer}, S., {L{\'o}pez-Morales}, M., {Rojo}, P., {et~al.} 2013, \mnras, 434,
  46

\bibitem[{{Hurley} {et~al.}(2002){Hurley}, {Tout}, \& {Pols}}]{HurleyTP02}
{Hurley}, J.~R., {Tout}, C.~A., \& {Pols}, O.~R. 2002, \mnras, 329, 897

\bibitem[{{Hut}(1981)}]{Hut1981}
{Hut}, P. 1981, \aap, 99, 126

\bibitem[{{Jackson} {et~al.}(2009){Jackson}, {Barnes}, \&
  {Greenberg}}]{Jackson+09}
{Jackson}, B., {Barnes}, R., \& {Greenberg}, R. 2009, \apj, 698, 1357

\bibitem[{{Jackson} {et~al.}(2008){Jackson}, {Greenberg}, \&
  {Barnes}}]{Jackson+08}
{Jackson}, B., {Greenberg}, R., \& {Barnes}, R. 2008, \apj, 678, 1396

\bibitem[{{Jenkins} {et~al.}(2010){Jenkins}, {Borucki}, {Koch}, {Marcy},
  {Cochran}, {Welsh}, {Basri}, {Batalha}, {Buchhave}, {Brown}, {Caldwell},
  {Dunham}, {Endl}, {Fischer}, {Gautier}, {Geary}, {Gilliland}, {Howell},
  {Isaacson}, {Johnson}, {Latham}, {Lissauer}, {Monet}, {Rowe}, {Sasselov},
  {Howard}, {MacQueen}, {Orosz}, {Chandrasekaran}, {Twicken}, {Bryson},
  {Quintana}, {Clarke}, {Li}, {Allen}, {Tenenbaum}, {Wu}, {Meibom}, {Klaus},
  {Middour}, {Cote}, {McCauliff}, {Girouard}, {Gunter}, {Wohler}, {Hall},
  {Ibrahim}, {Kamal Uddin}, {Wu}, {Bhavsar}, {Van Cleve}, {Pletcher}, {Dotson},
  \& {Haas}}]{Jenkins+10}
{Jenkins}, J.~M., {Borucki}, W.~J., {Koch}, D.~G., {et~al.} 2010, \apj, 724,
  1108

\bibitem[{{Johnson} {et~al.}(2009){Johnson}, {Winn}, {Albrecht}, {Howard},
  {Marcy}, \& {Gazak}}]{Johnson+09}
{Johnson}, J.~A., {Winn}, J.~N., {Albrecht}, S., {et~al.} 2009, \pasp, 121,
  1104

\bibitem[{{Johnson} {et~al.}(2008){Johnson}, {Winn}, {Narita}, {Enya},
  {Williams}, {Marcy}, {Sato}, {Ohta}, {Taruya}, {Suto}, {Turner}, {Bakos},
  {Butler}, {Vogt}, {Aoki}, {Tamura}, {Yamada}, {Yoshii}, \&
  {Hidas}}]{Johnson+08}
{Johnson}, J.~A., {Winn}, J.~N., {Narita}, N., {et~al.} 2008, \apj, 686, 649

\bibitem[{{Johnson} {et~al.}(2011){Johnson}, {Winn}, {Bakos}, {Hartman},
  {Morton}, {Torres}, {Kov{\'a}cs}, {Latham}, {Noyes}, {Sato}, {Esquerdo},
  {Fischer}, {Marcy}, {Howard}, {Buchhave}, {F{\H u}r{\'e}sz}, {Quinn},
  {B{\'e}ky}, {Sasselov}, {Stefanik}, {L{\'a}z{\'a}r}, {Papp}, \&
  {S{\'a}ri}}]{Johnson+11}
{Johnson}, J.~A., {Winn}, J.~N., {Bakos}, G.~{\'A}., {et~al.} 2011, \apj, 735,
  24

\bibitem[{{Joshi} {et~al.}(2009){Joshi}, {Pollacco}, {Collier Cameron},
  {Skillen}, {Simpson}, {Steele}, {Street}, {Stempels}, {Christian}, {Hebb},
  {Bouchy}, {Gibson}, {H{\'e}brard}, {Keenan}, {Loeillet}, {Meaburn}, {Moutou},
  {Smalley}, {Todd}, {West}, {Anderson}, {Bentley}, {Enoch}, {Haswell},
  {Hellier}, {Horne}, {Irwin}, {Lister}, {McDonald}, {Maxted}, {Mayor},
  {Norton}, {Parley}, {Perrier}, {Pont}, {Queloz}, {Ryans}, {Smith}, {Udry},
  {Wheatley}, \& {Wilson}}]{Joshi+09}
{Joshi}, Y.~C., {Pollacco}, D., {Collier Cameron}, A., {et~al.} 2009, \mnras,
  392, 1532

\bibitem[{{Kipping} {et~al.}(2010){Kipping}, {Bakos}, {Hartman}, {Torres},
  {Shporer}, {Latham}, {Kov{\'a}cs}, {Noyes}, {Howard}, {Fischer}, {Johnson},
  {Marcy}, {B{\'e}ky}, {Perumpilly}, {Esquerdo}, {Sasselov}, {Stefanik},
  {L{\'a}z{\'a}r}, {Papp}, \& {S{\'a}ri}}]{Kipping+10}
{Kipping}, D.~M., {Bakos}, G.~{\'A}., {Hartman}, J., {et~al.} 2010, \apj, 725,
  2017

\bibitem[{{Knutson} {et~al.}(2010){Knutson}, {Howard}, \&
  {Isaacson}}]{Knutson+10}
{Knutson}, H.~A., {Howard}, A.~W., \& {Isaacson}, H. 2010, \apj, 720, 1569

\bibitem[{{Kov{\'a}cs} {et~al.}(2007){Kov{\'a}cs}, {Bakos}, {Torres},
  {Sozzetti}, {Latham}, {Noyes}, {Butler}, {Marcy}, {Fischer}, {Fern{\'a}ndez},
  {Esquerdo}, {Sasselov}, {Stefanik}, {P{\'a}l}, {L{\'a}z{\'a}r}, {Papp}, \&
  {S{\'a}ri}}]{Kovacs+07}
{Kov{\'a}cs}, G., {Bakos}, G.~{\'A}., {Torres}, G., {et~al.} 2007, \apjl, 670,
  L41

\bibitem[{{Kozai}(1962)}]{Kozai62}
{Kozai}, Y. 1962, \aj, 67, 591

\bibitem[{{Lai}(2012)}]{Lai12}
{Lai}, D. 2012, \mnras, 423, 486

\bibitem[{{Latham} {et~al.}(2009){Latham}, {Bakos}, {Torres}, {Stefanik},
  {Noyes}, {Kov{\'a}cs}, {P{\'a}l}, {Marcy}, {Fischer}, {Butler}, {Sip{\H
  o}cz}, {Sasselov}, {Esquerdo}, {Vogt}, {Hartman}, {Kov{\'a}cs},
  {L{\'a}z{\'a}r}, {Papp}, \& {S{\'a}ri}}]{Latham+09}
{Latham}, D.~W., {Bakos}, G.~{\'A}., {Torres}, G., {et~al.} 2009, \apj, 704,
  1107

\bibitem[{{Lecavelier des Etangs} {et~al.}(2013){Lecavelier des Etangs},
  {Sirothia}, {Gopal-Krishna}, \& {Zarka}}]{LecavelierSGKZ13}
{Lecavelier des Etangs}, A., {Sirothia}, S.~K., {Gopal-Krishna}, \& {Zarka}, P.
  2013, \aap, 552, A65

\bibitem[{{Lidov}(1962)}]{Lidov62}
{Lidov}, M.~L. 1962, \planss, 9, 719

\bibitem[{{Lin} {et~al.}(1996){Lin}, {Bodenheimer}, \& {Richardson}}]{Lin+96}
{Lin}, D.~N.~C., {Bodenheimer}, P., \& {Richardson}, D.~C. 1996, \nat, 380, 606

\bibitem[{{Lister} {et~al.}(2009){Lister}, {Anderson}, {Gillon}, {Hebb},
  {Smalley}, {Triaud}, {Collier Cameron}, {Wilson}, {West}, {Bentley},
  {Christian}, {Enoch}, {Haswell}, {Hellier}, {Horne}, {Irwin}, {Joshi},
  {Kane}, {Mayor}, {Maxted}, {Norton}, {Parley}, {Pepe}, {Pollacco}, {Queloz},
  {Ryans}, {Segransan}, {Skillen}, {Street}, {Todd}, {Udry}, \&
  {Wheatley}}]{Lister+09}
{Lister}, T.~A., {Anderson}, D.~R., {Gillon}, M., {et~al.} 2009, \apj, 703, 752

\bibitem[{{Mancini} {et~al.}(2013){Mancini}, {Southworth}, {Ciceri}, {Fortney},
  {Morley}, {Dittmann}, {Tregloan-Reed}, {Bruni}, {Barbieri}, {Evans}, {D'Ago},
  {Nikolov}, \& {Henning}}]{Mancini+13}
{Mancini}, L., {Southworth}, J., {Ciceri}, S., {et~al.} 2013, \aap, 551, A11

\bibitem[{{Matsumura} {et~al.}(2010){Matsumura}, {Peale}, \&
  {Rasio}}]{MatsumuraPR2010}
{Matsumura}, S., {Peale}, S.~J., \& {Rasio}, F.~A. 2010, \apj, 725, 1995

\bibitem[{{Maxted} {et~al.}(2011){Maxted}, {Koen}, \& {Smalley}}]{Maxted+11}
{Maxted}, P.~F.~L., {Koen}, C., \& {Smalley}, B. 2011, \mnras, 418, 1039

\bibitem[{{Maxted} {et~al.}(2010){Maxted}, {Anderson}, {Gillon}, {Hellier},
  {Queloz}, {Smalley}, {Triaud}, {West}, {Wilson}, {Bentley}, {Cegla}, {Collier
  Cameron}, {Enoch}, {Hebb}, {Horne}, {Irwin}, {Lister}, {Mayor}, {Parley},
  {Pepe}, {Pollacco}, {Segransan}, {Udry}, \& {Wheatley}}]{Maxted+10}
{Maxted}, P.~F.~L., {Anderson}, D.~R., {Gillon}, M., {et~al.} 2010, \aj, 140,
  2007

\bibitem[{{McCullough} {et~al.}(2008){McCullough}, {Burke}, {Valenti}, {Long},
  {Johns-Krull}, {Machalek}, {Janes}, {Taylor}, {Gregorio}, {Foote}, {Gary},
  {Fleenor}, {Garc{\'{\i}}a-Melendo}, \& {Vanmunster}}]{McCullough+08}
{McCullough}, P.~R., {Burke}, C.~J., {Valenti}, J.~A., {et~al.} 2008, ArXiv
  e-prints, arXiv:0805.2921

\bibitem[{{Morton} \& {Johnson}(2011)}]{MortonJohnson11}
{Morton}, T.~D., \& {Johnson}, J.~A. 2011, \apj, 729, 138

\bibitem[{{Moutou} {et~al.}(2011){Moutou}, {D{\'{\i}}az}, {Udry},
  {H{\'e}brard}, {Bouchy}, {Santerne}, {Ehrenreich}, {Arnold}, {Boisse},
  {Bonfils}, {Delfosse}, {Eggenberger}, {Forveille}, {Lagrange}, {Lovis},
  {Martinez}, {Pepe}, {Perrier}, {Queloz}, {Santos}, {S{\'e}gransan},
  {Toublanc}, {Troncin}, {Vanhuysse}, \& {Vidal-Madjar}}]{Moutou+11}
{Moutou}, C., {D{\'{\i}}az}, R.~F., {Udry}, S., {et~al.} 2011, \aap, 533, A113

\bibitem[{{Murray} {et~al.}(1998){Murray}, {Hansen}, {Holman}, \&
  {Tremaine}}]{MurrayHHT98}
{Murray}, N., {Hansen}, B., {Holman}, M., \& {Tremaine}, S. 1998, Science, 279,
  69

\bibitem[{{Nagasawa} {et~al.}(2008){Nagasawa}, {Ida}, \& {Bessho}}]{Nagasawa08}
{Nagasawa}, M., {Ida}, S., \& {Bessho}, T. 2008, \apj, 678, 498

\bibitem[{{Naoz} {et~al.}(2011){Naoz}, {Farr}, {Lithwick}, {Rasio}, \&
  {Teyssandier}}]{Naoz+11}
{Naoz}, S., {Farr}, W.~M., {Lithwick}, Y., {Rasio}, F.~A., \& {Teyssandier}, J.
  2011, \nat, 473, 187

\bibitem[{{Narita} {et~al.}(2010{\natexlab{a}}){Narita}, {Hirano},
  {Sanchis-Ojeda}, {Winn}, {Holman}, {Sato}, {Aoki}, \& {Tamura}}]{Narita+10b}
{Narita}, N., {Hirano}, T., {Sanchis-Ojeda}, R., {et~al.} 2010{\natexlab{a}},
  \pasj, 62, L61

\bibitem[{{Narita} {et~al.}(2011){Narita}, {Hirano}, {Sato}, {Harakawa},
  {Fukui}, {Aoki}, \& {Tamura}}]{Narita+11}
{Narita}, N., {Hirano}, T., {Sato}, B., {et~al.} 2011, \pasj, 63, L67

\bibitem[{{Narita} {et~al.}(2009){Narita}, {Sato}, {Hirano}, \&
  {Tamura}}]{Narita+09}
{Narita}, N., {Sato}, B., {Hirano}, T., \& {Tamura}, M. 2009, \pasj, 61, L35

\bibitem[{{Narita} {et~al.}(2010{\natexlab{b}}){Narita}, {Sato}, {Hirano},
  {Winn}, {Aoki}, \& {Tamura}}]{Narita+10}
{Narita}, N., {Sato}, B., {Hirano}, T., {et~al.} 2010{\natexlab{b}}, \pasj, 62,
  653

\bibitem[{{Narita} {et~al.}(2007){Narita}, {Enya}, {Sato}, {Ohta}, {Winn},
  {Suto}, {Taruya}, {Turner}, {Aoki}, {Yoshii}, {Yamada}, \&
  {Tamura}}]{Narita+07}
{Narita}, N., {Enya}, K., {Sato}, B., {et~al.} 2007, \pasj, 59, 763

\bibitem[{{Noyes} {et~al.}(2008){Noyes}, {Bakos}, {Torres}, {P{\'a}l},
  {Kov{\'a}cs}, {Latham}, {Fern{\'a}ndez}, {Fischer}, {Butler}, {Marcy},
  {Sip{\H o}cz}, {Esquerdo}, {Kov{\'a}cs}, {Sasselov}, {Sato}, {Stefanik},
  {Holman}, {L{\'a}z{\'a}r}, {Papp}, \& {S{\'a}ri}}]{Noyes+08}
{Noyes}, R.~W., {Bakos}, G.~{\'A}., {Torres}, G., {et~al.} 2008, \apjl, 673,
  L79

\bibitem[{{Nutzman} {et~al.}(2011){Nutzman}, {Gilliland}, {McCullough},
  {Charbonneau}, {Christensen-Dalsgaard}, {Kjeldsen}, {Nelan}, {Brown}, \&
  {Holman}}]{Nutzman+11}
{Nutzman}, P., {Gilliland}, R.~L., {McCullough}, P.~R., {et~al.} 2011, \apj,
  726, 3

\bibitem[{{Ogilvie} \& {Lin}(2007)}]{OgilvieLin2007}
{Ogilvie}, G.~I., \& {Lin}, D.~N.~C. 2007, \apj, 661, 1180

\bibitem[{{P{\'a}l} {et~al.}(2010){P{\'a}l}, {Bakos}, {Torres}, {Noyes},
  {Fischer}, {Johnson}, {Henry}, {Butler}, {Marcy}, {Howard}, {Sip{\H o}cz},
  {Latham}, \& {Esquerdo}}]{Pal+10}
{P{\'a}l}, A., {Bakos}, G.~{\'A}., {Torres}, G., {et~al.} 2010, \mnras, 401,
  2665

\bibitem[{{Papaloizou} \& {Larwood}(2000)}]{PapaloizouLarwood00}
{Papaloizou}, J.~C.~B., \& {Larwood}, J.~D. 2000, \mnras, 315, 823

\bibitem[{{Paxton} {et~al.}(2011){Paxton}, {Bildsten}, {Dotter}, {Herwig},
  {Lesaffre}, \& {Timmes}}]{PBDHLT2011}
{Paxton}, B., {Bildsten}, L., {Dotter}, A., {et~al.} 2011, \apjs, 192, 3

\bibitem[{{Paxton} {et~al.}(2013){Paxton}, {Cantiello}, {Arras}, {Bildsten},
  {Brown}, {Dotter}, {Mankovich}, {Montgomery}, {Stello}, {Timmes}, \&
  {Townsend}}]{Paxton+13}
{Paxton}, B., {Cantiello}, M., {Arras}, P., {et~al.} 2013, \apjs, 208, 4

\bibitem[{{Penev} \& {Sasselov}(2011)}]{PenevSasselov11}
{Penev}, K., \& {Sasselov}, D. 2011, \apj, 731, 67

\bibitem[{{Plavchan} \& {Bilinski}(2013)}]{PlavchanBilinski13}
{Plavchan}, P., \& {Bilinski}, C. 2013, \apj, 769, 86

\bibitem[{{Polfliet} \& {Smeyers}(1990)}]{PolflietSmeyers1990}
{Polfliet}, R., \& {Smeyers}, P. 1990, \aap, 237, 110

\bibitem[{{Pollacco} {et~al.}(2008){Pollacco}, {Skillen}, {Collier Cameron},
  {Loeillet}, {Stempels}, {Bouchy}, {Gibson}, {Hebb}, {H{\'e}brard}, {Joshi},
  {McDonald}, {Smalley}, {Smith}, {Street}, {Udry}, {West}, {Wilson},
  {Wheatley}, {Aigrain}, {Alsubai}, {Benn}, {Bruce}, {Christian}, {Clarkson},
  {Enoch}, {Evans}, {Fitzsimmons}, {Haswell}, {Hellier}, {Hickey}, {Hodgkin},
  {Horne}, {Hrudkov{\'a}}, {Irwin}, {Kane}, {Keenan}, {Lister}, {Maxted},
  {Mayor}, {Moutou}, {Norton}, {Osborne}, {Parley}, {Pont}, {Queloz}, {Ryans},
  \& {Simpson}}]{Pollacco+08}
{Pollacco}, D., {Skillen}, I., {Collier Cameron}, A., {et~al.} 2008, \mnras,
  385, 1576

\bibitem[{{Pont} {et~al.}(2010){Pont}, {Endl}, {Cochran}, {Barnes}, {Sneden},
  {MacQueen}, {Moutou}, {Aigrain}, {Alonso}, {Baglin}, {Bouchy}, {Deleuil},
  {Fridlund}, {H{\'e}brard}, {Hatzes}, {Mazeh}, \& {Shporer}}]{Pont+10}
{Pont}, F., {Endl}, M., {Cochran}, W.~D., {et~al.} 2010, \mnras, 402, L1

\bibitem[{{Queloz} {et~al.}(2010){Queloz}, {Anderson}, {Collier Cameron},
  {Gillon}, {Hebb}, {Hellier}, {Maxted}, {Pepe}, {Pollacco}, {S{\'e}gransan},
  {Smalley}, {Triaud}, {Udry}, \& {West}}]{Queloz+10}
{Queloz}, D., {Anderson}, D.~R., {Collier Cameron}, A., {et~al.} 2010, \aap,
  517, L1

\bibitem[{{Rasio} \& {Ford}(1996)}]{RasioFord96}
{Rasio}, F.~A., \& {Ford}, E.~B. 1996, Science, 274, 954

\bibitem[{{Rasio} {et~al.}(1996){Rasio}, {Tout}, {Lubow}, \&
  {Livio}}]{RasioTLL1996}
{Rasio}, F.~A., {Tout}, C.~A., {Lubow}, S.~H., \& {Livio}, M. 1996, \apj, 470,
  1187

\bibitem[{{Reimers}(1975)}]{Reimers75}
{Reimers}, D. 1975, {in Problems in stellar atmospheres and envelopes}, ed.
  B.~{Baschek}, W.~H. {Kegel}, \& G.~{Traving} (New York: Springer-Verlag),
  229--256

\bibitem[{{Rogers} \& {Lin}(2013)}]{RogersLin13}
{Rogers}, T.~M., \& {Lin}, D.~N.~C. 2013, \apjl, 769, L10

\bibitem[{{Sada} {et~al.}(2012){Sada}, {Deming}, {Jennings}, {Jackson},
  {Hamilton}, {Fraine}, {Peterson}, {Haase}, {Bays}, {Lunsford}, \&
  {O'Gorman}}]{Sada+12}
{Sada}, P.~V., {Deming}, D., {Jennings}, D.~E., {et~al.} 2012, \pasp, 124, 212

\bibitem[{{Sanchis-Ojeda} {et~al.}(2011){Sanchis-Ojeda}, {Winn}, {Holman},
  {Carter}, {Osip}, \& {Fuentes}}]{SanchisO+11}
{Sanchis-Ojeda}, R., {Winn}, J.~N., {Holman}, M.~J., {et~al.} 2011, \apj, 733,
  127

\bibitem[{{Santerne} {et~al.}(2012){Santerne}, {Moutou}, {Barros}, {Damiani},
  {D{\'{\i}}az}, {Almenara}, {Bonomo}, {Bouchy}, {Deleuil}, \&
  {H{\'e}brard}}]{Santerne+12}
{Santerne}, A., {Moutou}, C., {Barros}, S.~C.~C., {et~al.} 2012, \aap, 544, L12

\bibitem[{{Santos} {et~al.}(2004){Santos}, {Israelian}, \& {Mayor}}]{Santos+04}
{Santos}, N.~C., {Israelian}, G., \& {Mayor}, M. 2004, \aap, 415, 1153

\bibitem[{{Santos} {et~al.}(2006){Santos}, {Pont}, {Melo}, {Israelian},
  {Bouchy}, {Mayor}, {Moutou}, {Queloz}, {Udry}, \& {Guillot}}]{Santos+06b}
{Santos}, N.~C., {Pont}, F., {Melo}, C., {et~al.} 2006, \aap, 450, 825

\bibitem[{{Sato} {et~al.}(2005){Sato}, {Fischer}, {Henry}, {Laughlin},
  {Butler}, {Marcy}, {Vogt}, {Bodenheimer}, {Ida}, {Toyota}, {Wolf}, {Valenti},
  {Boyd}, {Johnson}, {Wright}, {Ammons}, {Robinson}, {Strader}, {McCarthy},
  {Tah}, \& {Minniti}}]{Sato+05}
{Sato}, B., {Fischer}, D.~A., {Henry}, G.~W., {et~al.} 2005, \apj, 633, 465

\bibitem[{{Schlaufman}(2010)}]{Schlaufman+10}
{Schlaufman}, K.~C. 2010, \apj, 719, 602

\bibitem[{{Shporer} {et~al.}(2009){Shporer}, {Bakos}, {Bouchy}, {Pont},
  {Kov{\'a}cs}, {Latham}, {Sip{\"o}cz}, {Torres}, {Mazeh}, {Esquerdo},
  {P{\'a}l}, {Noyes}, {Sasselov}, {L{\'a}z{\'a}r}, {Papp}, {S{\'a}ri}, \&
  {Kov{\'a}cs}}]{Shporer+09}
{Shporer}, A., {Bakos}, G.~{\'A}., {Bouchy}, F., {et~al.} 2009, \apj, 690, 1393

\bibitem[{{Shporer} {et~al.}(2011){Shporer}, {Jenkins}, {Rowe}, {Sanderfer},
  {Seader}, {Smith}, {Still}, {Thompson}, {Twicken}, \& {Welsh}}]{Shporer+11}
{Shporer}, A., {Jenkins}, J.~M., {Rowe}, J.~F., {et~al.} 2011, \aj, 142, 195

\bibitem[{{Simpson} {et~al.}(2011{\natexlab{a}}){Simpson}, {Barros}, {Brown},
  {Collier Cameron}, {Pollacco}, {Skillen}, {Stempels}, {Boisse}, {Faedi},
  {H{\'e}brard}, {McCormac}, {Sorensen}, {Street}, {Anderson}, {Bento},
  {Bouchy}, {Butters}, {Enoch}, {Haswell}, {Hebb}, {Hellier}, {Holmes},
  {Horne}, {Keenan}, {Lister}, {Maxted}, {Miller}, {Moulds}, {Moutou},
  {Norton}, {Parley}, {Santerne}, {Smalley}, {Smith}, {Todd}, {Watson}, {West},
  \& {Wheatley}}]{Simpson+11}
{Simpson}, E.~K., {Barros}, S.~C.~C., {Brown}, D.~J.~A., {et~al.}
  2011{\natexlab{a}}, \aj, 141, 161

\bibitem[{{Simpson} {et~al.}(2011{\natexlab{b}}){Simpson}, {Pollacco},
  {Cameron}, {H{\'e}brard}, {Anderson}, {Barros}, {Boisse}, {Bouchy}, {Faedi},
  {Gillon}, {Hebb}, {Keenan}, {Miller}, {Moutou}, {Queloz}, {Skillen},
  {Sorensen}, {Stempels}, {Triaud}, {Watson}, \& {Wilson}}]{Simpson+11b}
{Simpson}, E.~K., {Pollacco}, D., {Cameron}, A.~C., {et~al.}
  2011{\natexlab{b}}, \mnras, 414, 3023

\bibitem[{{Skumanich}(1972)}]{Skumanich72}
{Skumanich}, A. 1972, \apj, 171, 565

\bibitem[{{Smalley} {et~al.}(2010){Smalley}, {Anderson}, {Collier Cameron},
  {Gillon}, {Hellier}, {Lister}, {Maxted}, {Queloz}, {Triaud}, {West},
  {Bentley}, {Enoch}, {Pepe}, {Pollacco}, {Segransan}, {Smith}, {Southworth},
  {Udry}, {Wheatley}, {Wood}, \& {Bento}}]{Smalley+10}
{Smalley}, B., {Anderson}, D.~R., {Collier Cameron}, A., {et~al.} 2010, \aap,
  520, A56

\bibitem[{{Smith} {et~al.}(2011){Smith}, {Anderson}, {Skillen}, {Collier
  Cameron}, \& {Smalley}}]{Smith+11}
{Smith}, A.~M.~S., {Anderson}, D.~R., {Skillen}, I., {Collier Cameron}, A., \&
  {Smalley}, B. 2011, \mnras, 416, 2096

\bibitem[{{Smith} {et~al.}(2013){Smith}, {Anderson}, {Bouchy}, {Collier
  Cameron}, {Doyle}, {Fumel}, {Gillon}, {H{\'e}brard}, {Hellier}, {Jehin},
  {Lendl}, {Maxted}, {Moutou}, {Pepe}, {Pollacco}, {Queloz}, {Santerne},
  {Segransan}, {Smalley}, {Southworth}, {Triaud}, {Udry}, \& {West}}]{Smith+13}
{Smith}, A.~M.~S., {Anderson}, D.~R., {Bouchy}, F., {et~al.} 2013, \aap, 552,
  A120

\bibitem[{{Southworth}(2008)}]{Southworth+08}
{Southworth}, J. 2008, \mnras, 386, 1644

\bibitem[{{Southworth}(2010)}]{Southworth+10}
---. 2010, \mnras, 408, 1689

\bibitem[{{Southworth}(2011)}]{Southworth+11}
---. 2011, \mnras, 417, 2166

\bibitem[{{Southworth} {et~al.}(2009){Southworth}, {Hinse}, {Dominik},
  {Glitrup}, {J{\o}rgensen}, {Liebig}, {Mathiasen}, {Anderson}, {Bozza},
  {Browne}, {Burgdorf}, {Calchi Novati}, {Dreizler}, {Finet}, {Harps{\o}e},
  {Hessman}, {Hundertmark}, {Maier}, {Mancini}, {Maxted}, {Rahvar}, {Ricci},
  {Scarpetta}, {Skottfelt}, {Snodgrass}, {Surdej}, \& {Zimmer}}]{Southworth+09}
{Southworth}, J., {Hinse}, T.~C., {Dominik}, M., {et~al.} 2009, \apj, 707, 167

\bibitem[{{Southworth} {et~al.}(2011){Southworth}, {Dominik}, {J{\o}rgensen},
  {Rahvar}, {Snodgrass}, {Alsubai}, {Bozza}, {Browne}, {Burgdorf}, {Calchi
  Novati}, {Dodds}, {Dreizler}, {Finet}, {Gerner}, {Hardis}, {Harps{\o}e},
  {Hellier}, {Hinse}, {Hundertmark}, {Kains}, {Kerins}, {Liebig}, {Mancini},
  {Mathiasen}, {Penny}, {Proft}, {Ricci}, {Sahu}, {Scarpetta}, {Sch{\"a}fer},
  {Sch{\"o}nebeck}, \& {Surdej}}]{Southworth+11b}
{Southworth}, J., {Dominik}, M., {J{\o}rgensen}, U.~G., {et~al.} 2011, \aap,
  527, A8

\bibitem[{{Southworth} {et~al.}(2013){Southworth}, {Mancini}, {Browne},
  {Burgdorf}, {Calchi Novati}, {Dominik}, {Gerner}, {Hinse}, {J{\o}rgensen},
  {Kains}, {Ricci}, {Sch{\"a}fer}, {Sch{\"o}nebeck}, {Tregloan-Reed},
  {Alsubai}, {Bozza}, {Chen}, {Dodds}, {Dreizler}, {Fang}, {Finet}, {Gu},
  {Hardis}, {Harps{\o}e}, {Henning}, {Hundertmark}, {Jessen-Hansen}, {Kerins},
  {Kjeldsen}, {Liebig}, {Lund}, {Lundkvist}, {Mathiasen}, {Nikolov}, {Penny},
  {Proft}, {Rahvar}, {Sahu}, {Scarpetta}, {Skottfelt}, {Snodgrass}, {Surdej},
  \& {Wertz}}]{Southworth+13}
{Southworth}, J., {Mancini}, L., {Browne}, P., {et~al.} 2013, \mnras, 434, 1300

\bibitem[{{Sozzetti} {et~al.}(2009){Sozzetti}, {Torres}, {Charbonneau}, {Winn},
  {Korzennik}, {Holman}, {Latham}, {Laird}, {Fernandez}, {O'Donovan},
  {Mandushev}, {Dunham}, {Everett}, {Esquerdo}, {Rabus}, {Belmonte}, {Deeg},
  {Brown}, {Hidas}, \& {Baliber}}]{Sozzetti+09}
{Sozzetti}, A., {Torres}, G., {Charbonneau}, D., {et~al.} 2009, \apj, 691, 1145

\bibitem[{{Street} {et~al.}(2010){Street}, {Simpson}, {Barros}, {Pollacco},
  {Joshi}, {Todd}, {Collier Cameron}, {Enoch}, {Parley}, {Stempels}, {Hebb},
  {Triaud}, {Queloz}, {Segransan}, {Pepe}, {Udry}, {Lister}, {Depagne}, {West},
  {Norton}, {Smalley}, {Hellier}, {Anderson}, {Maxted}, {Bentley}, {Skillen},
  {Gillon}, {Wheatley}, {Bento}, {Cathaway-Kjontvedt}, \&
  {Christian}}]{Street+10}
{Street}, R.~A., {Simpson}, E., {Barros}, S.~C.~C., {et~al.} 2010, \apj, 720,
  337

\bibitem[{{Szab{\'o}} {et~al.}(2012){Szab{\'o}}, {P{\'a}l}, {Derekas}, {Simon},
  {Szalai}, \& {Kiss}}]{Szabo+12}
{Szab{\'o}}, G.~M., {P{\'a}l}, A., {Derekas}, A., {et~al.} 2012, \mnras, 421,
  L122

\bibitem[{{Szab{\'o}} {et~al.}(2011){Szab{\'o}}, {Szab{\'o}}, {Benk{\H o}},
  {Lehmann}, {Mez{\H o}}, {Simon}, {K{\H o}v{\'a}ri}, {Hodos{\'a}n},
  {Reg{\'a}ly}, \& {Kiss}}]{Szabo+11}
{Szab{\'o}}, G.~M., {Szab{\'o}}, R., {Benk{\H o}}, J.~M., {et~al.} 2011, \apjl,
  736, L4

\bibitem[{{Thies} {et~al.}(2011){Thies}, {Kroupa}, {Goodwin}, {Stamatellos}, \&
  {Whitworth}}]{Thies+11}
{Thies}, I., {Kroupa}, P., {Goodwin}, S.~P., {Stamatellos}, D., \& {Whitworth},
  A.~P. 2011, \mnras, 417, 1817

\bibitem[{{Todorov} {et~al.}(2012){Todorov}, {Deming}, {Knutson}, {Burrows},
  {Sada}, {Cowan}, {Agol}, {Desert}, {Fortney}, {Charbonneau}, {Laughlin},
  {Langton}, {Showman}, \& {Lewis}}]{Todorov+12}
{Todorov}, K.~O., {Deming}, D., {Knutson}, H.~A., {et~al.} 2012, \apj, 746, 111

\bibitem[{{Torres} {et~al.}(2012){Torres}, {Fischer}, {Sozzetti}, {Buchhave},
  {Winn}, {Holman}, \& {Carter}}]{Torres+12}
{Torres}, G., {Fischer}, D.~A., {Sozzetti}, A., {et~al.} 2012, \apj, 757, 161

\bibitem[{{Torres} {et~al.}(2008){Torres}, {Winn}, \& {Holman}}]{Torres+08}
{Torres}, G., {Winn}, J.~N., \& {Holman}, M.~J. 2008, \apj, 677, 1324

\bibitem[{{Tregloan-Reed} {et~al.}(2013){Tregloan-Reed}, {Southworth}, \&
  {Tappert}}]{TR+13}
{Tregloan-Reed}, J., {Southworth}, J., \& {Tappert}, C. 2013, \mnras, 428, 3671

\bibitem[{{Triaud} {et~al.}(2009){Triaud}, {Queloz}, {Bouchy}, {Moutou},
  {Collier Cameron}, {Claret}, {Barge}, {Benz}, {Deleuil}, {Guillot},
  {H{\'e}brard}, {Lecavelier Des {\'E}tangs}, {Lovis}, {Mayor}, {Pepe}, \&
  {Udry}}]{Triaud+09}
{Triaud}, A.~H.~M.~J., {Queloz}, D., {Bouchy}, F., {et~al.} 2009, \aap, 506,
  377

\bibitem[{{Triaud} {et~al.}(2010){Triaud}, {Collier Cameron}, {Queloz},
  {Anderson}, {Gillon}, {Hebb}, {Hellier}, {Loeillet}, {Maxted}, {Mayor},
  {Pepe}, {Pollacco}, {S{\'e}gransan}, {Smalley}, {Udry}, {West}, \&
  {Wheatley}}]{Triaud+10}
{Triaud}, A.~H.~M.~J., {Collier Cameron}, A., {Queloz}, D., {et~al.} 2010,
  \aap, 524, A25

\bibitem[{{Triaud} {et~al.}(2011){Triaud}, {Queloz}, {Hellier}, {Gillon},
  {Smalley}, {Hebb}, {Collier Cameron}, {Anderson}, {Boisse}, {H{\'e}brard},
  {Jehin}, {Lister}, {Lovis}, {Maxted}, {Pepe}, {Pollacco}, {S{\'e}gransan},
  {Simpson}, {Udry}, \& {West}}]{Triaud+11}
{Triaud}, A.~H.~M.~J., {Queloz}, D., {Hellier}, C., {et~al.} 2011, \aap, 531,
  A24

\bibitem[{{Triaud} {et~al.}(2013){Triaud}, {Anderson}, {Collier Cameron},
  {Doyle}, {Fumel}, {Gillon}, {Hellier}, {Jehin}, {Lendl}, {Lovis}, {Maxted},
  {Pepe}, {Pollacco}, {Queloz}, {S{\'e}gransan}, {Smalley}, {Smith}, {Udry},
  {West}, \& {Wheatley}}]{Triaud+13}
{Triaud}, A.~H.~M.~J., {Anderson}, D.~R., {Collier Cameron}, A., {et~al.} 2013,
  \aap, 551, A80

\bibitem[{{Tripathi} {et~al.}(2010){Tripathi}, {Winn}, {Johnson}, {Howard},
  {Halverson}, {Marcy}, {Holman}, {de Kleer}, {Carter}, {Esquerdo}, {Everett},
  \& {Cabrera}}]{Tripathi+10}
{Tripathi}, A., {Winn}, J.~N., {Johnson}, J.~A., {et~al.} 2010, \apj, 715, 421

\bibitem[{{Tutukov} \& {Fedorova}(2012)}]{TutukovFedorova2012}
{Tutukov}, A.~V., \& {Fedorova}, A.~V. 2012, Astronomy Reports, 56, 305

\bibitem[{{Valsecchi} {et~al.}(2013){Valsecchi}, {Farr}, {Willems}, {Rasio}, \&
  {Kalogera}}]{ValsecchiFWRK13}
{Valsecchi}, F., {Farr}, W.~M., {Willems}, B., {Rasio}, F.~A., \& {Kalogera},
  V. 2013, \apj, 773, 39

\bibitem[{{Valsecchi} \& {Rasio}(2014)}]{VR14}
{Valsecchi}, F., \& {Rasio}, F.~A. 2014, ArXiv e-prints, arXiv:1403.1870

\bibitem[{{Van Eylen} {et~al.}(2013){Van Eylen}, {Lindholm Nielsen}, {Hinrup},
  {Tingley}, \& {Kjeldsen}}]{VanEylen+13}
{Van Eylen}, V., {Lindholm Nielsen}, M., {Hinrup}, B., {Tingley}, B., \&
  {Kjeldsen}, H. 2013, \apjl, 774, L19

\bibitem[{{VanEylen} {et~al.}(2012){VanEylen}, {Kjeldsen},
  {Christensen-Dalsgaard}, \& {Aerts}}]{VanEylen+12}
{VanEylen}, V., {Kjeldsen}, H., {Christensen-Dalsgaard}, J., \& {Aerts}, C.
  2012, Astronomische Nachrichten, 333, 1088

\bibitem[{{Verbunt} \& {Phinney}(1995)}]{VerbuntPhinney1995}
{Verbunt}, F., \& {Phinney}, E.~S. 1995, \aap, 296, 709

\bibitem[{{Ward}(1997)}]{Ward97}
{Ward}, W.~R. 1997, Icarus, 126, 261

\bibitem[{{Winn} {et~al.}(2010{\natexlab{a}}){Winn}, {Fabrycky}, {Albrecht}, \&
  {Johnson}}]{WinnFAJ10}
{Winn}, J.~N., {Fabrycky}, D., {Albrecht}, S., \& {Johnson}, J.~A.
  2010{\natexlab{a}}, \apjl, 718, L145

\bibitem[{{Winn} {et~al.}(2009{\natexlab{a}}){Winn}, {Holman}, {Carter},
  {Torres}, {Osip}, \& {Beatty}}]{Winn+09b}
{Winn}, J.~N., {Holman}, M.~J., {Carter}, J.~A., {et~al.} 2009{\natexlab{a}},
  \aj, 137, 3826

\bibitem[{{Winn} {et~al.}(2009{\natexlab{b}}){Winn}, {Johnson}, {Albrecht},
  {Howard}, {Marcy}, {Crossfield}, \& {Holman}}]{Winn+09c}
{Winn}, J.~N., {Johnson}, J.~A., {Albrecht}, S., {et~al.} 2009{\natexlab{b}},
  \apjl, 703, L99

\bibitem[{{Winn} {et~al.}(2006){Winn}, {Johnson}, {Marcy}, {Butler}, {Vogt},
  {Henry}, {Roussanova}, {Holman}, {Enya}, {Narita}, {Suto}, \&
  {Turner}}]{Winn+06}
{Winn}, J.~N., {Johnson}, J.~A., {Marcy}, G.~W., {et~al.} 2006, \apjl, 653, L69

\bibitem[{{Winn} {et~al.}(2008{\natexlab{a}}){Winn}, {Johnson}, {Narita},
  {Suto}, {Turner}, {Fischer}, {Butler}, {Vogt}, {O'Donovan}, \&
  {Gaudi}}]{Winn+08}
{Winn}, J.~N., {Johnson}, J.~A., {Narita}, N., {et~al.} 2008{\natexlab{a}},
  \apj, 682, 1283

\bibitem[{{Winn} {et~al.}(2008{\natexlab{b}}){Winn}, {Holman}, {Torres},
  {McCullough}, {Johns-Krull}, {Latham}, {Shporer}, {Mazeh}, {Garcia-Melendo},
  {Foote}, {Esquerdo}, \& {Everett}}]{Winn+08b}
{Winn}, J.~N., {Holman}, M.~J., {Torres}, G., {et~al.} 2008{\natexlab{b}},
  \apj, 683, 1076

\bibitem[{{Winn} {et~al.}(2009{\natexlab{c}}){Winn}, {Johnson}, {Fabrycky},
  {Howard}, {Marcy}, {Narita}, {Crossfield}, {Suto}, {Turner}, {Esquerdo}, \&
  {Holman}}]{Winn+09}
{Winn}, J.~N., {Johnson}, J.~A., {Fabrycky}, D., {et~al.} 2009{\natexlab{c}},
  \apj, 700, 302

\bibitem[{{Winn} {et~al.}(2010{\natexlab{b}}){Winn}, {Johnson}, {Howard},
  {Marcy}, {Bakos}, {Hartman}, {Torres}, {Albrecht}, \& {Narita}}]{Winn+10}
{Winn}, J.~N., {Johnson}, J.~A., {Howard}, A.~W., {et~al.} 2010{\natexlab{b}},
  \apj, 718, 575

\bibitem[{{Winn} {et~al.}(2011){Winn}, {Howard}, {Johnson}, {Marcy},
  {Isaacson}, {Shporer}, {Bakos}, {Hartman}, {Holman}, {Albrecht}, {Crepp}, \&
  {Morton}}]{Winn+11}
{Winn}, J.~N., {Howard}, A.~W., {Johnson}, J.~A., {et~al.} 2011, \aj, 141, 63

\bibitem[{{Witte} \& {Savonije}(1999{\natexlab{a}})}]{WitteSavonije1999b}
{Witte}, M.~G., \& {Savonije}, G.~J. 1999{\natexlab{a}}, \aap, 341, 842

\bibitem[{{Witte} \& {Savonije}(1999{\natexlab{b}})}]{WitteSavonije1999}
---. 1999{\natexlab{b}}, \aap, 350, 129

\bibitem[{{Witte} \& {Savonije}(2002)}]{WitteSavonije02}
---. 2002, \aap, 386, 222

\bibitem[{{Wu} \& {Lithwick}(2011)}]{WuLithwick11}
{Wu}, Y., \& {Lithwick}, Y. 2011, \apj, 735, 109

\bibitem[{{Wu} \& {Murray}(2003)}]{WuMurray03}
{Wu}, Y., \& {Murray}, N. 2003, \apj, 589, 605

\bibitem[{{Xiang-Gruess} \& {Papaloizou}(2014)}]{XiangPapaloizou14}
{Xiang-Gruess}, M., \& {Papaloizou}, J.~C.~B. 2014, ArXiv e-prints,
  arXiv:1402.2792

\bibitem[{{Xue} {et~al.}(2014){Xue}, {Suto}, {Taruya}, {Hirano}, {Fujii}, \&
  {Masuda}}]{Xue+14}
{Xue}, Y., {Suto}, Y., {Taruya}, A., {et~al.} 2014, ArXiv e-prints,
  arXiv:1401.5876

\bibitem[{{Yi} {et~al.}(2001){Yi}, {Demarque}, {Kim}, {Lee}, {Ree}, {Lejeune},
  \& {Barnes}}]{Yi+01}
{Yi}, S., {Demarque}, P., {Kim}, Y.-C., {et~al.} 2001, \apjs, 136, 417

\bibitem[{{Zahn}(1975)}]{Zahn1975}
{Zahn}, J.-P. 1975, \aap, 41, 329

\bibitem[{{Zahn}(1977)}]{Zahn1977}
---. 1977, \aap, 57, 383

\bibitem[{{Zahn}(1989)}]{Zahn1989}
---. 1989, \aap, 220, 112

\end{thebibliography}

\appendix

\section{Validation of the Orbital Evolution Code}\label{Tests on the Orbital Evolution code}
We test the code developed to integrate the equations summarized in \S~\ref{Orbital Evolution Model} by reproducing some of the results presented in M10 and R13.

\begin{figure} [!h]
\epsscale{0.8}
\plotone{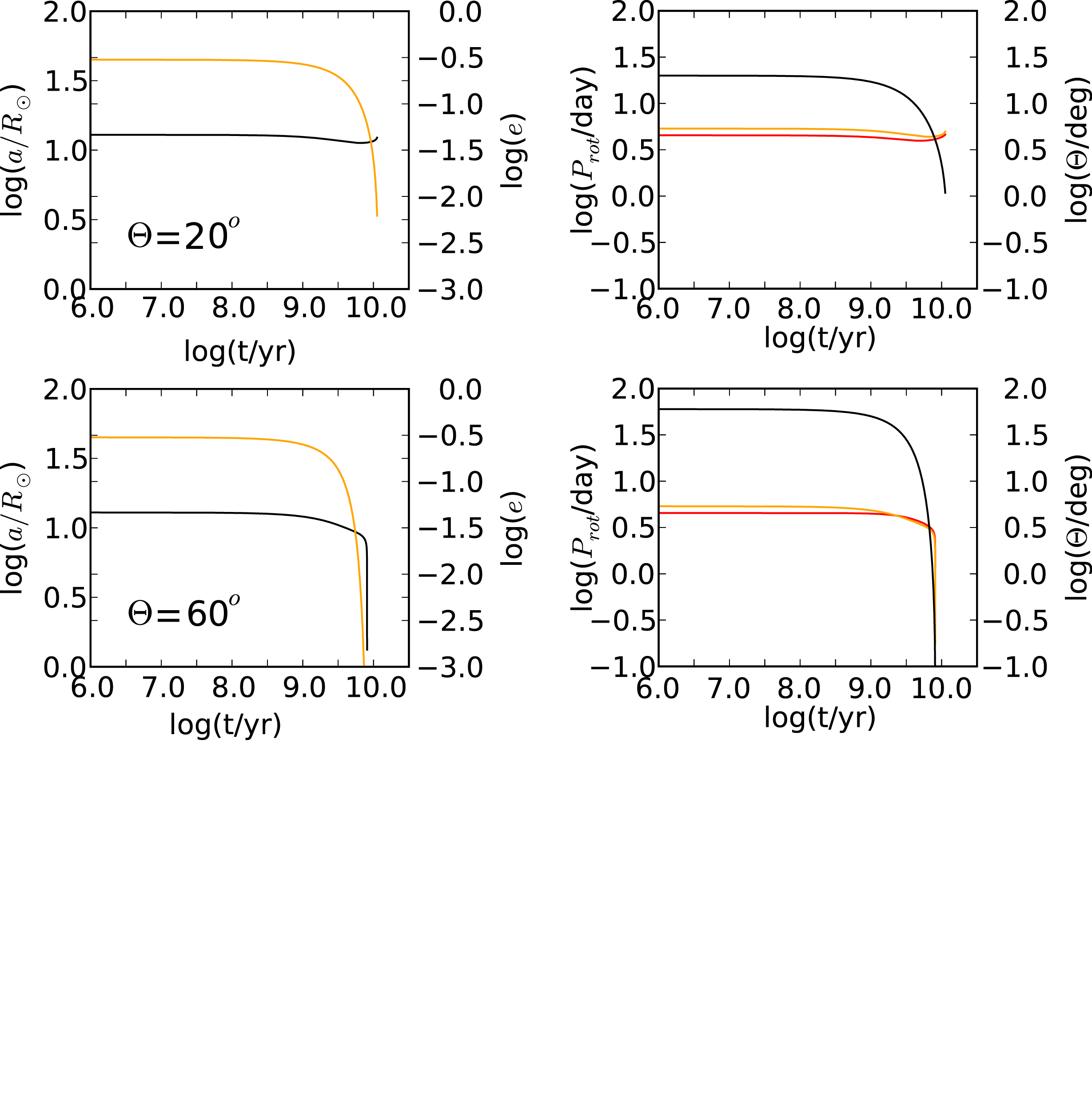}
\caption{Tidal evolution of a 3\,$\,M_{\rm{Jup}}$ at 0.06~AU and $e\,=\,$0.3. For ease of comparison with Fig.~4 in M10 we adopt the same color-scheme. Specifically, the top (bottom) plots represent the evolution of the orbital and stellar parameters for an initial stellar obliquity of 20$^{o}$ (60$^{o}$). In the left panels, black lines represent the evolution of  the orbital separation, while orange lines represent the evolution of the eccentricity. In the right panels, orange and red lines represent the evolution of the orbital and star's rotation period, respectively, while black lines show the evolution of the star's misalignment (which in M10 is denoted with $\epsilon_{*}$).}
\label{fig:MatsumuraTest.eps}
\end{figure}

First, we verify the equations describing the evolution of $a$, $\Omega_{*}$, and $\Theta_{*}$ due to tides in the weak friction approximation by reproducing the results presented in Fig.\,4 of M10. Specifically, we integrate Eqs.~(\ref{eq:dadtTidesStartrack})-(\ref{eq:dThetadtHut}) generalized for an eccentric orbit \citep{Hut1981}:

\begin{align}
(\dot{a})_{\rm wf} &= -\frac{1}{\tau_{\rm wf}}\frac{a}{(1-e^{2})^{15/2}}\left[f_{\rm 1}(e^{2})-(1-e^{2})^{3/2}f_{\rm 2}(e^{2})\rm{cos}\Theta_{*}\frac{\Omega_{*}}{\Omega_{orb}}\right], \label{eq:dadtTidesStartrackAppend}\\
(\dot{e})_{\rm wf} &= -\frac{9}{2}\frac{1}{\tau_{\rm wf}}\frac{e}{(1-e^{2})^{13/2}}\left[f_{\rm 3}(e^{2})-\frac{11}{18}(1-e^{2})^{3/2}f_{\rm 4}(e^{2})\rm{cos}\Theta_{*}\frac{\Omega_{*}}{\Omega_{orb}}\right], \label{eq:dedt}\\
(\dot{\Omega}_{*})_{\rm wf} & =\frac{1}{\tau_{\rm wf}}\frac{\Omega_{orb}}{(1-e^{2})^{6}}\left(\frac{L}{2S}\right)\left[f_{\rm 2}(e^{2})\rm{cos}\Theta_{*}-\frac{1}{2}(1+cos^{2}\Theta_{*})(1-e^{2})^{3/2}f_{5}(e^{2})\frac{\Omega_{*}}{\Omega_{orb}}\right], \label{eq:dwdtTidesStartrackAppend}\\
(\dot{\Theta}_{*})_{\rm wf} &= -\frac{1}{\tau_{\rm wf}}\frac{\rm{sin}\Theta_{*}}{(1-e^{2})^{6}}\left(\frac{L}{2S}\right)\left[f_{\rm 2}(e^{2})-\frac{1}{2}(\rm{cos}\Theta_{*}-\frac{S}{L})(1-e^{2})^{3/2}f_{5}(e^{2})\frac{\Omega_{*}}{\Omega_{orb}}\right]. \label{eq:dThetadtHutAppend}
\end{align}
Here we have added Eq.~(\ref{eq:dedt}) which describes the evolution of the eccentricity $e$. In the above equations the coefficients $f_{\rm i}$ ($i\,=\,1, 2, 3, 4, 5$) are given by
\begin{align} 
f_{\rm 1}(e^{2})& = 1 + (31/2)e^2+ (255/8)e^4 + (185/16)e^6 + (25/64)e^8,\\
f_{\rm 2}(e^{2})& = 1 + (15/2)e^2 + (45/8)e^4 + (5/16)e^6,\\
f_{\rm 3}(e^{2})& = 1 + (15/4)e^2 + (15/8)e^4 + (5/64)e^6, \\
f_{\rm 4}(e^{2})& = 1 + (3/2)e^2 + (1/8)e^4, \\
f_{\rm 5}(e^{2})& = 1 + 3e^2 + (3/8)e^4. 
\end{align}
Eqs.~(\ref{eq:dadtTidesStartrackAppend}), (\ref{eq:dwdtTidesStartrackAppend}), and (\ref{eq:dThetadtHutAppend}) fall back to Eqs.~(\ref{eq:dadtTidesStartrack})-(\ref{eq:dThetadtHut}) if $e\,=\,0$. M10 consider also tides in the planet which here are neglected. This assumption doesn't' affect the results significantly for the particular example presented here (see below). Neglecting the terms related to tidal dissipation in the planet and setting $F_{\rm tid}\,=\,1$ in Eq~(\ref{eq:tau_wf}), Eqs.~(\ref{eq:dadtTidesStartrackAppend})-(\ref{eq:dThetadtHutAppend}) are equivalent to Eqs.~(7)-(9), and (11) in M10, as the term $(k/T)$ in 
Eq.~(\ref{eq:tau_wf}) is equivalent to $k_{\rm 2, *}\Delta t_{*}(GM_{*}/R_{*}^{3})$. For the calculation presented in Fig. 4 of M10, $Q_{*}\sim1/(\Delta t_{*}\,\Omega_{o})$, therefore $k_{\rm 2, *}\Delta t_{*} = \frac{3}{2}\frac{1}{Q'_{*}}\frac{1}{\Omega_{\rm o}}$, where $Q'_{*}\,=\,\frac{3}{2}\frac{k_{\rm 2,*}}{Q_{*}}$ is the star's modified tidal quality factor.  Closely following M10, we consider a 1\,$M_{\odot}$ star at solar metallicity coupled with a 3\,$\,M_{\rm{Jup}}$ companion. We neglect stellar wind mass loss, the effect of magnetic braking, and changes in the stellar spin related to the evolution of the star's moment of inertia. Furthermore, we take $i_{*}\,=\,i_{o}\,=\,90^{o}$. With this assumption the sky-projected misalignment is equal to the true one ($\Theta_{*}\,=\,\lambda$). We consider an initial semimajor axis, eccentricity and stellar velocity of $a\,=\,0.06\,$AU, $e\,=\,0.3$, and $v_{\rm rot}\,=\,10\,$km\,s$^{-1}$, respectively, and integrate Eqs.~(\ref{eq:dadtTidesStartrackAppend})-(\ref{eq:dThetadtHutAppend}) forward in time for an initial obliquity $\Theta_{*}$ of 20$^{o}$ and 60$^{o}$. As in M10, the star's tidal quality factor scales as $Q'_{*} = Q'_{\rm *,in}\Omega_{\rm o,in}/\Omega_{\rm o}$, where the subscript ``in'' denotes the initial values and $Q'_{\rm *,in}$ is set to $10^{6}$. The star's squared radius of gyration is set to 0.06. In Fig.~\ref{fig:MatsumuraTest.eps} we show the result on this calculation, to be compared with Fig.~4 in M10. Similarly, we find that for the smaller stellar obliquity (20$^{o}$) the system arrives at a stable tidal equilibrium state as synchronization, circularization and alignment are reached. Instead, for the larger stellar obliquity (60$^o$), the system evolves towards a state of unstable tidal equilibrium in which the planets spirals onto the star on a $\simeq\,10\,$Gyr timescale.
Here we remind the reader that our calculation neglects tidal dissipation in the planet, while M10 parametrizes it in term of a modified tidal $Q'$ for the planet ($Q'_{\rm pl}$). As it is clear from Fig.\,4 in M10, this would affect mostly the evolution of the eccentricity. In agreement with M10's results for the highest $Q'_{\rm pl}$ value ($10^{7}$), we find that the orbit quickly circularizes right before the orbital separation drops. 

Next, we verify the tidal prescription proposed by \cite{Lai12} by reproducing the results presented in Fig.\,2 by R13.
Specifically, we integrate Eq.~(\ref{eq:59Lai12}), which is equivalent to Eq.~(7) in R13. In fact, it is straightforward to show that $\tau_{\rm wf}$ defined in Eq.~(\ref{eq:tau_wf}) is equivalent to $\tau_{\rm e}$ in R13, if $Q_{*}\sim1/(2\Delta t_{*}\,\Omega_{o})$. 
We start from a random distribution of 50 obliquities and integrate Eq.~(\ref{eq:59Lai12}) forward in time, keeping the orbital separation and stellar spin fixed during the integration. The time variable $t$ in R13 is defined as a multiple of $\tau_{\rm 10}$ [Eq.~\ref{eq:53Lai12}]. Defining $t\,=\, \tilde{t}\tau_{\rm 10}$, the terms in Eq.~(\ref{eq:59Lai12}) are give by
\begin{align}
&\left(\frac{d\Theta_{*}}{d\tilde{t}}\right)_{\rm wf} = -\frac{\tau_{\rm 10}}{\tau_{\rm wf}}{\rm sin\,}\Theta_{*}\left(\frac{L}{2S}\right)\left[1-\left(\frac{\Omega_{*}}{2\Omega_{o}}\right)\left({\rm cos\,}\Theta_{*}-\frac{S}{L}\right)\right], \label{eq:dThetadtHutAppendRogers}\\
&\left(\frac{d\Theta_{*}}{d\tilde{t}}\right)_{\rm 10}  = -\rm{sin\,}\Theta_{*}\,\rm{cos}^{2}\,\Theta_{*}\,(\rm{cos}\,\Theta_{*}+\frac{S}{L}),\label{eq:52Lai12AppndRogers}\\
&\left(\frac{d\Theta_{*}}{d\tilde{t}}\right)_{\rm 10, wf} = \frac{\tau_{\rm 10}}{\tau_{\rm wf}}\frac{L}{4S}\left(\frac{d\Theta_{*}}{ds}\right)_{\rm 10}.\label{eq:CrossLaiTermAppendRogers}
\end{align}
As in Fig.~2 in R13, we set $\tau_{\rm 10}/\tau_{\rm wf}\,=\,0.001$, consider initial values of $S/L$\,=\,0.1, 0.5, and 2 and compute the evolution of the obliquity until 30$\tau_{\rm 10}$\footnote{Here we note a TYPO in the caption of Fig.~2 in R13, as the final integration time is not 30$\,\tau_{\rm wf}$.}. The ratio $\Omega_{*}/\Omega_{o}$ is set to 0.1.

\begin{figure} [!h]
\epsscale{0.9}
\plotone{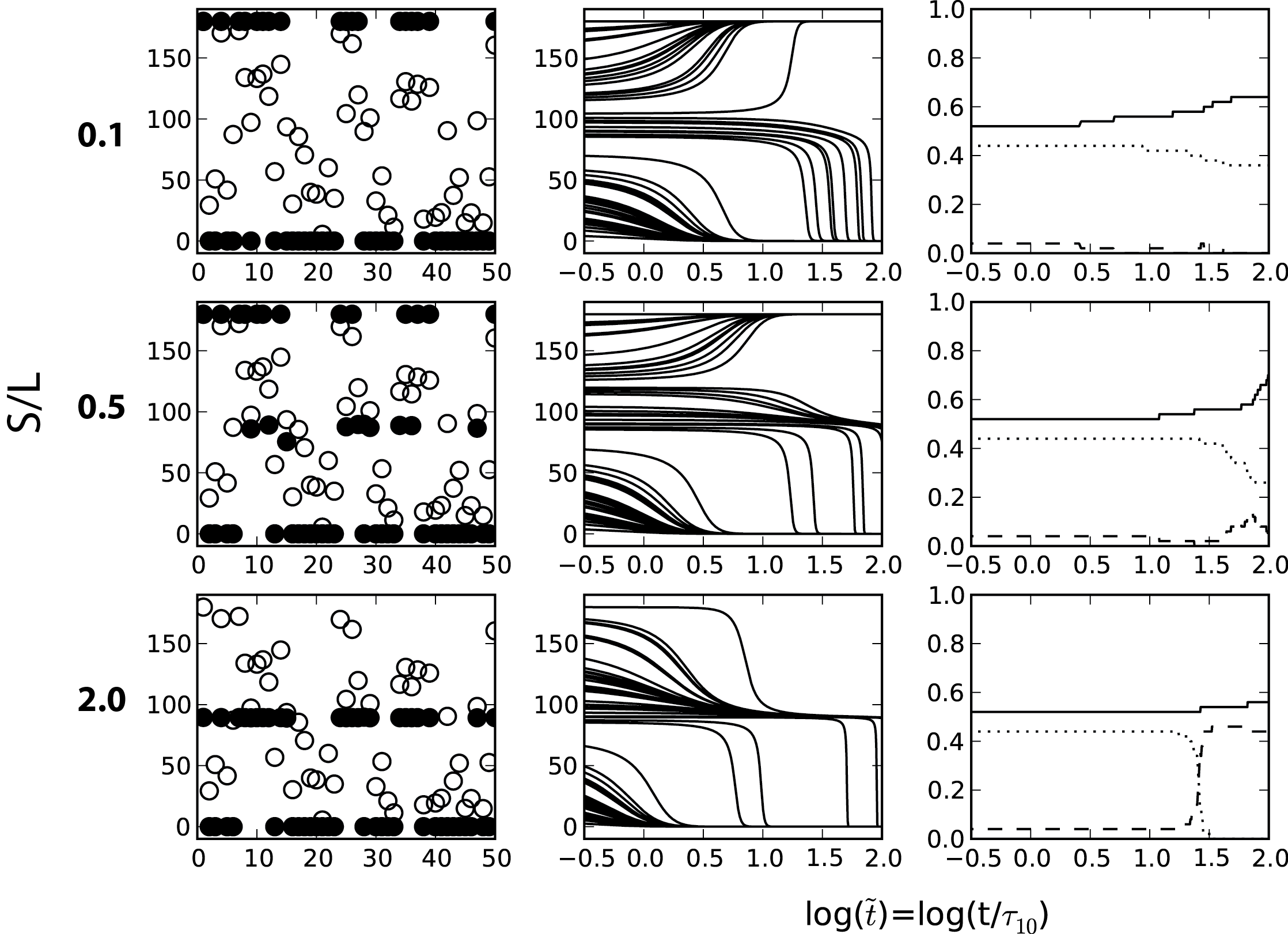}
\caption{Obliquity evolution of a population of 50 objects with initially random obliquities. The calculation was performed with $\Omega_{*}/\Omega_{o}\,=\,$0.1 and $\tau_{\rm 10}/\tau_{\rm wf}\,=\,0.001$. Similarly to R13, the left panels show the distribution of initial $\Theta_{*}$ with open circles and the distribution of $\Theta_{*}$ after 30$\,\tau_{\rm wf}$ with filled circles. The middle panels shows the time evolution of $\Theta_{*}$ and the right panels shows the fraction of objects as a function of time with $\Theta\textless\,$89$^{o}$ (solid line), $\Theta\textgreater\,$91$^{o}$ (dotted line), and 89$^{o}\leq\Theta_{*}\leq$91$^{o}$ (dashed line). The top, middle, and bottom row of panels are for initial $S/L\,=\,$0.1, 0.5, and 2,  respectively. In all scenarios the objects evolve to prograde, retrograde, or 90$^{o}$ orbits.}
\label{fig:orbitalPropertiesEnd_Fig2Rogers}
\end{figure}

In Fig.~\ref{fig:orbitalPropertiesEnd_Fig2Rogers} we show the result of this calculation, to be compared with Fig.~2 in R13. In agreement with R13, we find that in all different configurations of $S/L$, the objects evolve to prograde, retrograde, or 90$^{o}$ orbits. The majority of hot Jupiters with initial $\Theta_{*}\textless$90$^{o}$ evolves toward alignment regardless the value of $S/L$. Instead, most hot Jupiters with initial $\Theta_{*}\textgreater$90$^{o}$ either evolve towards $\Theta_{*}\,=\,$180$^{o}$ (for small $S/L$) or $\Theta_{*}\,=\,$90$^{o}$ (if $S\textgreater L$).
\section{Dynamic Tides in WASP-71}\label{Dynamic Tides in WASP-71}
In this section, we investigate whether dynamic tides might have been significant in the past orbital evolution of a WASP-71-type system for a broad spectrum of tidal forcing frequencies. First, we select two stellar profiles from the evolutionary sequence shown in the top left panel of Fig.~\ref{fig:HR_detailed_orbEv}. The two profiles are a snapshot of the structure of the star at 10\% and 50\% of its main sequence lifetime $t_{\rm MS}$ (according to the detailed stellar modeling presented in \S~\ref{Detailed orbital evolution for four representative systems}, \mbox{WASP-71} is at $\sim\,90\%\,t_{\rm MS}$). Next, we compute the timescales associated with radiative damping of dynamic tides with {\tt CAFein} \citep{ValsecchiFWRK13}.
This code computes the star's response to the tidal action of a companion solving the full set of non-adiabatic and non-radial forced stellar oscillation equations. 

As described in \cite{ValsecchiFWRK13}, we assume that the star rotates uniformly around an axis orthogonal to the orbital plane with angular velocity $\Omega_{*}$ in the sense of the orbital motion. We assume $\Omega_{*}$ to be small enough so that the Coriolis force and the centrifugal force can be neglected. We treat the planet as a point mass. 
The coupling of convection and pulsations, and the perturbation of the convective flux are neglected. We consider a circular orbit and fix the orbital period to the currently observed one, while varying the stellar spin. We focus on the leading quadrupole order in the spherical harmonic expansion of the tide-generating potential (e.g., \citealt{PolflietSmeyers1990}).

\begin{figure} [!h]
\epsscale{0.8}
\plotone{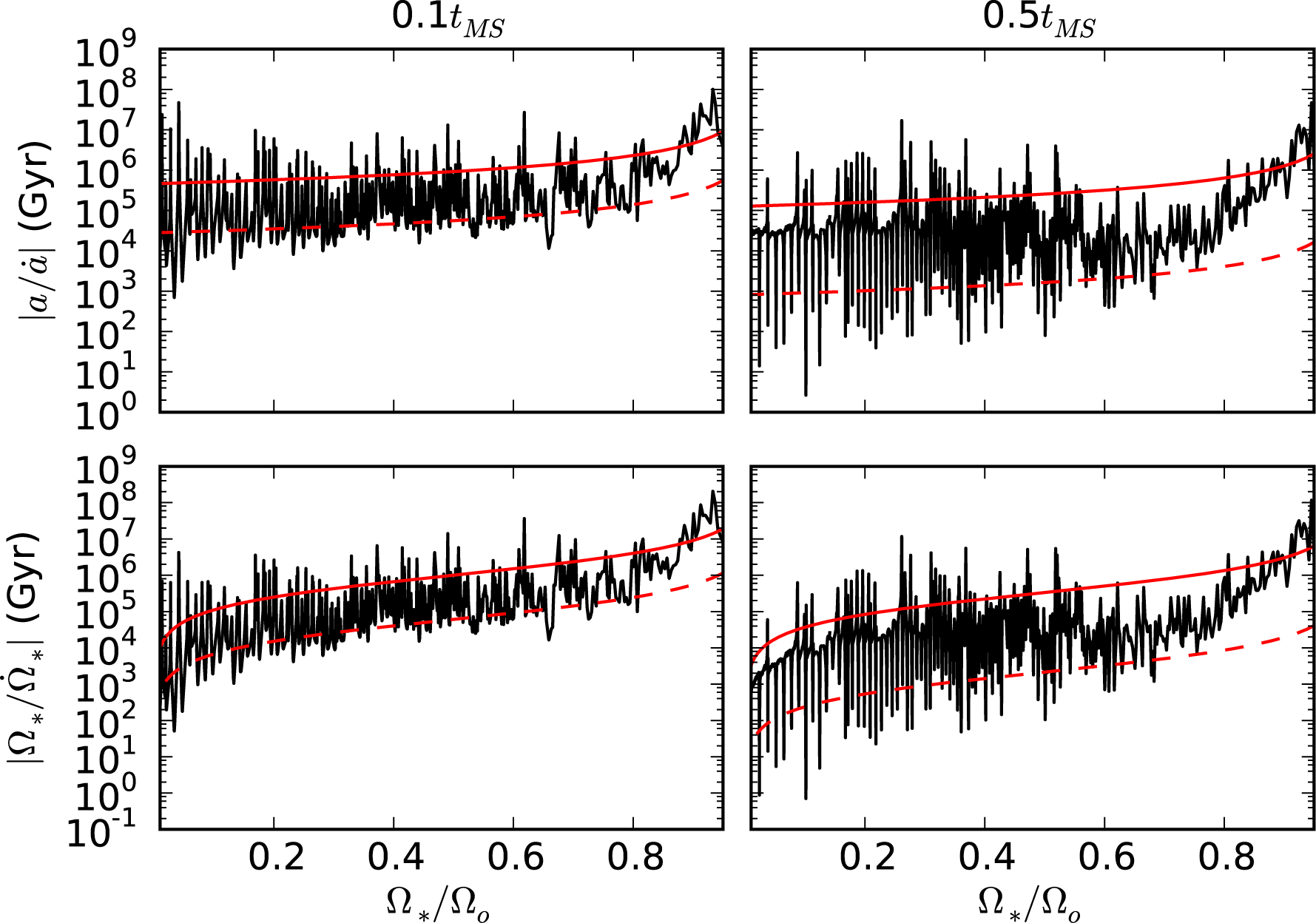}
\caption{Timescales for the evolution of the orbital separation (top) and stellar spin (bottom) due to tides for a WASP-71-type system. Here we take the star to be at 10\% (left) and 50\% (right) of its main sequence lifetime, and $\Theta_{*}\,=\,0$. The black solid line represents the timescales related to radiative dissipation of the dynamical tide computed with {\tt CAFein}. Their complex behavior is due to resonances between the star's eigenfrequencies and the tidal forcing frequencies. For comparison, the red solid and dashed lines are the tidal timescales computed from Eqs.\,\ref{eq:dadtTidesStartrack} and \ref{eq:dwdtTidesStartrack} associated with radiative damping of the dynamical tide [$k/T$ from Eq.\,(\ref{eq:kOverTRad})] and convective damping of the equilibrium tide [$k/T$ from Eq.\,(\ref{eq:kOverTConv})], respectively, in the weak friction approximation. }
\label{fig:CAFein_and_Zahn}
\end{figure}

The orbital evolution timescales computed with {\tt CAFein} are shown in Fig.~\ref{fig:CAFein_and_Zahn} (black line), together with the weak-friction tides prescription adopted in this work (red lines). The complex behavior of the dynamic tides timescales is due to resonances between the star's eigenfrequencies and the tidal forcing frequencies. Fixing the orbital period to the currently observed value and varying the stellar spin produces a spectrum of tidal forcing frequencies $\omega_{\rm T}\,=\,2(\Omega_{o}-\Omega_{*})$. For an asynchronous star, radiative dissipation of dynamic tides in the weak friction approximation (red solid line) yields timescales that are almost an order of magnitude longer than those computed accounting for the full spectrum of tidal forcing frequencies. As expected, the two prescriptions agree as $\Omega_{*}$ approaches $\Omega_{o}$ (in the limit of small tidal forcing frequencies). Convective dissipation of the equilibrium tides leads to overall shorter timescales. 

Here we note that, for the orbital configurations and stellar ages considered, the timescales computed with {\tt CAFein} and presented in Fig.~\ref{fig:CAFein_and_Zahn} should be taken as upper limits. In fact, both convection and rotation could affect the star's tidal response. Convection could excite non-radial stellar pulsations \citep{CantielloLBdKSVVLY2009}, while rotation enriches the spectrum of eigenfrequencies, leading to more resonances \citep{WitteSavonije1999b,WitteSavonije1999}. Furthermore, WASP-71's misalignment is constrained to be $\textless 90^{o}$ \citep{Smith+13}. The detailed orbital evolution described in \S~\ref{WASP-71} accounts mainly for convective damping, as it is more efficient than radiative damping in the limit of small tidal forcing frequencies. However, the calculation presented here suggests that radiative dissipation of dynamic tides might be significant, once the fully dynamical tidal response of the star is taken into account. Therefore, a detailed calculation accounting for the interaction between tides and the free oscillation modes of the star, and its effect on the evolution of the orbital separation and stellar spin might be more appropriate (e.g., \citealt{WitteSavonije02}).

\clearpage
\LongTables
\begin{deluxetable}{ccccccccc}
\tablecaption{Hot Jupiters Sample.}
\tablewidth{0pt}
\tablehead{
\\[0.5em]
\colhead{Planet Name} & \colhead{$M_{\rm pl}$} & \colhead{$P_{\rm orb}$} & \colhead{$M_{*}$} & \colhead{$R_{*}$} &
\colhead{$T_{\rm eff *}$} & \colhead{Fe/H} & \colhead{$\lambda\,$} &
\colhead{$v_{{\rm rot}}{\rm sin}~i_{*}$}
\\
 & \colhead{$(M_{\rm{Jup}})$} & \colhead{(d)} & \colhead{$(M_{\rm{\odot}})$} & \colhead{$(R_{\rm{\odot}})$} &
\colhead{(K)} &  & \colhead{(deg)} &
\colhead{(km\,s$^{-1}$)}
}
\startdata
\\[0.6em]
\smallskip
\textcolor{white}{1}&&&&&&&&\\
CoRoT-1 b& $^{1}$1.030 & $^{1}$1.509 & $^{1}$0.950$^{+0.150}_{-0.150}$ & $^{1}$1.110$^{+0.050}_{-0.050}$ & $^{1}$6298$^{+150}_{-150}$  & $^{1}$-0.300$^{+0.250}_{-0.250}$   & $^{2}$77.0$^{+11.0}_{-11.0}$ & $^{1}$5.20$^{+1.00}_{-1.00}$\\
\textcolor{white}{1}&&&&&&&&\\
CoRoT-2 b& $^{3}$3.310 & $^{3}$1.743 & $^{3}$0.970$^{+0.060}_{-0.060}$ & $^{3}$0.902$^{+0.018}_{-0.018}$ & $^{4}$5625$^{+120}_{-120}$  & $^{4}$0.000$^{+0.100}_{-0.100}$   & $^{5}$4.0$^{+6.1}_{-5.9}$ & $^{4}$11.85$^{+0.50}_{-0.50}$\\
\textcolor{white}{1}&&&&&&&&\\
CoRoT-3 b& $^{6}$21.230 & $^{6}$4.257 & $^{6}$1.359$^{+0.059}_{-0.043}$ & $^{6}$1.540$^{+0.083}_{-0.078}$ & $^{7}$6740$^{+140}_{-140}$  & $^{7}$-0.020$^{+0.060}_{-0.060}$   & $^{6}$-37.6$^{+10.0}_{-22.3}$ & $^{6}$35.80$^{+8.20}_{-8.30}$\\
\textcolor{white}{1}&&&&&&&&\\
CoRoT-18 b& $^{8}$3.470 & $^{8}$1.900 & $^{8}$0.950$^{+0.150}_{-0.150}$ & $^{8}$1.000$^{+0.130}_{-0.130}$ & $^{8}$5440$^{+100}_{-100}$  & $^{8}$-0.100$^{+0.100}_{-0.100}$   & $^{8}$-10.0$^{+20.0}_{-20.0}$ & $^{8}$8.00$^{+1.00}_{-1.00}$\\
\textcolor{white}{1}&&&&&&&&\\
CoRoT-19 b& $^{9}$1.110 & $^{9}$3.897 & $^{9}$1.210$^{+0.050}_{-0.050}$ & $^{9}$1.650$^{+0.040}_{-0.040}$ & $^{9}$6090$^{+70}_{-70}$  & $^{9}$-0.020$^{+0.100}_{-0.100}$   & $^{9}$-52.0$^{+27.0}_{-22.0}$ & $^{9}$6.00$^{+1.00}_{-1.00}$\\
\textcolor{white}{1}&&&&&&&&\\
HAT-P-1 b& $^{10}$0.532 & $^{11}$4.465 & $^{10}$1.133$^{+0.075}_{-0.079}$ & $^{10}$1.135$^{+0.048}_{-0.048}$ & $^{10}$5975$^{+120}_{-120}$  & $^{10}$0.130$^{+0.080}_{-0.080}$   & $^{11}$3.7$^{+2.1}_{-2.1}$ & $^{11}$3.75$^{+0.58}_{-0.58}$\\
\textcolor{white}{1}&&&&&&&&\\
HAT-P-4 b& $^{12}$0.680 & $^{14}$3.056 & $^{12}$1.271$^{+0.120}_{-0.070}$ & $^{12}$1.600$^{+0.117}_{-0.042}$ & $^{13}$5860$^{+80}_{-80}$  & $^{13}$0.240$^{+0.080}_{-0.080}$   & $^{15}$-4.9$^{+11.9}_{-11.9}$ & $^{13}$5.50$^{+0.50}_{-0.50}$\\
\textcolor{white}{1}&&&&&&&&\\
HAT-P-6 b& $^{16}$1.057 & $^{16}$3.853 & $^{16}$1.290$^{+0.060}_{-0.060}$ & $^{16}$1.460$^{+0.060}_{-0.060}$ & $^{16}$6570$^{+80}_{-80}$  & $^{16}$-0.130$^{+0.080}_{-0.080}$   & $^{17}$165.0$^{+6.0}_{-6.0}$ & $^{17}$7.80$^{+0.60}_{-0.60}$\\
\textcolor{white}{1}&&&&&&&&\\
HAT-P-7 b& $^{18}$1.741 & $^{19}$2.205 & $^{18}$1.361$^{+0.021}_{-0.021}$ & $^{18}$1.904$^{+0.010}_{-0.010}$ & $^{18}$6259$^{+32}_{-32}$  & $^{18}$0.130   & $^{17, I}$155.0$^{+37.0}_{-37.0}$ & $^{17}$2.70$^{+0.50}_{-0.50}$\\
\textcolor{white}{1}&&&&&&&&\\
HAT-P-8 b& $^{20}$1.275 & $^{20}$3.076 & $^{20}$1.192$^{+0.075}_{-0.075}$ & $^{20}$1.475$^{+0.034}_{-0.034}$ & $^{22}$6130$^{+80}_{-80}$  & $^{21}$0.010$^{+0.080}_{-0.080}$   & $^{23}$-17.0$^{+9.2}_{-11.5}$ & $^{22}$2.60$^{+0.50}_{-0.50}$\\
\textcolor{white}{1}&&&&&&&&\\
HAT-P-9 b& $^{24}$0.780 & $^{24}$3.923 & $^{24}$1.280$^{+0.130}_{-0.130}$ & $^{24}$1.320$^{+0.070}_{-0.070}$ & $^{24}$6350$^{+150}_{-150}$  & $^{24}$0.120$^{+0.200}_{-0.200}$   & $^{23}$-16.0$^{+8.0}_{-8.0}$ & $^{23}$12.50$^{+1.80}_{-1.80}$\\
\textcolor{white}{1}&&&&&&&&\\
HAT-P-13 b& $^{25}$0.851 & $^{25}$2.916 & $^{25}$1.220$^{+0.050}_{-0.100}$ & $^{25}$1.559$^{+0.080}_{-0.080}$ & $^{26}$5653$^{+90}_{-90}$  & $^{26}$0.410$^{+0.080}_{-0.080}$   & $^{25}$1.9$^{+8.6}_{-8.6}$ & $^{25}$1.66$^{+0.37}_{-0.37}$\\
\textcolor{white}{1}&&&&&&&&\\
HAT-P-14 b& $^{27}$2.200 & $^{27}$4.628 & $^{27}$1.300$^{+0.030}_{-0.030}$ & $^{27}$1.480$^{+0.050}_{-0.050}$ & $^{27}$6583$^{+100}_{-100}$  & $^{27}$0.080$^{+0.100}_{-0.100}$   & $^{15}$-170.9$^{+5.1}_{-5.1}$ & $^{27}$8.40$^{+1.00}_{-1.00}$\\
\textcolor{white}{1}&&&&&&&&\\
HAT-P-16 b& $^{28}$4.193 & $^{28}$2.776 & $^{28}$1.216$^{+0.055}_{-0.055}$ & $^{28}$1.158$^{+0.025}_{-0.025}$ & $^{29}$6140$^{+72}_{-72}$  & $^{29}$0.120$^{+0.080}_{-0.080}$   & $^{23}$-10.0$^{+16.0}_{-16.0}$ & $^{30}$3.50$^{+0.50}_{-0.50}$\\
\textcolor{white}{1}&&&&&&&&\\
HAT-P-23 b& $^{31}$2.090 & $^{31}$1.213 & $^{31}$1.130$^{+0.035}_{-0.035}$ & $^{31}$1.203$^{+0.074}_{-0.074}$ & $^{31}$5905$^{+80}_{-80}$  & $^{31}$0.150$^{+0.040}_{-0.040}$   & $^{23}$15.0$^{+22.0}_{-22.0}$ & $^{31}$8.10$^{+0.50}_{-0.50}$\\
\textcolor{white}{1}&&&&&&&&\\
HAT-P-24 b& $^{32}$0.685 & $^{32}$3.355 & $^{32}$1.191$^{+0.042}_{-0.042}$ & $^{32}$1.317$^{+0.068}_{-0.068}$ & $^{32}$6373$^{+80}_{-80}$  & $^{32}$-0.160$^{+0.080}_{-0.080}$   & $^{17}$20.0$^{+16.0}_{-16.0}$ & $^{17}$11.20$^{+0.90}_{-0.90}$\\
\textcolor{white}{1}&&&&&&&&\\
HAT-P-30 b& $^{33}$0.711 & $^{33}$2.811 & $^{33}$1.242$^{+0.041}_{-0.041}$ & $^{33}$1.215$^{+0.051}_{-0.051}$ & $^{33}$6304$^{+88}_{-88}$  & $^{33}$0.130$^{+0.080}_{-0.080}$   & $^{33}$73.5$^{+9.0}_{-9.0}$ & $^{33}$3.07$^{+0.24}_{-0.24}$\\
\textcolor{white}{1}&&&&&&&&\\
HAT-P-32 b$^{II}$& $^{34}$0.860 & $^{34}$2.150 & $^{34}$1.160$^{+0.041}_{-0.041}$ & $^{34}$1.219$^{+0.016}_{-0.016}$ & $^{34}$6207$^{+88}_{-88}$  & $^{34}$-0.040$^{+0.080}_{-0.080}$   & $^{17}$85.0$^{+1.5}_{-1.5}$ & $^{17}$20.60$^{+1.50}_{-1.50}$\\
\textcolor{white}{1}&&&&&&&&\\
& $^{34}$0.941 & $^{34}$2.150 & $^{34}$1.176$^{+0.043}_{-0.070}$ & $^{34}$1.387$^{+0.067}_{-0.067}$ & $^{34}$6001$^{+88}_{-88}$  & $^{34}$-0.160$^{+0.080}_{-0.080}$   & $^{17}$85.0$^{+1.5}_{-1.5}$ & $^{17}$20.60$^{+1.50}_{-1.50}$\\
\textcolor{white}{1}&&&&&&&&\\
HD189733Ab b& $^{35}$1.130 & $^{6}$2.219 & $^{35}$0.820$^{+0.030}_{-0.030}$ & $^{6}$0.766$^{+0.007}_{-0.013}$ & $^{36}$5050$^{+50}_{-50}$  & $^{36}$-0.030$^{+0.040}_{-0.040}$   & $^{37}$-0.5$^{+0.4}_{-0.4}$ & $^{37}$3.10$^{+0.03}_{-0.03}$\\
\textcolor{white}{1}&&&&&&&&\\
HD209458 b& $^{38}$0.714 & $^{40}$3.525 & $^{38}$1.148$^{+0.040}_{-0.040}$ & $^{38}$1.162$^{+0.014}_{-0.014}$ & $^{39}$6117$^{+50}_{-50}$  & $^{39}$0.020$^{+0.050}_{-0.050}$   & $^{17}$-5.0$^{+7.0}_{-7.0}$ & $^{17}$4.40$^{+0.20}_{-0.20}$\\
\textcolor{white}{1}&&&&&&&&\\
Kepler-8 b& $^{41}$0.603 & $^{41}$3.522 & $^{41}$1.213$^{+0.067}_{-0.063}$ & $^{41}$1.486$^{+0.053}_{-0.062}$ & $^{41}$6213$^{+150}_{-150}$  & $^{41}$-0.055$^{+0.030}_{-0.030}$   & $^{17}$5.0$^{+7.0}_{-7.0}$ & $^{17}$8.90$^{+1.00}_{-1.00}$\\
\textcolor{white}{1}&&&&&&&&\\
Kepler-13 b& $^{44}$14.800 & $^{45}$1.764 & $^{42}$2.050$^{+0.000}_{-0.000}$ & $^{43}$1.756$^{+0.014}_{-0.014}$ & $^{42}$8500$^{+400}_{-400}$  & $^{42}$0.200$^{+0.000}_{-0.000}$   & $^{43}$23.0$^{+4.0}_{-4.0}$ & $^{42}$65.00$^{+10.00}_{-10.00}$\\
\textcolor{white}{1}&&&&&&&&\\
Kepler-17 b& $^{46}$2.470 & $^{47}$1.486 & $^{46}$1.160$^{+0.060}_{-0.060}$ & $^{46}$1.050$^{+0.030}_{-0.030}$ & $^{46}$5781$^{+85}_{-85}$  & $^{46}$0.260$^{+0.100}_{-0.100}$   & $^{47}$0.0$^{+15.0}_{-0.0}$ & $^{46}$6.00$^{+2.00}_{-2.00}$\\
\textcolor{white}{1}&&&&&&&&\\
TrES-1 b& $^{38}$0.761 & $^{40}$3.030 & $^{38}$0.892$^{+0.049}_{-0.049}$ & $^{38}$0.818$^{+0.021}_{-0.021}$ & $^{48}$5226$^{+38}_{-38}$  & $^{48}$0.060$^{+0.050}_{-0.050}$   & $^{49}$30.0$^{+21.0}_{-21.0}$ & $^{49}$1.30$^{+0.30}_{-0.30}$\\
\textcolor{white}{1}&&&&&&&&\\
TrES-2 b& $^{38}$1.253 & $^{38}$2.471 & $^{38}$1.049$^{+0.062}_{-0.062}$ & $^{38}$1.002$^{+0.031}_{-0.031}$ & $^{50}$5795$^{+73}_{-73}$  & $^{50}$0.060$^{+0.080}_{-0.080}$   & $^{51}$-9.0$^{+12.0}_{-12.0}$ & $^{51}$1.00$^{+0.60}_{-0.60}$\\
\textcolor{white}{1}&&&&&&&&\\
TrES-4 b& $^{52}$0.917 & $^{52}$3.554 & $^{52}$1.388$^{+0.042}_{-0.042}$ & $^{52}$1.798$^{+0.052}_{-0.052}$ & $^{53}$6200$^{+75}_{-75}$  & $^{53}$0.140$^{+0.090}_{-0.090}$   & $^{54}$6.3$^{+4.7}_{-4.7}$ & $^{52}$9.50$^{+1.00}_{-1.00}$\\
\textcolor{white}{1}&&&&&&&&\\
WASP-1 b& $^{38}$0.860 & $^{55}$2.520 & $^{38}$1.243$^{+0.036}_{-0.040}$ & $^{38}$1.445$^{+0.052}_{-0.079}$ & $^{55}$6213$^{+51}_{-51}$  & $^{55}$0.170$^{+0.050}_{-0.050}$   & $^{55}$-59.0$^{+99.0}_{-26.0}$ & $^{55}$0.70$^{+1.40}_{-0.50}$\\
\textcolor{white}{1}&&&&&&&&\\
WASP-3 b& $^{38}$2.060 & $^{38}$1.847 & $^{38}$1.260$^{+0.100}_{-0.100}$ & $^{38}$1.377$^{+0.085}_{-0.085}$ & $^{56}$6400$^{+100}_{-100}$  & $^{56}$0.000$^{+0.200}_{-0.200}$   & $^{57}$3.3$^{+2.5}_{-4.4}$ & $^{57}$14.10$^{+1.50}_{-1.30}$\\
\textcolor{white}{1}&&&&&&&&\\
WASP-4 b& $^{58}$1.237 & $^{60}$1.338 & $^{58}$0.925$^{+0.040}_{-0.040}$ & $^{58}$0.912$^{+0.013}_{-0.013}$ & $^{59}$5500$^{+100}_{-100}$  & $^{59}$-0.030$^{+0.090}_{-0.090}$   & $^{61}$-1.0$^{+14.0}_{-12.0}$ & $^{59}$2.00$^{+1.00}_{-1.00}$\\
\textcolor{white}{1}&&&&&&&&\\
WASP-5 b& $^{62}$1.555 & $^{63}$1.628 & $^{62}$1.000$^{+0.063}_{-0.064}$ & $^{62}$1.060$^{+0.076}_{-0.028}$ & $^{59}$5700$^{+100}_{-100}$  & $^{59}$0.090$^{+0.090}_{-0.090}$   & $^{62}$12.1$^{+10.0}_{-8.0}$ & $^{62}$3.24$^{+0.35}_{-0.27}$\\
\textcolor{white}{1}&&&&&&&&\\
WASP-6 b& $^{64}$0.503 & $^{64}$3.361 & $^{64}$0.880$^{+0.050}_{-0.080}$ & $^{64}$0.870$^{+0.025}_{-0.036}$ & $^{64}$5450$^{+100}_{-100}$  & $^{64}$-0.200$^{+0.090}_{-0.090}$   & $^{64}$-11.0$^{+14.0}_{-18.0}$ & $^{64}$1.60$^{+0.27}_{-0.17}$\\
\textcolor{white}{1}&&&&&&&&\\
WASP-7 b& $^{65}$0.960& $^{65}$4.955 & $^{65}$1.276$^{+0.065}_{-0.065}$ & $^{65}$1.432$^{+0.092}_{-0.092}$ & $^{66}$6400$^{+100}_{-100}$  & $^{66}$0.000$^{+0.100}_{-0.100}$   & $^{67}$86.0$^{+6.0}_{-6.0}$ & $^{67}$14.00$^{+2.00}_{-2.00}$\\
\textcolor{white}{1}&&&&&&&&\\
WASP-12 b& $^{52}$1.404 & $^{52}$1.091 & $^{52}$1.350$^{+0.140}_{-0.140}$ & $^{52}$1.599$^{+0.071}_{-0.071}$ & $^{52}$6300$^{+150}_{-150}$  & $^{52}$0.300$^{+0.100}_{-0.100}$   & $^{17}$59.0$^{+15.0}_{-20.0}$ & $^{52}$0.00$^{+2.20}_{-0.00}$\\
\textcolor{white}{1}&&&&&&&&\\
WASP-14 b& $^{68}$7.341 & $^{69}$2.244 & $^{68}$1.211$^{+0.127}_{-0.122}$ & $^{68}$1.306$^{+0.066}_{-0.073}$ & $^{68}$6475$^{+100}_{-100}$  & $^{68}$0.000$^{+0.200}_{-0.200}$   & $^{70}$-33.1$^{+7.4}_{-7.4}$ & $^{70}$2.80$^{+0.57}_{-0.57}$\\
\textcolor{white}{1}&&&&&&&&\\
WASP-15 b& $^{71}$0.592 & $^{71}$3.752 & $^{71}$1.305$^{+0.051}_{-0.051}$ & $^{71}$1.522$^{+0.044}_{-0.044}$ & $^{72}$6405$^{+80}_{-80}$  & $^{72}$0.000$^{+0.100}_{-0.100}$   & $^{62}$-139.6$^{+5.2}_{-4.3}$ & $^{72}$4.90$^{+0.40}_{-0.40}$\\
\textcolor{white}{1}&&&&&&&&\\
WASP-16 b& $^{71}$0.832 & $^{71}$3.119 & $^{71}$0.980$^{+0.054}_{-0.054}$ & $^{71}$1.087$^{+0.042}_{-0.042}$ & $^{72}$5630$^{+70}_{-70}$  & $^{72}$0.070$^{+0.100}_{-0.100}$   & $^{17}$11.0$^{+26.0}_{-19.0}$ & $^{72}$2.50$^{+0.40}_{-0.40}$\\
\textcolor{white}{1}&&&&&&&&\\
WASP-18 b& $^{73}$10.430 & $^{73}$0.941 & $^{73}$1.281$^{+0.069}_{-0.069}$ & $^{73}$1.230$^{+0.047}_{-0.047}$ & $^{74}$6400$^{+100}_{-100}$  & $^{74}$0.000$^{+0.090}_{-0.090}$   & $^{17}$13.0$^{+7.0}_{-7.0}$ & $^{17}$11.20$^{+0.60}_{-0.60}$\\
\textcolor{white}{1}&&&&&&&&\\
WASP-19 b& $^{75}$1.114 & $^{75}$0.789 & $^{75}$0.904$^{+0.045}_{-0.045}$ & $^{75}$1.004$^{+0.018}_{-0.018}$ & $^{77}$5440$^{+60}_{-60}$  & $^{76}$0.020$^{+0.090}_{-0.090}$   & $^{17}$15.0$^{+11.0}_{-11.0}$ & $^{75, III}$4.30$^{+0.15}_{-0.15}$\\
\textcolor{white}{1}&&&&&&&&\\
WASP-22 b& $^{78}$0.588 & $^{78}$3.533 & $^{78}$1.109$^{+0.026}_{-0.026}$ & $^{78}$1.219$^{+0.052}_{-0.033}$ & $^{79}$6000$^{+100}_{-100}$  & $^{79}$0.050$^{+0.080}_{-0.080}$   & $^{78}$22.0$^{+16.0}_{-16.0}$ & $^{78}$4.42$^{+0.34}_{-0.34}$\\
\textcolor{white}{1}&&&&&&&&\\
WASP-24 b& $^{80}$1.071 & $^{80}$2.341 & $^{80}$1.184$^{+0.027}_{-0.027}$ & $^{80}$1.331$^{+0.032}_{-0.032}$ & $^{80}$6075$^{+100}_{-100}$  & $^{80}$0.070$^{+0.100}_{-0.100}$   & $^{81}$-4.7$^{+4.0}_{-4.0}$ & $^{80}$7.00$^{+1.00}_{-1.00}$\\
\textcolor{white}{1}&&&&&&&&\\
WASP-25 b& $^{82}$0.580 & $^{82}$3.765 & $^{82}$1.000$^{+0.030}_{-0.030}$ & $^{82}$0.920$^{+0.040}_{-0.040}$ & $^{82}$5703$^{+100}_{-100}$  & $^{82}$-0.070$^{+0.100}_{-0.100}$   & $^{83}$14.6$^{+6.7}_{-6.7}$ & $^{83}$2.90$^{+0.30}_{-0.30}$\\
\textcolor{white}{1}&&&&&&&&\\
WASP-26 b& $^{84}$1.028 & $^{84}$2.757 & $^{84}$1.111$^{+0.028}_{-0.028}$ & $^{84}$1.303$^{+0.059}_{-0.059}$ & $^{84}$5939$^{+100}_{-100}$  & $^{85}$-0.020$^{+0.091}_{-0.091}$   & $^{17}$-34.0$^{+36.0}_{-26.0}$ & $^{85}$2.40$^{+1.30}_{-1.30}$\\
\textcolor{white}{1}&&&&&&&&\\
WASP-33 b& $^{86}<$4.590 & $^{86}$1.220 & $^{86}$1.512$^{+0.040}_{-0.040}$ & $^{86}$1.512$^{+0.060}_{-0.054}$ & $^{87}$7430$^{+100}_{-100}$  & $^{87}$0.100$^{+0.200}_{-0.200}$   & $^{87}$-107.7$^{+1.6}_{-1.6}$ & $^{87}$90.00$^{+10.00}_{-10.00}$\\
\textcolor{white}{1}&&&&&&&&\\
WASP-52 b& $^{88}$0.460 & $^{88}$1.750 & $^{88}$0.870$^{+0.030}_{-0.030}$ & $^{88}$0.790$^{+0.020}_{-0.020}$ & $^{88}$5000$^{+100}_{-100}$  & $^{88}$0.030$^{+0.120}_{-0.120}$   & $^{88}$24.0$^{+17.0}_{-9.0}$ & $^{88}$2.50$^{+1.00}_{-1.00}$\\
\textcolor{white}{1}&&&&&&&&\\
WASP-71 b& $^{89}$2.242 & $^{89}$2.904 & $^{89}$1.559$^{+0.070}_{-0.070}$ & $^{89}$2.260$^{+0.170}_{-0.170}$ & $^{89}$6059$^{+98}_{-98}$  & $^{89}$0.140$^{+0.080}_{-0.080}$   & $^{89}$20.1$^{+9.7}_{-9.7}$ & $^{89}$9.89$^{+0.48}_{-0.48}$\\
\textcolor{white}{1}&&&&&&&&\\
WASP-80 b& $^{90}$0.554 & $^{90}$3.068 & $^{90}$0.570$^{+0.050}_{-0.050}$ & $^{90}$0.571$^{+0.016}_{-0.016}$ & $^{90}$4145$^{+100}_{-100}$  & $^{90}$-0.140$^{+0.160}_{-0.160}$   & $^{90}|75.0|^{+4.0}_{-4.3}$ & $^{90}$3.46$^{+0.34}_{-0.35}$\\
\textcolor{white}{1}&&&&&&&&\\
XO-2 b& $^{91}$0.570 & $^{91}$2.616 & $^{91}$0.980$^{+0.020}_{-0.020}$ & $^{91}$0.970$^{+0.020}_{-0.010}$ & $^{91}$5340$^{+32}_{-32}$  & $^{91}$0.450$^{+0.020}_{-0.020}$   & $^{92}$10.0$^{+72.0}_{-72.0}$ & $^{92}$1.45$^{+2.73}_{-0.14}$\\
\textcolor{white}{1}&&&&&&&&\\
XO-3 b& $^{93}$11.790 & $^{93}$3.191 & $^{93}$1.213$^{+0.066}_{-0.066}$ & $^{93}$1.377$^{+0.083}_{-0.083}$ & $^{93}$6429$^{+100}_{-100}$  & $^{93}$-0.177$^{+0.080}_{-0.080}$   & $^{94}$37.3$^{+3.7}_{-3.7}$ & $^{93}$18.54$^{+0.17}_{-0.17}$\\
\textcolor{white}{1}&&&&&&&&\\
XO-4 b& $^{95}$1.720 & $^{96}$4.125 & $^{95}$1.320$^{+0.020}_{-0.020}$ & $^{95}$1.560$^{+0.050}_{-0.050}$ & $^{95}$6397$^{+70}_{-70}$  & $^{95}$-0.040$^{+0.030}_{-0.030}$   & $^{97}$-46.7$^{+8.1}_{-6.1}$ & $^{95}$8.80$^{+0.50}_{-0.50}$\\
\hline
\\[0.5em]
\smallskip
HAT-P-2 b& $^{98}$9.090 & $^{98}$5.633 & $^{98}$1.360$^{+0.040}_{-0.040}$ & $^{98}$1.640$^{+0.090}_{-0.080}$ & $^{98}$6290$^{+60}_{-60}$  & $^{98}$0.140$^{+0.080}_{-0.080}$   & $^{17}$9.0$^{+10.0}_{-10.0}$ & $^{17}$19.50$^{+1.40}_{-1.40}$\\
\textcolor{white}{1}&&&&&&&&\\
HAT-P-34 b& $^{99}$3.328 & $^{99}$5.453 & $^{99}$1.392$^{+0.047}_{-0.047}$ & $^{99}$1.535$^{+0.135}_{-0.102}$ & $^{99}$6442$^{+88}_{-88}$  & $^{99}$0.220$^{+0.040}_{-0.040}$   & $^{17}$0.0$^{+14.0}_{-14.0}$ & $^{17}$24.30$^{+1.20}_{-1.20}$\\
\textcolor{white}{1}&&&&&&&&\\
WASP-8 b& $^{100}$2.244 & $^{100}$8.159 & $^{100}$1.030$^{+0.054}_{-0.061}$ & $^{100}$0.945$^{+0.051}_{-0.036}$ & $^{100}$5600$^{+80}_{-80}$  & $^{100}$0.170$^{+0.070}_{-0.070}$   & $^{100}$-123.0$^{+4.4}_{-3.4}$ & $^{100}$1.59$^{+0.08}_{-0.09}$\\
\textcolor{white}{1}&&&&&&&&\\
WASP-17 b& $^{101}$0.486& $^{101}$3.735 & $^{101}$1.306$^{+0.026}_{-0.026}$ & $^{101}$1.572$^{+0.056}_{-0.056}$ & $^{101}$6650$^{+80}_{-80}$  & $^{101}$-0.190$^{+0.090}_{-0.090}$   & $^{101}$-148.7$^{+7.7}_{-6.7}$ & $^{101}$10.05$^{+0.88}_{-0.79}$\\
\textcolor{white}{1}&&&&&&&&\\
WASP-31 b& $^{102}$0.478 & $^{102}$3.406 & $^{102}$1.163$^{+0.026}_{-0.026}$ & $^{102}$1.252$^{+0.033}_{-0.033}$ & $^{102}$6302$^{+102}_{-102}$  & $^{102}$-0.200$^{+0.090}_{-0.090}$   & $^{17}$-6.0$^{+6.0}_{-6.0}$ & $^{17}$6.80$^{+0.60}_{-0.60}$\\
\textcolor{white}{1}&&&&&&&&\\
WASP-38 b& $^{103}$2.691 & $^{103}$6.872 & $^{103}$1.203$^{+0.036}_{-0.036}$ & $^{103}$1.331$^{+0.030}_{-0.025}$ & $^{103}$6150$^{+80}_{-80}$  & $^{103}$-0.120$^{+0.070}_{-0.070}$   & $^{81}$15.0$^{+33.0}_{-43.0}$ & $^{103}$8.60$^{+0.40}_{-0.40}$\\
\textcolor{white}{1}&&&&&&&&\\
\enddata
\tablecomments{See Table~\ref{Tab:ParamsDefinition} for an explanation of the various symbols. The systems were selected from \href{http://www.openexoplanetcatalogue.com}{\it The Open Exoplanet Catalogue} on 2013 August 31. We required an observationally inferred best-fit planet's mass and orbital period $M_{\rm pl}\,\textgreater\,0.5\,M_{\rm Jup}$ and $P_{\rm orb}\,\textless\,5\,$d, respectively The last six systems were taken from \cite{Albrecht+12} (see \S~\ref{The Sample}). For the precise planet's mass and orbital period, see the corresponding reference (or the ApJ version of this manuscript). Orbital periods are generally known to a precision of $10^{-5}\,d$ or better.}
\tablerefs{The references are taken from \href{http://exoplanet.eu/catalog/}{http://exoplanet.eu/} and are as follows;
$(1)$: \citealt{Barge2008};
$(2)$: \citealt{Pont+10};
$(3)$: \citealt{Alonso+08};
$(4)$: \citealt{Bouchy+08};
$(5)$: \citealt{Gillon+10};
$(6)$: \citealt{Triaud+09};
$(7)$: \citealt{Deleuil+08};
$(8)$: \citealt{Hebrard+11};
$(9)$: \citealt{Guenther+12};
$(10)$: \citealt{Torres+08};
$(11)$: \citealt{Johnson+08};
$(12)$: \citealt{Southworth+11};
$(13)$: \citealt{Kovacs+07};
$(14)$: \citealt{Sada+12};
$(15)$: \citealt{Winn+11};
$(16)$: \citealt{Noyes+08};
$(17)$: \citealt{Albrecht+12};
$(18)$: \citealt{VanEylen+12};
$(19)$: \citealt{VanEylen+13};
$(20)$: \citealt{Mancini+13};
$(21)$: \citealt{Latham+09};
$(22)$: \citealt{Knutson+10};
$(23)$: \citealt{Moutou+11};
$(24)$: \citealt{Shporer+09};
$(25)$: \citealt{Winn+10};
$(26)$: \citealt{Bakos+09};
$(27)$: \citealt{Simpson+11};
$(28)$: \citealt{Ciceri+13};
$(29)$: \citealt{Torres+12};
$(30)$: \citealt{Buchhave+10};
$(31)$: \citealt{Bakos+11};
$(32)$: \citealt{Kipping+10};
$(33)$: \citealt{Johnson+11};
$(34)$: \citealt{Hartman+11};
$(35)$: \citealt{Winn+06};
$(36)$: \citealt{Bouchy+05};
$(37)$: \citealt{CollierC+10b};
$(38)$: \citealt{Southworth+10};
$(39)$: \citealt{Santos+04};
$(40)$: \citealt{Southworth+08};
$(41)$: \citealt{Jenkins+10};
$(42)$: \citealt{Szabo+11};
$(43)$: \citealt{Barnes+11};
$(44)$: \citealt{Santerne+12};
$(45)$: \citealt{Shporer+11};
$(46)$: \citealt{Bonomo+12};
$(47)$: \citealt{Desert+11};
$(48)$: \citealt{Santos+06b};
$(49)$: \citealt{Narita+07};
$(50)$: \citealt{AmmlerVE+09};
$(51)$: \citealt{Winn+08};
$(52)$: \citealt{Chan+11};
$(53)$: \citealt{Sozzetti+09};
$(54)$: \citealt{Narita+10};
$(55)$: \citealt{Albrecht+11};
$(56)$: \citealt{Pollacco+08};
$(57)$: \citealt{Tripathi+10};
$(58)$: \citealt{Winn+09b};
$(59)$: \citealt{Gillon+09};
$(60)$: \citealt{Hoyer+13};
$(61)$: \citealt{SanchisO+11};
$(62)$: \citealt{Triaud+10};
$(63)$: \citealt{Hoyer+12};
$(64)$: \citealt{Gillon+09b};
$(65)$: \citealt{Southworth+11b};
$(66)$: \citealt{Hellier+09b};
$(67)$: \citealt{Albrecht+12b};
$(68)$: \citealt{Joshi+09};
$(69)$: \citealt{Blecic+11};
$(70)$: \citealt{Johnson+09};
$(71)$: \citealt{Southworth+13};
$(72)$: \citealt{Doyle+13};
$(73)$: \citealt{Southworth+09};
$(74)$: \citealt{Hellier+09};
$(75)$: \citealt{TR+13};
$(76)$: \citealt{Hebb+10};
$(77)$: \citealt{Maxted+11};
$(78)$: \citealt{Anderson+11};
$(79)$: \citealt{Maxted+10};
$(80)$: \citealt{Street+10};
$(81)$: \citealt{Simpson+11b};
$(82)$: \citealt{Enoch+11};
$(83)$: \citealt{Brown+12};
$(84)$: \citealt{Anders+11};
$(85)$: \citealt{Smalley+10};
$(86)$: \citealt{Smith+11};
$(87)$: \citealt{CollierC+10};
$(88)$: \citealt{Hebrard+13};
$(89)$: \citealt{Smith+13};
$(90)$: \citealt{Triaud+13};
$(91)$: \citealt{Burke+07};
$(92)$: \citealt{Narita+11};
$(93)$: \citealt{Winn+08b};
$(94)$: \citealt{Winn+09};
$(95)$: \citealt{McCullough+08};
$(96)$: \citealt{Todorov+12};
$(97)$: \citealt{Narita+10b};
$(98)$: \citealt{Pal+10};
$(99)$: \citealt{Bakos+12};
$(100)$: \citealt{Queloz+10};
$(101)$: \citealt{Anderson+11b};
$(102)$: \citealt{Anderson+11d};
$(103)$: \citealt{Barros+11}.}
\tablenotetext{I}{For HAT-P-7 we list the obliquity reported by \cite{Albrecht+12}, but independent measurements of its $\lambda$ yielded results which disagree significantly \citep{Winn+09c,Narita+09}. We note however that these measurements are all consistent with a retrograde orbit.}
\tablenotetext{II}{For HAT-P-32, the components masses and radii depend on the eccentricity of the system, which is poorly constrained due to the star's high-velocity jitter. Here we provide the constraints at the two different eccentricities quoted in \cite{Hartman+11}.}
\tablenotetext{III}{The rotational velocity for WASP-19 is the {\it true} equatorial velocity.}
\label{Table:hotJupiterSample}
\end{deluxetable}
\clearpage

\clearpage
\LongTables
\begin{deluxetable}{ccccccc}
\tablecaption{Host star properties from detailed modeling.}
\tablewidth{-10pt}
\tablehead{
\\[0.6em]
\colhead{Host Star Name} & \colhead{$M_{*}$} &\colhead{$R_{*}$} &
\colhead{$T_{\rm eff *}$} & \colhead{Fe/H} & \colhead{$\Delta M_{\rm CZ}/M_{*}$} & \colhead{$\Delta R_{\rm CZ}/R_{*}$}
\\
 & \colhead{$(M_\odot)$} &\colhead{$(R_\odot)$} &
\colhead{(K)} &  &  &  
}
\startdata
\\[0.6em]
\smallskip
CoRoT-1 & 1.026~$\pm$~0.069 & 1.110~$\pm$~0.050 & 6298~$\pm$~150 & -0.355~$\pm$~0.168 & (1.002~$\pm$~0.909)$\times\,10^{-3}$ & (1.282~$\pm$~0.366)$\times\,10^{-1}$ \\
\smallskip
CoRoT-2 & 0.972~$\pm$~0.058 & 0.902~$\pm$~0.018 & 5624~$\pm$~119 & 0.000~$\pm$~0.097 & (2.491~$\pm$~0.807)$\times\,10^{-2}$ & (2.705~$\pm$~0.235)$\times\,10^{-1}$ \\
\smallskip
CoRoT-3 & 1.374~$\pm$~0.040 & 1.542~$\pm$~0.080 & 6740~$\pm$~139 & -0.025~$\pm$~0.046 & (1.129~$\pm$~1.129)$\times\,10^{-3}$ & (3.594~$\pm$~2.107)$\times\,10^{-2}$ \\
\smallskip
CoRoT-18 & 0.891~$\pm$~0.089 & 1.000~$\pm$~0.130 & 5440~$\pm$~100 & -0.094~$\pm$~0.094 & (4.689~$\pm$~1.879)$\times\,10^{-2}$ & (3.250~$\pm$~0.538)$\times\,10^{-1}$ \\
\smallskip
CoRoT-19 & 1.209~$\pm$~0.048 & 1.650~$\pm$~0.040 & 6090~$\pm$~70 & -0.009~$\pm$~0.088 & (1.767~$\pm$~1.096)$\times\,10^{-3}$ & (1.733~$\pm$~0.249)$\times\,10^{-1}$ \\
\smallskip
HAT-P-1 & 1.132~$\pm$~0.073 & 1.135~$\pm$~0.048 & 5975~$\pm$~120 & 0.132~$\pm$~0.072 & (7.743~$\pm$~4.841)$\times\,10^{-3}$ & (2.162~$\pm$~0.370)$\times\,10^{-1}$ \\
\smallskip
HAT-P-4 & 1.275~$\pm$~0.073 & 1.637~$\pm$~0.079 & 5874~$\pm$~66 & 0.237~$\pm$~0.075 & (7.411~$\pm$~3.174)$\times\,10^{-3}$ & (2.376~$\pm$~0.256)$\times\,10^{-1}$ \\
\smallskip
HAT-P-6 & 1.294~$\pm$~0.055 & 1.460~$\pm$~0.060 & 6570~$\pm$~80 & -0.129~$\pm$~0.058 & (1.639~$\pm$~1.637)$\times\,10^{-3}$ & (6.315~$\pm$~1.625)$\times\,10^{-2}$ \\
\smallskip
HAT-P-7$^{a}$ & 1.352~$\pm$~0.010 & 1.904~$\pm$~0.009 & 6255~$\pm$~20 & 0.122~$\pm$~0.025 & (1.882~$\pm$~0.490)$\times\,10^{-4}$ & (1.185~$\pm$~0.058)$\times\,10^{-1}$ \\
\smallskip
HAT-P-8 & 1.198~$\pm$~0.066 & 1.475~$\pm$~0.034 & 6130~$\pm$~80 & 0.017~$\pm$~0.062 & (3.166~$\pm$~2.533)$\times\,10^{-3}$ & (1.665~$\pm$~0.251)$\times\,10^{-1}$ \\
\smallskip
HAT-P-9 & 1.272~$\pm$~0.108 & 1.320~$\pm$~0.070 & 6350~$\pm$~150 & 0.121~$\pm$~0.191 & (1.396~$\pm$~1.392)$\times\,10^{-2}$ & (1.280~$\pm$~0.487)$\times\,10^{-1}$ \\
\smallskip
HAT-P-13 & 1.142~$\pm$~0.020 & 1.559~$\pm$~0.080 & 5697~$\pm$~46 & 0.408~$\pm$~0.076 & (1.906~$\pm$~0.443)$\times\,10^{-2}$ & (2.948~$\pm$~0.162)$\times\,10^{-1}$ \\
\smallskip
HAT-P-14 & 1.327~$\pm$~0.003 & 1.445~$\pm$~0.012 & 6493~$\pm$~7 & 0.000~$\pm$~0.000 & (3.869~$\pm$~0.393)$\times\,10^{-5}$ & (7.926~$\pm$~0.176)$\times\,10^{-2}$ \\
\smallskip
HAT-P-16 & 1.192~$\pm$~0.028 & 1.158~$\pm$~0.025 & 6119~$\pm$~51 & 0.116~$\pm$~0.074 & (2.649~$\pm$~1.021)$\times\,10^{-3}$ & (1.742~$\pm$~0.153)$\times\,10^{-1}$ \\
\smallskip
HAT-P-23 & 1.129~$\pm$~0.031 & 1.203~$\pm$~0.074 & 5905~$\pm$~80 & 0.145~$\pm$~0.031 & (9.373~$\pm$~3.991)$\times\,10^{-3}$ & (2.376~$\pm$~0.264)$\times\,10^{-1}$ \\
\smallskip
HAT-P-24 & 1.191~$\pm$~0.038 & 1.316~$\pm$~0.067 & 6373~$\pm$~80 & -0.159~$\pm$~0.062 & (2.576~$\pm$~1.956)$\times\,10^{-4}$ & (1.076~$\pm$~0.196)$\times\,10^{-1}$ \\
\smallskip
HAT-P-30 & 1.255~$\pm$~0.025 & 1.240~$\pm$~0.026 & 6264~$\pm$~48 & 0.132~$\pm$~0.072 & (7.428~$\pm$~6.989)$\times\,10^{-3}$ & (1.499~$\pm$~0.268)$\times\,10^{-1}$ \\
\smallskip
HAT-P-32 & 1.162~$\pm$~0.038 & 1.219~$\pm$~0.016 & 6206~$\pm$~87 & -0.038~$\pm$~0.059 & (1.380~$\pm$~0.870)$\times\,10^{-3}$ & (1.514~$\pm$~0.237)$\times\,10^{-1}$ \\
\smallskip
 & 1.135~$\pm$~0.028 & 1.389~$\pm$~0.064 & 6062~$\pm$~26 & -0.111~$\pm$~0.014 & (2.324~$\pm$~0.450)$\times\,10^{-3}$ & (1.863~$\pm$~0.098)$\times\,10^{-1}$ \\
\smallskip
HD189733Ab & 0.822~$\pm$~0.028 & 0.763~$\pm$~0.010 & 5050~$\pm$~49 & -0.023~$\pm$~0.023 & (5.420~$\pm$~0.495)$\times\,10^{-2}$ & (3.017~$\pm$~0.106)$\times\,10^{-1}$ \\
\smallskip
HD209458 & 1.160~$\pm$~0.025 & 1.162~$\pm$~0.014 & 6117~$\pm$~50 & 0.019~$\pm$~0.041 & (2.544~$\pm$~0.911)$\times\,10^{-3}$ & (1.739~$\pm$~0.148)$\times\,10^{-1}$ \\
\smallskip
Kepler-8 & 1.215~$\pm$~0.063 & 1.482~$\pm$~0.057 & 6213~$\pm$~149 & -0.058~$\pm$~0.012 & (2.064~$\pm$~1.931)$\times\,10^{-3}$ & (1.453~$\pm$~0.409)$\times\,10^{-1}$ \\
\smallskip
Kepler-13$^{b}$ & 2.050~$\pm$~0.000 & 1.764~$\pm$~0.005 & 8766~$\pm$~7 & 0.200~$\pm$~0.000 & $\simeq$\,1.5$\,\times\,10^{-9}$ & (4.742~$\pm$~0.009)$\times\,10^{-3}$  \\
\smallskip
Kepler-17 & 1.137~$\pm$~0.033 & 1.050~$\pm$~0.030 & 5781~$\pm$~85 & 0.257~$\pm$~0.095 & (1.681~$\pm$~0.554)$\times\,10^{-2}$ & (2.519~$\pm$~0.161)$\times\,10^{-1}$ \\
\smallskip
TrES-1 & 0.885~$\pm$~0.040 & 0.818~$\pm$~0.021 & 5226~$\pm$~38 & 0.059~$\pm$~0.038 & (4.730~$\pm$~0.433)$\times\,10^{-2}$ & (3.008~$\pm$~0.123)$\times\,10^{-1}$ \\
\smallskip
TrES-2 & 1.049~$\pm$~0.061 & 1.002~$\pm$~0.031 & 5795~$\pm$~73 & 0.065~$\pm$~0.065 & (1.570~$\pm$~0.462)$\times\,10^{-2}$ & (2.531~$\pm$~0.223)$\times\,10^{-1}$ \\
\smallskip
TrES-4 & 1.387~$\pm$~0.041 & 1.798~$\pm$~0.052 & 6200~$\pm$~75 & 0.139~$\pm$~0.078 & (6.284~$\pm$~4.353)$\times\,10^{-4}$ & (1.358~$\pm$~0.216)$\times\,10^{-1}$ \\
\smallskip
WASP-1 & 1.259~$\pm$~0.015 & 1.419~$\pm$~0.053 & 6207~$\pm$~45 & 0.174~$\pm$~0.044 & (1.751~$\pm$~1.141)$\times\,10^{-3}$ & (1.484~$\pm$~0.127)$\times\,10^{-1}$ \\
\smallskip
WASP-3 & 1.262~$\pm$~0.098 & 1.377~$\pm$~0.085 & 6400~$\pm$~100 & 0.002~$\pm$~0.189 & (1.395~$\pm$~1.392)$\times\,10^{-2}$ & (1.211~$\pm$~0.459)$\times\,10^{-1}$ \\
\smallskip
WASP-4 & 0.927~$\pm$~0.038 & 0.912~$\pm$~0.013 & 5504~$\pm$~94 & -0.028~$\pm$~0.069 & (3.247~$\pm$~0.723)$\times\,10^{-2}$ & (2.889~$\pm$~0.185)$\times\,10^{-1}$ \\
\smallskip
WASP-5 & 0.998~$\pm$~0.061 & 1.084~$\pm$~0.052 & 5700~$\pm$~100 & 0.088~$\pm$~0.088 & (2.264~$\pm$~0.763)$\times\,10^{-2}$ & (2.827~$\pm$~0.278)$\times\,10^{-1}$ \\
\smallskip
WASP-6 & 0.867~$\pm$~0.063 & 0.864~$\pm$~0.030 & 5451~$\pm$~98 & -0.192~$\pm$~0.067 & (3.505~$\pm$~0.818)$\times\,10^{-2}$ & (2.920~$\pm$~0.261)$\times\,10^{-1}$ \\
\smallskip
WASP-7 & 1.277~$\pm$~0.063 & 1.432~$\pm$~0.092 & 6400~$\pm$~100 & 0.000~$\pm$~0.097 & (1.395~$\pm$~1.392)$\times\,10^{-2}$ & (1.229~$\pm$~0.491)$\times\,10^{-1}$ \\
\smallskip
WASP-12 & 1.402~$\pm$~0.088 & 1.599~$\pm$~0.071 & 6300~$\pm$~150 & 0.301~$\pm$~0.097 & (1.056~$\pm$~1.015)$\times\,10^{-3}$ & (1.192~$\pm$~0.386)$\times\,10^{-1}$ \\
\smallskip
WASP-14 & 1.259~$\pm$~0.076 & 1.302~$\pm$~0.069 & 6475~$\pm$~100 & -0.013~$\pm$~0.174 & (1.391~$\pm$~1.390)$\times\,10^{-2}$ & (1.113~$\pm$~0.464)$\times\,10^{-1}$ \\
\smallskip
WASP-15 & 1.306~$\pm$~0.048 & 1.522~$\pm$~0.044 & 6405~$\pm$~80 & 0.000~$\pm$~0.097 & (1.334~$\pm$~1.308)$\times\,10^{-3}$ & (9.486~$\pm$~2.035)$\times\,10^{-2}$ \\
\smallskip
WASP-16 & 0.973~$\pm$~0.046 & 1.087~$\pm$~0.042 & 5630~$\pm$~70 & 0.070~$\pm$~0.092 & (2.807~$\pm$~0.636)$\times\,10^{-2}$ & (2.995~$\pm$~0.203)$\times\,10^{-1}$ \\
\smallskip
WASP-18 & 1.250~$\pm$~0.035 & 1.239~$\pm$~0.038 & 6382~$\pm$~82 & 0.004~$\pm$~0.075 & (9.478~$\pm$~9.389)$\times\,10^{-3}$ & (1.311~$\pm$~0.408)$\times\,10^{-1}$ \\
\smallskip
WASP-19 & 0.885~$\pm$~0.023 & 1.004~$\pm$~0.018 & 5446~$\pm$~54 & 0.037~$\pm$~0.060 & (4.217~$\pm$~0.541)$\times\,10^{-2}$ & (3.237~$\pm$~0.119)$\times\,10^{-1}$ \\
\smallskip
WASP-22 & 1.109~$\pm$~0.026 & 1.228~$\pm$~0.042 & 5991~$\pm$~91 & 0.046~$\pm$~0.068 & (6.015~$\pm$~3.301)$\times\,10^{-3}$ & (2.123~$\pm$~0.297)$\times\,10^{-1}$ \\
\smallskip
WASP-24 & 1.183~$\pm$~0.026 & 1.331~$\pm$~0.032 & 6075~$\pm$~100 & 0.070~$\pm$~0.092 & (3.897~$\pm$~2.802)$\times\,10^{-3}$ & (1.857~$\pm$~0.316)$\times\,10^{-1}$ \\
\smallskip
WASP-25 & 1.002~$\pm$~0.028 & 0.920~$\pm$~0.040 & 5703~$\pm$~100 & -0.067~$\pm$~0.088 & (1.958~$\pm$~0.643)$\times\,10^{-2}$ & (2.580~$\pm$~0.207)$\times\,10^{-1}$ \\
\smallskip
WASP-26 & 1.108~$\pm$~0.025 & 1.303~$\pm$~0.059 & 5957~$\pm$~82 & -0.018~$\pm$~0.079 & (6.282~$\pm$~3.213)$\times\,10^{-3}$ & (2.224~$\pm$~0.277)$\times\,10^{-1}$ \\
\smallskip
WASP-33 & 1.520~$\pm$~0.030 & 1.515~$\pm$~0.057 & 7430~$\pm$~99 & -0.048~$\pm$~0.048 & (2.401~$\pm$~0.404)$\times\,10^{-9}$ & (5.039~$\pm$~0.323)$\times\,10^{-3}$ \\
\smallskip
WASP-52 & 0.867~$\pm$~0.023 & 0.790~$\pm$~0.020 & 5020~$\pm$~79 & 0.062~$\pm$~0.084 & (5.638~$\pm$~0.637)$\times\,10^{-2}$ & (3.051~$\pm$~0.114)$\times\,10^{-1}$ \\
\smallskip
WASP-71 & 1.541~$\pm$~0.050 & 2.207~$\pm$~0.117 & 6117~$\pm$~40 & 0.166~$\pm$~0.052 & (6.696~$\pm$~3.532)$\times\,10^{-4}$ & (1.428~$\pm$~0.159)$\times\,10^{-1}$ \\
\smallskip
WASP-80 & 0.572~$\pm$~0.048 & 0.571~$\pm$~0.016 & 4145~$\pm$~100 & -0.130~$\pm$~0.130 & (1.217~$\pm$~0.444)$\times\,10^{-1}$ & (3.884~$\pm$~0.884)$\times\,10^{-1}$ \\
\smallskip
XO-2$^c$ & 0.982~$\pm$~0.058 & 0.985~$\pm$~0.045 & 5405~$\pm$~31 & 0.451~$\pm$~0.019 & (4.256~$\pm$~0.391)$\times\,10^{-2}$ & (3.108~$\pm$~0.175)$\times\,10^{-1}$ \\
\smallskip
XO-3 & 1.214~$\pm$~0.060 & 1.377~$\pm$~0.083 & 6429~$\pm$~100 & -0.173~$\pm$~0.048 & (1.827~$\pm$~1.808)$\times\,10^{-3}$ & (9.406~$\pm$~2.421)$\times\,10^{-2}$ \\
\smallskip
XO-4 & 1.321~$\pm$~0.018 & 1.560~$\pm$~0.050 & 6397~$\pm$~68 & -0.034~$\pm$~0.012 & (1.227~$\pm$~1.196)$\times\,10^{-3}$ & (9.367~$\pm$~1.581)$\times\,10^{-2}$ \\
\hline
\\[0.6em]
\smallskip
HAT-P-2 & 1.361~$\pm$~0.038 & 1.645~$\pm$~0.085 & 6290~$\pm$~60 & 0.139~$\pm$~0.078 & (9.847~$\pm$~8.785)$\times\,10^{-4}$ & (1.179~$\pm$~0.176)$\times\,10^{-1}$ \\
\smallskip
HAT-P-34 & 1.392~$\pm$~0.043 & 1.551~$\pm$~0.118 & 6442~$\pm$~88 & 0.223~$\pm$~0.032 & (7.552~$\pm$~7.355)$\times\,10^{-4}$ & (9.009~$\pm$~2.135)$\times\,10^{-2}$ \\
\smallskip
WASP-8 & 1.025~$\pm$~0.055 & 0.952~$\pm$~0.043 & 5600~$\pm$~80 & 0.172~$\pm$~0.058 & (2.726~$\pm$~0.567)$\times\,10^{-2}$ & (2.805~$\pm$~0.177)$\times\,10^{-1}$ \\
\smallskip
\smallskip
WASP-17 & 1.306~$\pm$~0.023 & 1.572~$\pm$~0.056 & 6649~$\pm$~79 & -0.192~$\pm$~0.067 & (1.846~$\pm$~1.846)$\times\,10^{-3}$ & (4.370~$\pm$~1.467)$\times\,10^{-2}$ \\
\smallskip
WASP-31 & 1.162~$\pm$~0.023 & 1.252~$\pm$~0.033 & 6302~$\pm$~101 & -0.192~$\pm$~0.067 & (5.765~$\pm$~4.333)$\times\,10^{-4}$ & (1.263~$\pm$~0.242)$\times\,10^{-1}$ \\
\smallskip
WASP-38 & 1.178~$\pm$~0.010 & 1.334~$\pm$~0.026 & 6198~$\pm$~31 & -0.084~$\pm$~0.013 & (8.788~$\pm$~2.180)$\times\,10^{-4}$ & (1.481~$\pm$~0.078)$\times\,10^{-1}$ \\
\smallskip
\enddata
\tablecomments{Host star properties derived with MESA as described in \S~\ref{Detailed Stellar modeling with MESA}. See Table~\ref{Tab:ParamsDefinition} for an explanation of the various symbols. The uncertainties quoted are not 1$\sigma$ errors, but are computed as follows. For each system, we collect $Z$, $M_{*}$, $R_{*}$, $T_{\rm eff *}$, $\Delta M_{\rm CZ}/M_{*}$, and $\Delta R_{\rm CZ}/R_{*}$ from all the {\it successful models} (see \S~\ref{Detailed Stellar modeling with MESA}). Next, for each of these parameters we extract the maximum ($Max$) and minimum ($Min$) values. The mean value is given by $Mean = \frac{Max + Min}{2}$, while the uncertainty is given by $Max - Mean$.}
\tablenotetext{a}{For HAT-P-7 we arbitrarily assume a small error in $Fe/H$ of 0.05 to find matching evolutionary tracks in our grid of models.}
\tablenotetext{b}{Since for Kepler-13 the 1$\sigma$ errors on $M_{*}$ and $Z$ are not known, we computed one evolutionary track with the observationally inferred $M_{*}$ and $Z$.}
\tablenotetext{c}{For XO-2 we are able to match the observationally inferred $M_{*}$, $R_{*}$, and $T_{\rm eff *}$ only within 3$\sigma$.}
\label{Table:hostStarModeling}
\end{deluxetable}
\clearpage

\end{document}